\documentclass[sigconf]{acmart}
\AtBeginDocument{%
  }

\setcopyright{acmlicensed}
\copyrightyear{2026}
\acmYear{2026}
\acmDOI{XXXXXXX.XXXXXXX}
\acmConference[SIGMOD ’26]{ In Proceedings of ACM Management of Data}{May 31 - June 5, 2026}{Bengaluru, India}

\usepackage{tikz}
\usepackage{amsmath}

\usepackage{subfiles}

\usepackage{microtype}
\usepackage{url}

\usepackage{multirow}
\usepackage{booktabs}
\usepackage{graphicx}
\usepackage{xspace}
\usepackage[linesnumbered,ruled,vlined]{algorithm2e}
\usepackage[noend]{algpseudocode}
\usepackage{balance}
\usepackage{mathtools}
\usepackage{amsfonts}
\usepackage{amsthm}
\usepackage{threeparttable}
\usepackage{xcolor}
\usepackage{colortbl}
\usepackage{enumitem,kantlipsum}
\usepackage{siunitx}
\usepackage{xparse}
\usepackage[compact]{titlesec}
\usepackage[nosort]{cleveref} 
\usepackage{subcaption}
\usepackage{multicol}
\usepackage{wrapfig}
\usepackage{pifont}%
\usepackage[skip=2pt]{caption}
\usepackage{ulem}

\NewDocumentCommand{\var}{O{s} m O{}}{%
  \ensuremath{#1_{#2}^{#3}}
}

\newcommand{\commentout}[1]{}

\definecolor{light-gray}{gray}{0.80}

\newcommand\fref{Fig.~\ref}

\newcommand\sref{\S~\ref}

\newtheorem{mytheorem}{Theorem}
\newtheorem{assumption}[mytheorem]{Assumption}
\newtheorem{lemma}[mytheorem]{Lemma}
\newtheorem{remk}[mytheorem]{Remark}
\newtheorem{definition}{Definition}

\newcommand{\minjia}[1]{\grumbler{Minjia}{#1}}
\newcommand{\jingyi}[1]{\grumbler{Jingyi}{#1}}
\newcommand{\chenghao}[1]{\grumbler{Chenghao}{#1}}
\newcommand{\ben}[1]{\grumbler{Ben}{#1}}

\newcommand{\originalgrumbler}[2]{\begin{quote}\textcolor{blue}{\sl{\bf #1 says:} #2}\end{quote}}
\newcommand{\grumbler}[2]{\originalgrumbler{#1}{#2}}
\newcommand{\hide}[1]{}

\newcommand{\name}{VecFlow\xspace}
\newcommand{\highspecificity}{high-specificity\xspace}
\newcommand{\lowspecificity}{low-specificity\xspace}
\newcommand{\hs}{HS\xspace}
\newcommand{\ls}{LS\xspace}

\newcommand{\specificityivf}{Label-Centric IVF\xspace}

\newcommand{\ivfgraph}{IVF-Graph\xspace}
\newcommand{\ivfbfs}{IVF-BFS\xspace}

\newcommand{\ivftwo}{IVF\textsuperscript{2}\xspace}

\newcommand{\revisionA}[1]{#1}
\newcommand{\revisionB}[1]{#1}
\newcommand{\revisionC}[1]{#1}

\newcommand{\revision}[1]{{\color{blue} #1}}

\renewcommand{\emph}[1]{\textit{#1}}
\renewcommand{\grumbler}[2]{}
\renewcommand{\href}[2]{}

\newcommand{\outline}[1]{\grumbler{outline}{#1}}
\renewcommand{\outline}[1]{}

\usepackage[skip=2pt]{caption}
\usepackage{setspace}
\usepackage{enumitem}

\setlist{nosep} 
\newcommand{\setvspace}[2]{%
  #1 = #2
  \advance #1 by -1\parskip}

\makeatletter 
\def\thm@space@setup{%
  \thm@preskip=3pt
  \thm@postskip=\thm@preskip 
}
\makeatother

\makeatletter
\g@addto@macro\normalsize{%
  \setlength\abovedisplayskip{1pt}
  \setlength\belowdisplayskip{1pt}
  \setlength\abovedisplayshortskip{1pt}
  \setlength\belowdisplayshortskip{1pt}
}
\makeatother

\makeatletter 
\newenvironment{proof}[1][\proofname]{\par
  \vspace{1pt}
  \pushQED{\qed}%
  \normalfont
  \topsep0pt \partopsep0pt 
  \trivlist
  \item[\hskip\labelsep
        \itshape
    #1\@addpunct{.}]\ignorespaces
}{%
  \popQED\endtrivlist\@endpefalse
  \addvspace{3pt plus 3pt} 
}
\makeatother

\begin{document}

\date{}

\title{\name: A High-Performance Vector Data Management System for Filtered-Search on GPUs}

\author{Jingyi Xi}
\authornote{Both authors contributed equally to this research.}
\authornote{Work done while intern at UIUC.}
\author{Chenghao Mo}
\authornotemark[1]
\affiliation{%
  \institution{SSAIL Lab, UIUC}
  \country{USA}
}
\email{flotherxi@gmail.com}
\email{cmo8@illinois.edu}

\author{Benjamin Karsin}
\affiliation{%
  \institution{NVIDIA}
  \country{USA}
}
\email{bkarsin@nvidia.com}

\author{Artem Chirkin}
\affiliation{%
  \institution{NVIDIA}
  \country{Switzerland}
}
\email{achirkin@nvidia.com}

\author{Mingqin Li}
\affiliation{%
 \institution{Microsoft}
 \country{USA}
}
\email{mingqli@microsoft.com}

\author{Minjia Zhang}
\affiliation{%
  \institution{SSAIL Lab, UIUC}
  \country{USA}
}
\email{minjiaz@illinois.edu}


\begin{abstract}
Vector search and database systems have become a keystone component in many AI applications. While many prior research has investigated how to accelerate the performance of generic vector search, emerging AI applications require running more sophisticated vector queries efficiently, such as vector search with attribute filters. Unfortunately, recent filtered-ANNS solutions are primarily designed for CPUs, with few exploration and limited performance of filtered-ANNS that take advantage of the massive parallelism offered by GPUs.
In this paper, we present \name, a novel high-performance vector \revisionB{filtered search system} that achieves unprecedented high throughput and recall while obtaining low latency for filtered-ANNS on GPUs. We propose a novel label-centric indexing and search algorithm that significantly improves the selectivity of ANNS with filters. In addition to algorithmic level optimization, we provide architecture-aware optimizations for \name's functional modules, 
effectively supporting both small batch and large batch queries, and single-label and multi-label query processing. 
Experimental results on NVIDIA A100 GPU over several public available datasets validate that \name achieves 5 million QPS for recall 90\%, outperforming state-of-the-art CPU-based solutions such as Filtered-DiskANN by up to 135 times. Alternatively, \name can easily extend its support to high recall 99\% regime, whereas strong GPU-based baselines plateau at around 80\% recall. The source code is available at \url{https://github.com/Supercomputing-System-AI-Lab/VecFlow}.
\end{abstract}





\maketitle

\section{Introduction}
\label{sec:intro}

\minjia{Format: I created \sref{} and \fref{}. You can directly use them  to reference sections and figures, so we can easily switch to different style later on. Therefore, please do not use "Figure \fref{fig:build-time}" as that will create duplicated references.}

Dense vectors have become a key form of data representation in the era of AI. Deep learning models encode various entities, such as text~\cite{word2vec,roberta}, images~\cite{clip,blip}, and code snippets~\cite{code-embedding}, into dense continuous vectors, enabling a growing number of vector-based online applications, including image search~\cite{billion-scale-search-on-gpus}, web search~\cite{dssm,multi-field-neural-ranking},
question and answering~\cite{dl-for-qa}, ad-hoc retrieval~\cite{learn-to-match,nrm-weak-supervision,dl-matching-model}, mobile search~\cite{mobile-search}, and product search~\cite{product-search}. More recently, vector search has also become an integral component for building generative AI systems, such as retrieval-augmented generation for large language model (LLM)-based applications~\cite{long-chain,rag}. 
To provide an interactive user experience at scale, these applications require high-performance deployment to achieve high throughput and recall with low query latency. As a result, there have been numerous studies on optimizing vector search performance by developing compute and memory-efficient approximate nearest neighbor search (ANNS) algorithms~\cite{kd-tree,r-star-tree,flann,lsh,product-quantization, opq, cartesian-kmeans,lopq,hnsw,spread-out-graph}. 

Despite many advancements, prior work often focuses on optimizing \emph{unconstrained top-K retrieval}, where the algorithm searches a vector dataset for the K closest vectors based purely on their geometric distance from the query. However, real-world and emerging applications often require finding Top-K results under certain \emph{filtering} or \emph{label} constraints. For example, Google Multisearch~\cite{google-multisearch} allows users to search images with additional text hints, advertising engine apply filters to display region-relevant ads, and enterprise search places access control to displayable documents based on users' permission levels. 

One straightforward approach to enable ANNS with filters is to combine ANNS with \emph{post-processing},
where one first runs a standard vector search over the entire dataset to obtain a list of candidates and then applies a filter operator on the given label to select the top-K elements with the given filter. This approach allows one to leverage existing high-performance solutions developed for unconstrained top-K ANNS. However, it is very challenging to predict exactly how many returned candidates will pass the filter operation. Therefore, the number of final candidates that match the query's labels can be much less than K, or in the worst-case scenario, none at all. 

In practice, post-processing methods perform poorly. As such, academia and industry companies such as Milvus~\cite{milvus}, 
Weaviate~\cite{weaviate} and Pinecone~\cite{pinecone} that offer ANNS-as-a-service have explored alternative methods to support ANNS with filters. 
Alternatively, Filtered-DiskANN~\cite{filtered-diskann} proposes building a graph index with enhanced navigability for filtered searches and achieves state-of-the-art results with tens of thousands of query-per-second (QPS) at $>$0.9 recall. One gap in this approach is that  Filtered-DiskANN~\cite{filtered-diskann} assume that a query only has one label while, in practice, queries can contain multiple labels, involving \emph{OR} and \emph{AND} operations. This is an important use case, so new filtered ANNS approaches should consider multiple query labels in their design.

ANNS index construction and search are computationally expensive operations, so modern GPU architectures offer an opportunity to significantly improve performance. 
Researchers have developed several GPU-based vector search algorithms for unconstrained top-K ANNS~\cite{cagra,bang,rummy}. The state-of-the-art GPU-based vector search system from 
NVidia, CAGRA~\cite{cagra},
builds GPU-friendly graph indices to accelerate the vector search on GPUs, offering over an order of magnitude higher QPS than CPU-based solutions.  
Despite achieving high ANNS speed on GPUs, few studies have explored high-performance design and implementations optimized for GPU-based filtered ANNS, due to several challenges. First, the hardware architecture of GPUs is very different from CPU. While the GPU has a much higher parallelism and memory bandwidth than CPU, one has to redesign the parallelism strategy for an algorithm to map it to the GPU architecture to extract sufficient parallelism to fully utilize the hardware resources. 
Second, directly applying filter-aware partitioning algorithms leads to huge memory redundancy, yet GPU global memory is a scarce resource. Therefore, it is important to optimize the memory efficiency of GPU-based filtered-ANNS. 
Third, existing GPU-based ANNS algorithms \revisionC{primarily} focus on maximizing throughput when the batch size is large. However, in online serving scenarios, queries are often coming in through small batch sizes, making it difficult to achieve high GPU utilization while maintaining a low query latency. 

\begin{figure}[t]
    \centering
    \includegraphics[width=\linewidth]{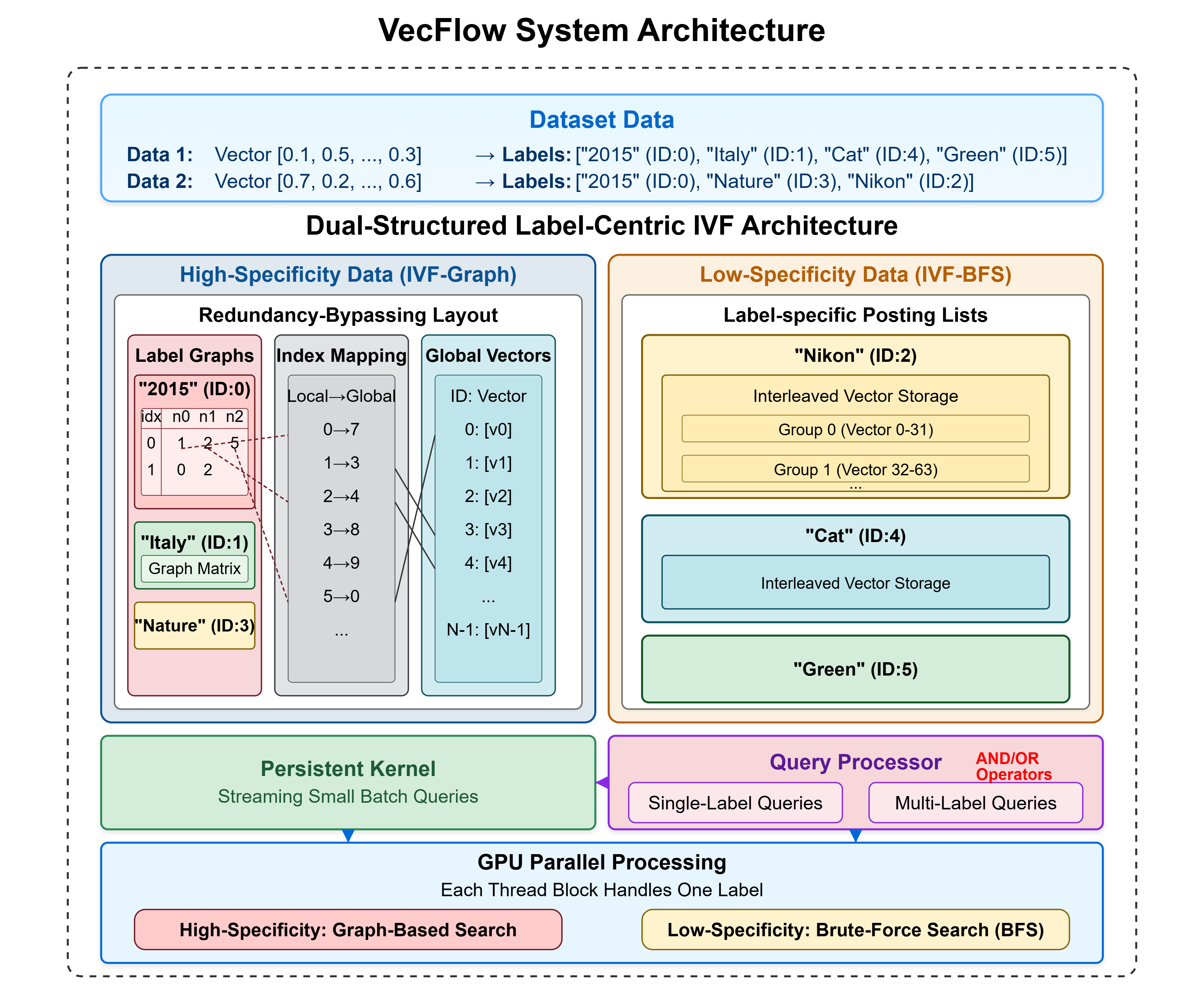}
    \caption{\revisionA{System overview of \name. The architecture shows how vectors with labels are managed through \name's Dual-Structured Label-Centric IVF design. High-specificity data are handled via graph-based indexing with a redundancy-bypassing layout (IVF-Graph), while low-specificity data use an optimized brute-force index (IVF-BFS). The system includes GPU-tailored optimizations such as GPU-centric parallel search, persistent kernels for streaming search of small batch sizes, supports for both efficient single-label and multi-label queries with AND/OR logic.}}
    \label{fig:overview}
    \minjia{This is much better. However, it is not very clear how the design is connected to GPUs. The only thing that indicates \name is a GPU-based solution is the small GPU icon. Can you give some thoughts if you can add the GPU architecture information (e.g., CTA, threads) in it?\jingyi{I think persistent kernel deserves an expansion in this Figure, let's first finish the persistent kernel Figure. Additionally, add a little GPU-related details for both IVF-Graph and IVF-BFS.}\minjia{Agreed. Who is working on the figure for persistent kernels?}\ben{Agreed that more emphasis can be put on the GPU side of things. Maybe something that completely separates what is run on GPU vs. CPU wit different colored background? Is the persistent kernel figure going to be here in intro as well?}]}
    
    \jingyi{TODO: separate color for algorithm and GPU.}
    
    \jingyi{TODO: integrate like thumbnails of IVF\_Graph, IVF\_BFS and persitent kernel into this figure. more about GPU things.}
\end{figure}

In this paper, we investigate the following research question: \emph{How to build a high-performance system for filtered ANNS that can effectively leverage GPU hardware?}
To address the aforementioned challenges, 
we present \name, a high-performance vector filtered-search system for GPUs. \revisionA{\fref{fig:overview} shows the overview of \name's design.} 
Different from recent graph-based solutions~\cite{filtered-diskann,cagra}, \revisionA{\name\ adopts a label-centric IVF (inverted file indexing) design to improve the selectivity of ANNS with filters, where data points are grouped based on shared labels instead of spatial proximity or distance.} \revisionA{\name\ partitions the IVF posting lists into high-specificity and low-specificity groups, each characterized by distinct label distribution patterns. \sref{subsubsec:specificity-def} provides detailed definition of specificity, which is a measure of how uniquely a label identifies data points. We design a dual-structured index along with tailored search algorithms for the two groups, maximizing GPU utilization to enable fast and accurate filtered-ANNS (\sref{subsec:dual-structure}).
}  
In addition, \name designs high-performance GPU kernels to achieve 
\revisionA{(1) redundancy-bypassing IVF-Graph (\sref{subsubsec:ivf-graph}) that reduces memory usage by avoiding redundant vector replication, (2) interleaved-scan-based IVF-BFS (\sref{subsubsec:ivf-bfs}) that maximizes GPU memory bandwidth efficiency for brute-force distance computations, and (3) persistent kernels for streaming small batches (\sref{subsubsec:persistent-kernel}) that eliminate kernel launch overhead and maintain high GPU utilization. Together, these optimizations deliver unprecedented performance for filtered ANNS.
Finally, unlike prior work that assumes each query only has one label, \name introduces a high-performance algorithm to efficiently tackle queries with multiple labels with \emph{OR} and \emph{AND} conditions (\sref{subsec:multi-label}).} 

Our experiments show that \name achieves significantly speedups over state-of-the-art filtered-ANNS solutions on both CPU and GPU over both \revisionA{real-world} and semi-synthetic filtered search datasets \revisionA{(\sref{sec:evaluation})}. In particular, \name achieves 5 million QPS for recall 90\%, outperforming state-of-the-art CPU-based solutions such as Filtered-DiskANN by up to 135 times. \name can easily extend its support to reach the high recall regime ($>$99\%), whereas strong GPU-based baselines plateau at around 80\% recall for these use cases. 

\minjia{TODO: Double check the correctness of the last statement.}
\minjia{TODO: Consider to add the GH200 results, given its high QPS.}

\section{Background}
\label{sec:background}

\subsection{Filtered ANNS}

\begin{figure}[t]
    \centering
    \includegraphics[width=1.0\linewidth]{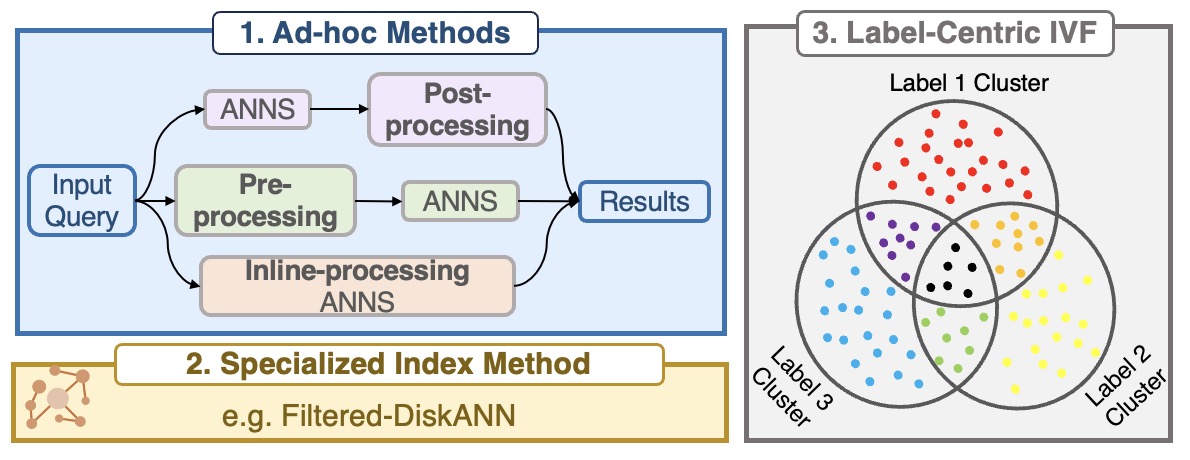}
    \caption{Three common approaches to enable filtered-ANNS.}
    \label{fig:common-methods}
    \minjia{Minor: The figure overall looks good. On a second thought, 2 and 3 are not at the same level as 1, e.g., 2 can also be considered as inline-processing + ANNS (what you said in the comment), and 3 can be considered as preprocessing + ANNS(?). Another option: having 1 at the top, which serves as an overview of how different filtered-ANNS methods can be built, and 2 and 3 are at the bottom (left and right). For 2, you can consider if it is possible to draw a figure that shows graph traversal + inline filtering.\jingyi{good point, 1 can include 2 and 3. I will try redraw, also, I will try to draw graph traversal + inline filtering later.}}
    \jingyi{I find graph traversal + inline filtering is not easy to draw.}
    \jingyi{we need have some text associate with this figure.}
\end{figure}

As one of the core components for data management, there are numerous algorithms for ANNS with diverse index construction methods and a range of trade-offs concerning indexing time, search time, and recall~\cite{kd-tree,r-star-tree,flann,lsh,hnsw,spread-out-graph,product-quantization, opq, cartesian-kmeans,lopq}. Several work have done an excellent survey and comparison of ANNS algorithms~\cite{ann-survey,li2020approximate,wang2021comprehensive,vector-search-survey}. However, most existing research addresses the unconstrained ANNS problem from the perspective of improving search efficiency~\cite{hnsw,spread-out-graph}, reducing the memory and compute costs~\cite{product-quantization,hm-ann}, enabling fresh updates to the index~\cite{fresh-diskann}. With the increasingly central role of ANNS in dense retrievers, new capabilities are required for ANNS, including \emph{search with filters} (filtered-ANNS). 

\revisionA{
    Filtered-ANNS extends ANNS by allowing users to search for similar vectors that also satisfy specific metadata constraints, combining both vector similarity and attribute-based filtering to deliver more precise and contextually relevant results.
}
\revisionA{
    In filtered-ANNS, each data point contains both a vector embedding and metadata labels. As shown in \fref{fig:overview}, the data point (a vector, e.g., [0.1, 0.5, ..., 0.3]) is associated with a list of labels (e.g., ["2015" (ID:0), "Italy" (ID:1), "Cat" (ID:4), "Green" (ID:5)]). During search, the system combines vector similarity with label constraints, e.g., "Find images visually similar to this reference image, but only consider those taken in `Italy' in `2015'." Here, vector similarity determines visual resemblance, while labels "2015" and "Italy" constrain the search space to return only results matching both labels.


}

There has been several recent works on filtered-ANNS, such as vbase~\cite{vbase}, NHQ~\cite{nhq}, \ivftwo~\cite{ivf2}, and Filtered-DiskANN~\cite{filtered-diskann}.
\revisionB{
We categorize existing approaches to filtered-ANNS into three major types, as illustrated in \fref{fig:common-methods}.
Ad-hoc methods integrate filtering through pre-/post-processing or inline filtering on top of existing ANNS systems. For example, FAISS-IVF~\cite{faiss} supports inline filtering by associating metadata with index entries to skip mismatched points during traversal.
Weaviate~\cite{weaviate} and Milvus~\cite{milvus} support inline-processing by maintaining inverted file indexing (IVF) where each label constraint has an "approved list" containing data IDs that fulfill the label constraints. 
Systems like pgvector~\cite{pgvector}, PASE~\cite{pase}, Analytic DB-V~\cite{analytic-db-v} and SingleStore-V~\cite{singlestore-v} use similar mechanism on top of general-purpose ANNS and build unified SQL and vector search backends. However, all ad-hoc methods just modify the search process, do not modify the index building.
} 

\revisionB{Filtered-DiskANN~\cite{filtered-diskann} builds a specialized filtered-ANNS index.}
It achieves excellent results by building filtered search on top of Vamana graph-based ANNS solution DiskANN~\cite{diskann}. In particular, Filtered-DiskANN builds separate proximity graph for data points associated to each label and stitches multiple small graphs together to eliminate redundant edges. By leveraging the geometric relation between points and incorporating label information during index construction, Filtered-DiskANN achieves much better search efficiency than prior filtered-ANNS methods based on IVF/LSH indices. 
More recently, \ivftwo introduces a label-centric IVF-based (inverted file indexing) method that wins the NeurIPS'23 Big-ANN Competition Filter Track\cite{bigann}.
Unlike these existing filtered-ANNS solutions, ours is the first to explore achieving optimal filtered-ANN performance on GPUs. Furthermore, different from prior work that only considers simple filters, such as exact matches with one filter in Filtered-DiskANN, we address more complex filtering conditions, including efficient search with the disjunction (OR) and conjunction (AND) of multiple filters, which have been less explored in previous research.




\subsection{GPU-based Vector Search}

The high computational throughput of modern GPU architectures has made it a key resource to solve computationally difficult problems~\cite{gupta2021introduction}. In recent years, increasingly computationally difficult operations, such as ANNS, have become commonplace in database systems, necessitating hardware acceleration~\cite{fang2020memory,sharma2024comprehensive}. Early work to efficiently accelerate ANNS using GPUs focused on developing fast k-selection algorithms based on product quantization~\cite{billion-scale-search-on-gpus,vlq-adc}. Subsequently, several studies proposed to build graph-based search algorithm on GPUs~\cite{song,ganns,ggnn,bang,rummy}. For example, GGNN builds hierarchical kNN graphs on GPUs~\cite{ggnn}. BANG~\cite{bang} employs a heterogeneous CPU-GPU vector search solution, using compressed data for distance computations on GPUs while maintaining the graph and actual data points on the CPU. Although these methods achieve promising speedups, they primarily focus on adapting or optimizing GPU resource utilization over existing CPU-designed graphs, without fully leveraging the GPU's capabilities.  
More recently, Ootomo et al.~\cite{cagra} proposed CAGRA, an enhanced GPU algorithm for solving k-ANNS. CAGRA designs hardware-friendly proximity graph structures to fully leverage the compute and memory bandwidth of modern GPU, outperforming both well-known CPU-based ANNS such as HNSW~\cite{hnsw} and state-of-the-art GPU-based ANNS, such as GGNN~\cite{ggnn} and GANNS~\cite{ganns}, achieving the state-of-the-art results for k-ANNS. 
Different from those work, \name focuses on solving the filtered-ANNS problem on GPUs. In addition, although \name leverages CAGRA graphs, it is fundamentally different. First, CAGRA can only perform k-ANNS without filters. Although CAGRA can be combined with post-processing to support filtered-ANNS, the performance is quite sub-optimal. 
In contrast, \name's novel index and search algorithms are specifically designed to enable significantly higher QPS and recall on filtered-ANNS. Second, while CAGRA achieves high throughput with very large batch sizes, \name introduces a persistent kernel-based method to achieve filtered-ANNS high throughput in streaming scenarios with small incoming query batch sizes.

\section{Problem Formulation}
\label{sec:problem}

\begin{definition}[\textbf{Filtered Nearest Neighbor Search (Filtered-NNS)}]\label{def:filtered-NNS}
    Given a finite data point set $X$ of $N$ points in a vector space $\mathcal{R}^D$, each data point (a.k.a. vector) $x \in X$ has an associated set of labels $L_x \subseteq \mathcal{L}$, where $\mathcal{L}$ is a finite set of labels. The goal of filtered-NNS is to answer a given query point $q \in \mathcal{R}^D$ with labels $L_q \subseteq \mathcal{L}$ by finding the closest point $x \in X$ with $L_q$, i.e., points in $X$ that have the label $L_q$ associated with them.
\end{definition}

Note that the query point $q$ may or may not be in the point set $X$, and the above definition generalizes naturally to \emph{Top-K Filtered-NNS}, i.e., finding {A$_{topk}$ = \{x| top-K nearest points to q in $X$\}}. Given that $N$ is often quite large and the service level objective (SLO) is stringent, we define an approximated form of the above problem. 

\begin{definition}[\textbf{Filtered Approximated Nearest Neighbor Search} (Filtered-ANNS)]\label{def:filtered-ANNS}
   The goal of $\epsilon$-filtered-NNS is to design an algorithm that answers a given query point $q \in \mathcal{R}^D$ with labels $L_q \subseteq \mathcal{L}$ by maximizing the recall of finding the top-k closest points $A_{topk} \in X$ with $L_q$ while minimizing the search time spent to return $A_{topk}$. 
\end{definition}

Assume the ground truth top-$K$ nearest neighbors of $q$ is $GT_{topK}$, the \emph{recall} is
defined as follows:
\begin{align} \label{formula:recall}
    Recall = \frac{\left | A_{topK} \cap GT_{topK} \right |}{\left | GT_{topK} \right |} = \frac{\left | A_{topK} \cap GT_{topK} \right |}{K}
\end{align}

    

\section{Design and Implementation of \name}
\label{sec:design}


\revisionA{
    As described in \sref{sec:intro}, the core of \name is the Dual-Structured Label-Centric IVF index and search algorithm, which processes data differently based on label specificity.  This design enables \name to flexibly and efficiently handle the full spectrum of label distributions. 
    The remainder of this section details the key design components of \name: 
    \sref{subsec:dual-structure} presents the dual-structured IVF approach and motivates the separation of high- and low-specificity labels.
    \sref{subsec:gpu-index-search} describes our GPU-optimized indexing and search kernel design, including redundancy-bypassing data layout for IVF-Graph, interleaved memory organization for IVF-BFS and techniques for streaming small-batch queries via persistent kernels.
    \sref{subsec:multi-label} extends the design to multi-label queries with support for both AND and OR operations utilizing parallel GPU search.
}

\subsection{Dual-Structured \specificityivf}
\label{subsec:dual-structure}

We start by considering two popular approaches for ANNS with filters, namely the graph traversal-based approach such as Filtered-DiskANN~\cite{filtered-diskann}, and the clustering-based method employed in Milvus~\cite{milvus} and Weaviate~\cite{weaviate}, which relies on inverted file indexing (IVF)~\cite{inverted-multi-index}. 
Filtered-DiskANN proposes to exploit both geometric relationship between data points and the labels that each point has to construct a navigable graph index and perform graph traversal with inline-filtering, which achieves excellent performance in Filtered-ANNS and outperforms filtered-ANNS methods based on IVF/LSH indices on CPU.

\begin{figure}[t]
    \centering
    \begin{minipage}
        {0.5\linewidth}
        \centering
        \includegraphics[width=\linewidth]{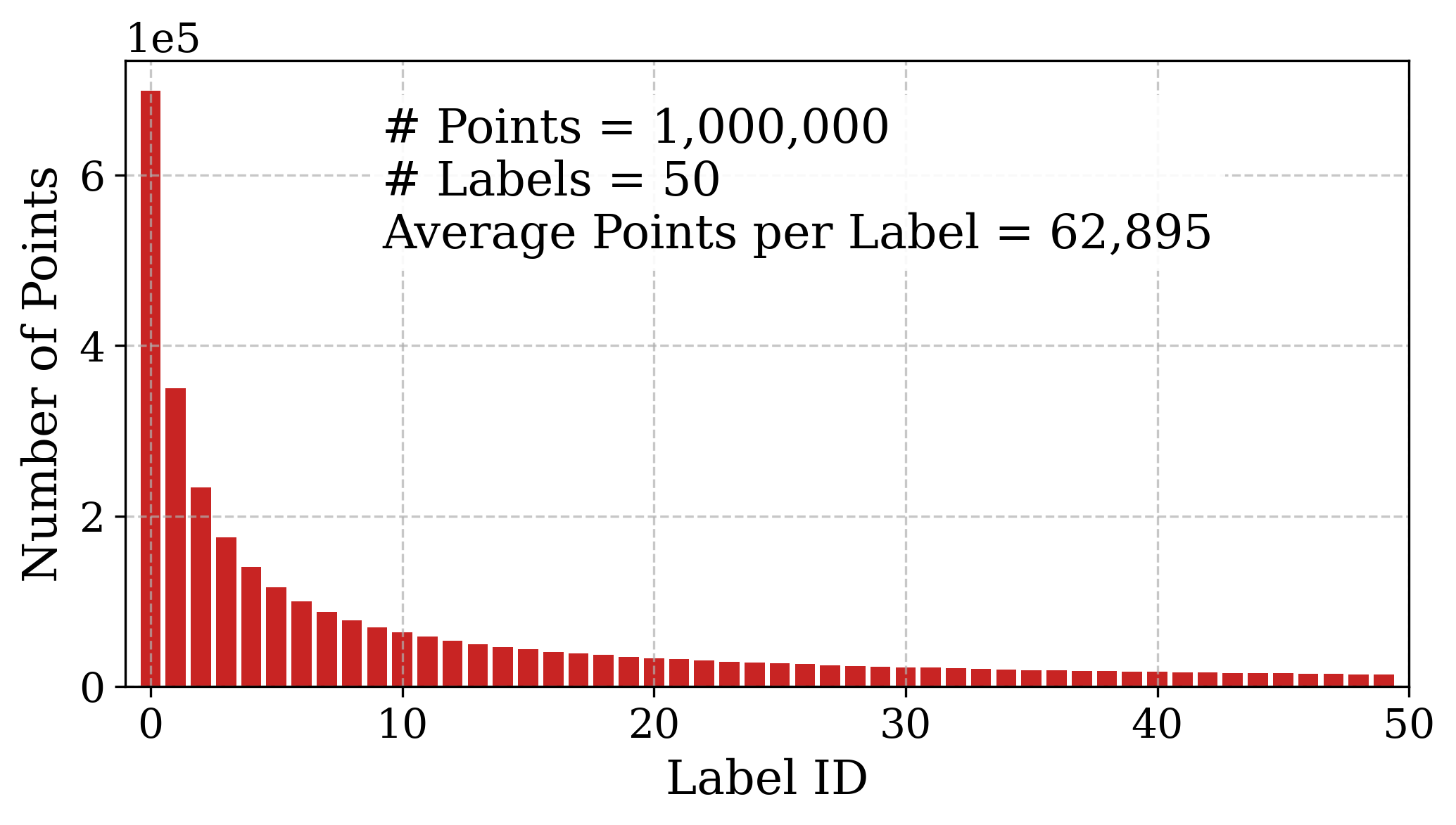}
        \caption*{(a) SIFT-1M}
    \end{minipage}
    \begin{minipage}
        {0.48\linewidth}
        \centering
        \includegraphics[width=\linewidth]{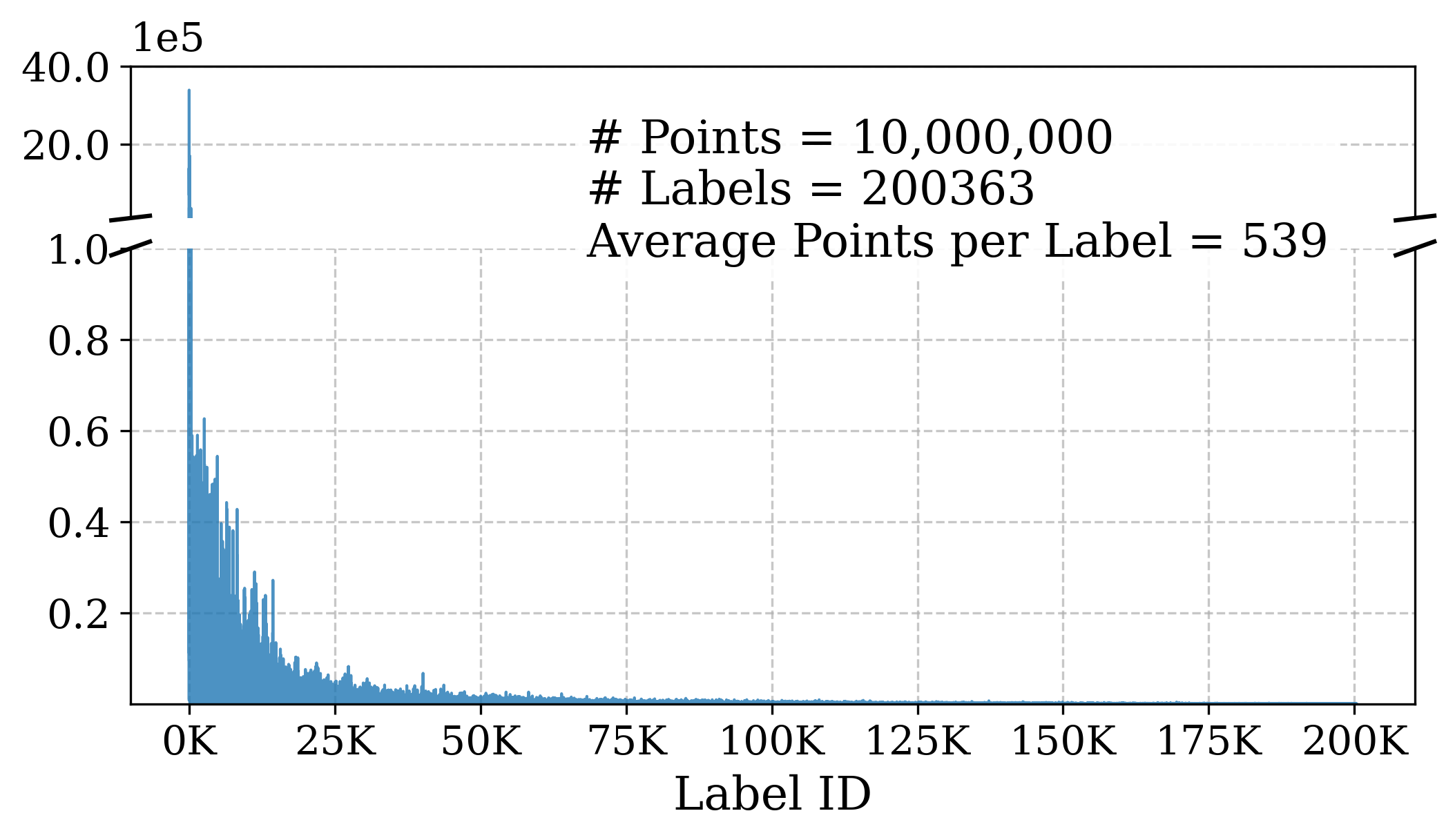}
        \caption*{(b) YFCC-10M}
    \end{minipage} 
    \caption{Label specificity distribution in different datasets.} 
    \label{fig:specificity}
\end{figure}

Despite its promising results, Filtered-DiskANN uses a single graph index for all points, meaning that multiple queries with different labels must use the same method to search this single graph index.
When the number of data points with associated labels is high, the single index approach performs similarly to unfiltered search. However, we observe that the label distribution in many datasets exhibits a \emph{long tail} distribution, with a few frequent labels and many rare ones, as shown in Figure \ref{fig:specificity}.  Challenges arise when data points with certain rare labels fall into the long tail, as most distance computations in the index search process become unnecessary, negatively impacting query performance. In addition, it is difficult to ignore vectors that do not pass the filter during graph traversal, as this can break the connectivity that the graph-based index relies on to achieve high recall. When the graph is large and the frequency of data points corresponding to a given label is very low, the search performance of Filtered-DiskANN becomes problematically low. As we later show in the evaluation,
FilteredVamana achieves a maximum recall of only 50\%, while StitchedVamana reaches at most just 70\% on the YFCC dataset. Both methods struggle to reach high recall because YFCC has many rare labels, making it challenging to predict how many top-K returned candidates will include those rare labels. Consequently, the number of final candidates that match the query's labels is less than K. Moreover, a large graph degree is required to maintain search quality (96 for FilteredVamana and 64 for StitchedVamana, as reported in \cite{filtered-diskann}). This further impacts search efficiency, as a higher graph degree results in slower search performance. This motivates us to consider algorithms that leverage the high computational throughput of GPUs to achieve high recall even on these cases with rare labels, which may be prohibitively costly on CPU systems.

Different from graph-based methods, IVF indexing is another fundamental technique in ANNS that partitions vectors into clusters to reduce the search space~\cite{product-quantization,sptag}. In traditional IVF methods, vectors are clustered based on their distance to the cluster centroids (typically using k-means clustering), creating "inverted lists" where the cluster centroid becomes the keyword, i.e., the descriptor of the cluster, and the posting list comprises all vectors in the corresponding cluster. During a search, the system first identifies the most promising clusters by comparing the query vector with centroids, then performs exact searches only within those selected posting lists. 
In contrast to traditional IVF, one can construct an inverted index where the keyword is the label, and the posting list contains vectors that share the same label~\cite{pinecone,ivf2}. This label-centric IVF knows exactly which posting list to search for each query based on its labels, eliminating the cluster selection overhead. This approach works effectively like setting \( n\_probes = 1 \) in traditional IVF search, but without the accuracy trade-off, as it guarantees searching the exact cluster containing the matching vectors. Prior studies show that IVF-based filtered-ANNS solutions can also achieve competitive performance compared to graph-based solutions such as Filtered-DiskANN~\cite{big-ann-result}.

To address the issue, we introduce a \emph{dual-structured label-centric IVF} that leverages the high computational throughput and parallelism of GPUs for high-performance filtered-ANNS. 
\name adopts a label-centric IVF design to improve the selectivity of ANNS with filters, because a query only needs to search the posting list that has the query's label. 
In addition, we define label specificity, a metric for quantifying the density of a given label in a dataset, and we divide the IVF posting lists into distinct groups, each with its own characteristics of label distribution. \name employs a dual-structured index and search algorithms to achieve fast and accurate filtered ANNS. For each group, we design high-performance GPU kernel optimizations to maximize GPU resource utilization.
We first describe this algorithm and then introduce several optimizations to map the algorithm to GPUs. 

\subsubsection{Specificity Definition} 
\label{subsubsec:specificity-def}

We start with the definition of \emph{specificity}, a property that measures the fraction of data points in $X$ that have the label $l \in \mathcal{L}$ associated with them. More specifically:
$$
\text{specificity}(l) = \frac{\text{number of data points with label }l}{\text{total number of data points }N}
$$
This definition is similar to prior work~\cite{diskann,acorn}. Different from prior work, we further classify a label as:
\begin{itemize}
    \item High-specificity (\hs): $\text{specificity}(l) \geq T / N$
    \item Low-specificity (\ls): $\text{specificity}(l) < T / N$
\end{itemize}

where T is a chosen specificity threshold. A data point $x$ therefore can also be categorized according to its labels:
\begin{itemize}
    \item High-specificity data: $x$ has at least one high-specificity label
    \item Low-specificity data: $x$ has at least one low-specificity label
\end{itemize}
Note that in scenarios where a data point can have multiple labels, a data point can have both high- and low-specificity labels simultaneously. 

\subsubsection{Decoupling High/Low-Specificity Labels with Label-Centric IVF and Dual-Structured Indices} 
\label{subsubsec:decoupling}

We construct an inverted index where the \emph{keyword} is the label, 
\revisionB{and the posting list contains data point indices. Specifically, for each label $l \in \mathcal{L}$, we define $C_l = \{i \mid l \in L_{X_i}, 0 \leq i < |X|\}$, which stores the global indices of points in $X$ associated with label $l$. Using $C_l$, we can efficiently retrieve the actual vectors for label $l$ as $X_l = \{X_i \mid i \in C_l\}$ without scanning the entire data point set.}
Once we build the \specificityivf, we divide the posting lists into two groups: the \highspecificity group (\hs) and the \lowspecificity group (\ls), and develop a dual-structured index construction and search algorithm to enable efficient filtered-ANNS for both \hs and \ls. 
\revisionB{We note that classifying data based on labels is a long-standing idea in database systems~\cite{bi-focal, sketches}. However, our approach mainly focuses on optimizing filtered-ANNS on GPUs, given the long-tailed distribution of real-world filtered search scenarios.}
For the \hs partition, we build a separate GPU-friendly graph index (e.g., CAGRA~\cite{cagra}) over an individual posting list $C_l$ (IVF-Graph). This allows \name to avoid an exhaustive search within a posting list while taking advantage of the hardware-friendly GPU-based ANNS solution, achieving performance similar to unfiltered search. For the \ls partition, we directly resort to brute force search (IVF-BFS).

\begin{figure}[t]
    \centering
    \begin{minipage}{0.49\linewidth}
        \includegraphics[width=\linewidth]{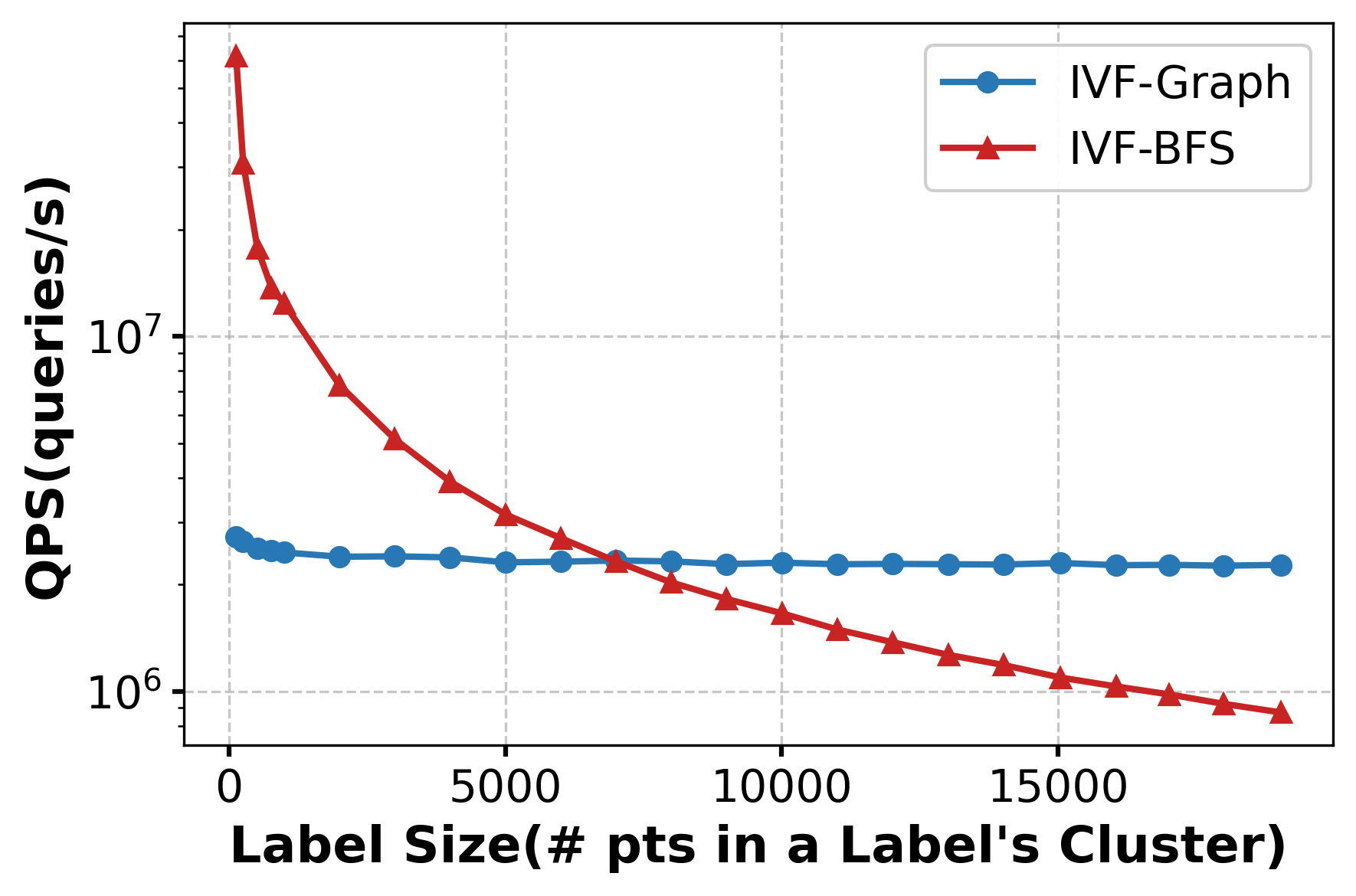}
        \caption*{(a)}
        \label{fig:cagra-scaling}
    \end{minipage}
    \begin{minipage}{0.49\linewidth}
        \includegraphics[width=\linewidth]{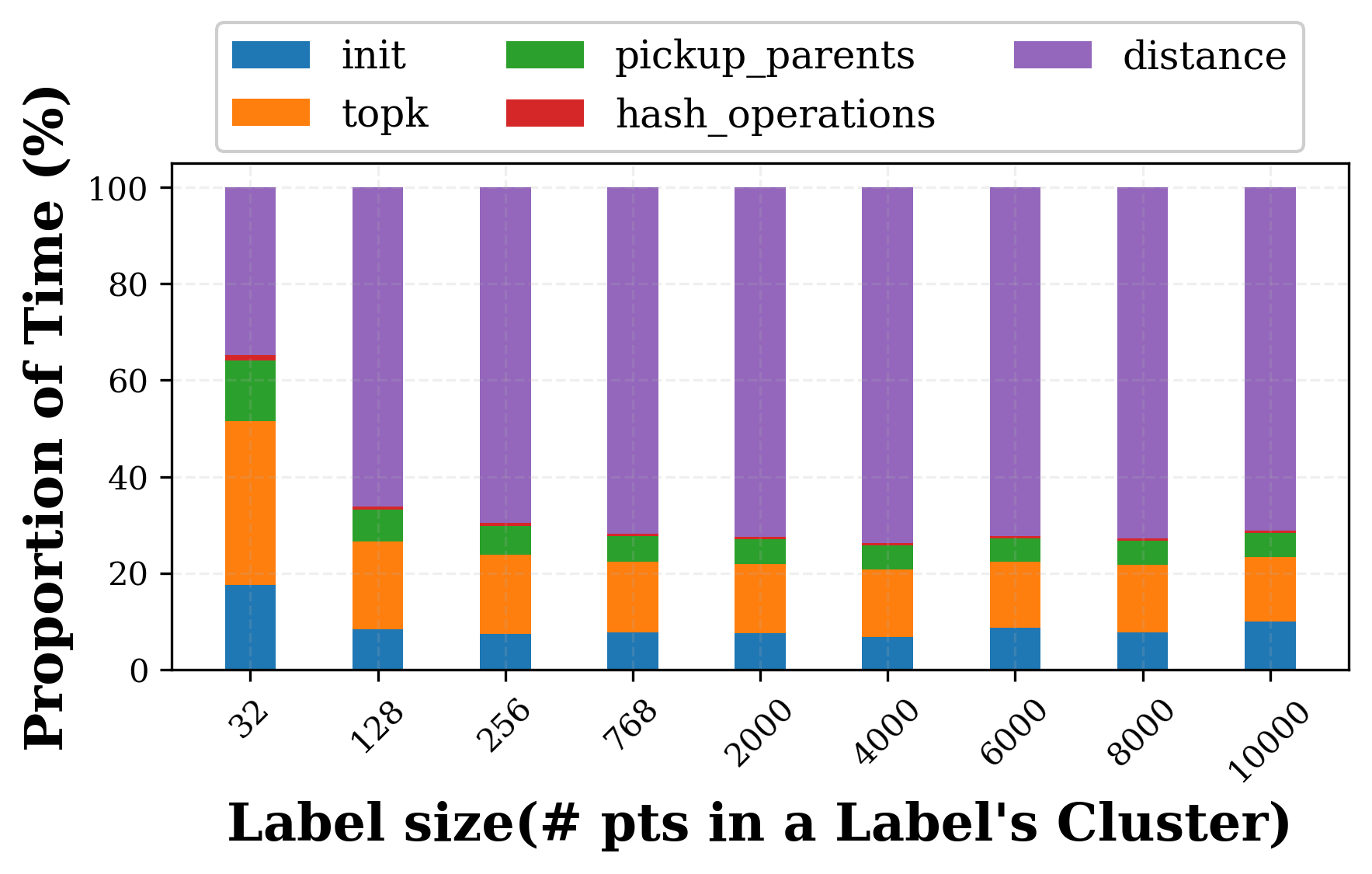}
        \caption*{(b)}
        \label{fig:cagra-breakdown}
    \end{minipage}
    \caption{(a) Scaling efficiency of \ivfgraph and \ivfbfs varying the number of data points within a cluster. (b) Time decomposition of CAGRA search process.}
    \label{fig:scaling}
\end{figure}

The rationale for adopting this dual-structured index is that for \ls data, each IVF list contains a relatively small number of points, making the construction and maintenance of a graph-based index unnecessarily complex and potentially counterproductive. For example, CAGRA's search behavior is primarily governed by its internal top-k (\emph{itopk}) parameter rather than the actual dataset size. This parameter determines both the number of iterations and the number of distance computations performed, regardless of cluster size. Our experimental analysis demonstrates this scaling inefficiency -- when testing across different cluster sizes from 100 to 20,000 points with constant itopk(32), CAGRA's throughput remains relatively flat (2.6M QPS for 100-point clusters vs. 2.3M QPS for 20,000-point clusters), as shown in \fref{fig:scaling}(a).

Despite showing excellent performance, graph traversal incurs iterative neighbor expansion that leads to additional synchronization overhead.  \fref{fig:scaling}(b) demonstrates that for queries with low-specificity labels in the YFCC dataset, an average of 71.29\% of the CAGRA search time (itopk=32) is allocated to actual distance computations, while the remaining time is distributed across graph traversal overhead, including initialization (8.04\%), parent node selection (5.41\%), and internal top-k maintenance (14.79\%). Consequently, when a label corresponds to only a small number of points, it becomes more efficient to directly compare the query vector against all points in the IVF list, avoiding the overhead of initializing and traversing a graph.

For low-specificity data, \revision{brute-force search (BFS)} offers advantages over graph-based methods because 
BFS can be more easily parallelized to efficiently leverage modern GPUs, especially when the number of data points is relatively small. Essentially, the distance computation between a query and a set of vectors can be effectively formulated as a \emph{batched matrix-vector multiplication (GeMV)} operation, which can be tiled across Streaming Multiprocessors (SMs) on a GPU. This approach helps to maximize the utilization of the GPU's compute and memory bandwidth. GeMV is primarily a memory bandwidth bound operation and the peak memory bandwidth for a GPU (e.g., Nvidia A100) is often over 2 Terabytes per second (2TB/s), which is over 20x higher than modern CPU bandwidth (often $<$100GB/s). Thus, BFS on GPUs offers significant speedups over CPUs, on which BFS may be prohibitively slow. By employing BFS, execution time is optimized by focusing on the distance computation, eliminating graph traversal overhead entirely. This makes BFS a more efficient choice for low-specificity scenarios where the benefits of graph traversal are minimal.


During the search, whether a query searches the \hs or \ls partition is determined by the specificity threshold $T$. Given a query with label \revisionC{$l$}, we route the query based on the size of its corresponding cluster $|C_l|$:

\begin{equation}
    \text{method} = \begin{cases} 
    \text{\ivfbfs} & \text{if } |C_l| < T \\
    \text{\ivfgraph} & \text{otherwise}
    \end{cases}
\end{equation}

Finding the optimal threshold $T$ analytically is non-trivial as it depends on many architectural parameters and specificity distribution. However, it is not necessary in practice. $T$ can be carefully chosen to represent the crossover point where the brute force approach becomes less efficient than graph traversal for a given number of data points. This is determined through offline profiling query performance across different cluster sizes and auto-tuning, e.g., by comparing the performance of \hs and \ls search to identify a threshold \( T \) that maximizes the overall query throughput. 

\subsection{Index and Search Strategies of \name on GPUs}
\label{subsec:gpu-index-search}

\subsubsection{Bottom-Level IVF-Graph with Redundancy-Bypassing for \hs}
\label{subsubsec:ivf-graph}
\paragraph{Indexing.}
While combining IVF with graphs significantly improves the search speed, building an individual graph \revisionB{index} for each IVF list leads to one major challenge -- the memory consumption becomes much higher compared to a single index for all data points. We observe that directly building IVF-Graph index for each posting list leads to a 10.8× increase in memory compared to a single graph index over the entire YFCC dataset. 
Why does the memory consumption \revisionA{increase} dramatically? We dive deep into how graph-based ANNS index \revisionA{is} constructed for GPUs and find out that this is caused by memory redundancy in IVF-Graph. Taking CAGRA as an example, for a given set of data points, CAGRA builds a k-NN graph with fixed out-degree using NN-descent~\cite{nn-descent} on GPUs and performs rank-based reordering to improve the graph's navigability. 
\revisionB{The final CAGRA-based IVF-Graph list for each label $l $, consists of two components (1) vectors $X_l $ requiring $|C_l| \times D $ memory, and (2) a graph structure consisting geometric relation requiring $|C_l| \times R $ memory, where $D$ is the vector dimension and $R$ is the graph out-degree. When a data point $X_i$ belongs to multiple labels (e.g., $i \in C_{l1} $ and $i \in C_{l2} $), its vector must be duplicated in both $X_{l1} $ and $X_{l2} $. Since vectors typically dominate memory consumption (e.g., $D=192 $ in YFCC dataset versus $R=16$), this redundant storage of vectors across multiple IVF lists causes the substantial memory increase.}

To address the redundancy issue, we propose an approach called \ivfgraph with Redundancy-Bypassing. 
\revisionB{For each label $l$, \name builds a \emph{local virtual} graph $G_l$, where each virtual graph only contains the graph structure information expressed via local vertex IDs. Alongside these graphs, \name has a single \emph{global} vectors $X$ that contains all data points.}
In addition, \name maintains a \textbf{local-global} index mapping table \revisionB{$M_{HS}$} that translates these local vertex IDs (within each \revisionB{$G_l$}) to their corresponding global indices in the \revisionB{global vectors}. This mapping mechanism allows \name to maintain only a single copy of \revisionB{data point set} while having multiple virtual graphs to share the same \revisionB{global vectors}. 

Current graph-based ANNS algorithms typically use a single graph structure represented by adjacency lists, where each row uniquely corresponds to the neighbors of a specific data point. However, \name introduces multiple virtual graphs to enable filtered search. If these virtual graphs were stored separately, processing queries with different labels would require frequent index switching, hindering the parallel performance of methods like CAGRA for large batches. On the other hand, storing multiple graphs together in a single index is not supported by existing ANNS libraries.
To address the issue, 
we design a readily integrable and GPU-efficient index structure to implement our approach. In particular, we compact all virtual graphs into a single, continuous memory space, organized and ordered by their label \revisionB{$l$}. This compacted structure not only avoids the overhead of switching between indices but also represents multiple virtual graphs as a unified structure. As a result, it seamlessly integrates into any graph-based ANNS algorithm by replacing the original single-graph structure, enabling filtered searches without requiring significant modifications to the index structure. Furthermore, since our combined index is compatible with single-graph structures, any optimization techniques designed for single indices can be directly applied to our unified graph.


\begin{figure}[t]
    \centering
    \includegraphics[width=\linewidth]{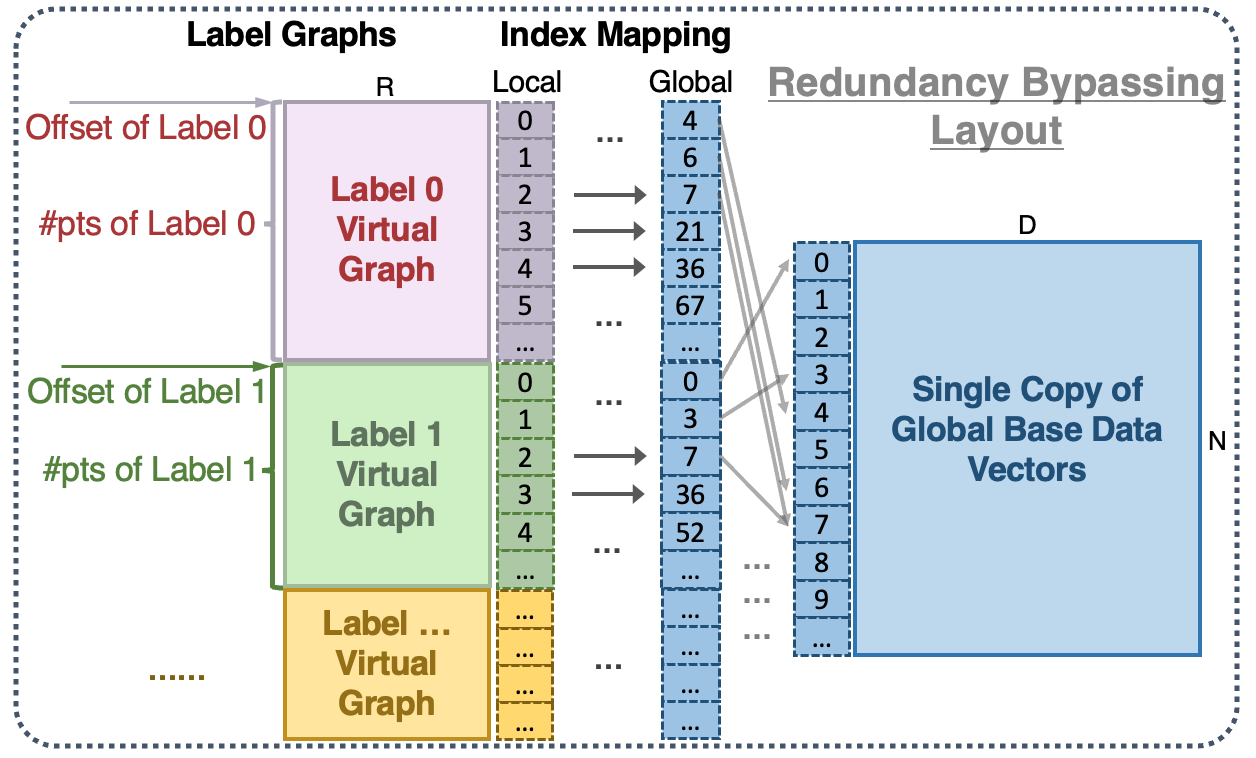}
    \caption{The data layout used in \name’s
redundancy-bypassing IVF-Graph index and search.}
    \label{fig:graph-index}
\end{figure}

\revisionB{
    The layout of \name's compacted Label Graphs $G_{HS}$ is illustrated in \fref{fig:graph-index}. To make this structure functional, \name maintains three key metadata components: (1) Label Sizes $S_{HS}$: the number of data points in each posting list $C_l$, (2) Label Offsets $O_{HS}$: the starting position of label's virtual graph $G_l$ within the compacted Label Graphs $G_{HS}$, which together help \name efficiently locate $G_l$, and (3) Index Mapping $M_{HS}$: which translates local vertex IDs in $G_l$ to their global indices in $X$. As shown in Algorithm~\ref{alg:dual_structured_index_construction}, to manage labels, \name's indexing follows a two-pass process: first, \name collects all data points' indices associated with each label $l$ into posting list $C_l$. Then, \name builds separate virtual graphs for each high-specificity label while establishing metadata for each label that includes size, offset, and mapping information. This approach allows \name index to construct multiple virtual graphs into a compacted Label Graphs $G_{HS}$ referencing a single global vectors $X$, thereby reducing memory consumption while preserving search efficiency.
}

\begin{algorithm}[t]
\newcommand{\BuildCAGRAGraph}{\text{BuildCAGRAGraph}}
\newcommand{\vstack}{\text{vstack}} 
\newcommand{\SetMetadata}{\text{SetMetadata}}
\SetCommentSty{textit}
\SetKwComment{Comment}{$\triangleright$\ }{}
\newcommand{\sizeOf}[1]{\left| #1 \right|}
\caption{\revisionC{VecFlow Index Construction}}
\label{alg:dual_structured_index_construction}
\DontPrintSemicolon 
\KwIn{Data point set $X$; $L_x$ for each $x \in X$; Graph Degree $R$; 
 specificity-threshold $T$.}
\KwOut{IVF-Graph-Index; IVF-BFS-Index.}
$\mathcal{L} \gets \bigcup_{x \in X} L_x$ \Comment*[l]{Set of unique labels }
$C_l = \{i \mid l \in L_{X_i}, 0 \leq i < \sizeOf{X}\}$ for all $l \in \mathcal{L}$\;
\ForEach{$l \in \mathcal{L}$}{
    $X_l = \{X_i | i \in C_l \}$\;
  \uIf{$\sizeOf{C_l} \ge T$}{
    \SetMetadata($C_l, S_{HS}[l], O_{HS}[l], M_{HS}[l]$)\;
    $G_{HS} \gets \vstack(G_{HS}, \BuildCAGRAGraph(X_l, R))$\; 
  }
  \Else{
    \SetMetadata($C_l, S_{LS}[l], O_{LS}[l], M_{LS}[l]$)\;
    $X_{LS} \gets \text{StoreInterleaved}(X_{LS}, X_l)$\;
  }
}
\KwRet $(X, G_{HS}, S_{HS}, O_{HS}, M_{HS}), (X_{LS}, S_{LS}, O_{LS},  M_{LS})$\;
\end{algorithm}

\begin{algorithm}[t]
\SetCommentSty{textit}
\SetKwComment{Comment}{$\triangleright$\ }{}
\newcommand{\sizeOf}[1]{\left| #1 \right|}
\newcommand{\DIS}{\text{GetDist}}
\newcommand{\BFS}{\text{BruteForceSearch}}
\caption{\revisionC{VecFlow Search Algorithm}}
\label{alg:dual_structured_search}
\DontPrintSemicolon
\KwIn{Query batches $Q$; $L_q$ for each $q \in Q$; $\text{IVF-Graph-Index}(X, G_{HS}, S_{HS}, O_{HS}, M_{HS}$); $\text{IVF-BFS-Index}(X_{LS}, S_{LS}, O_{LS}, M_{LS}$). specificity-threshold $T$;}
\KwOut{Top-$k$ nearest neighbors for each query.}
\BlankLine
$(Q_{HS}, L_{HS}), (Q_{LS}, L_{LS}) \gets \text{ClassifyQueries}(Q, L_q, T)$\;
\ForEach{$(q, l)$ from $(Q_{HS}, L_q)$ in parallel}{
  $G_l \gets G_{HS}[O_{HS}[l] : O_{HS}[l] + S_{HS}[l] - 1]$\;
    $topM, candidates \gets \text{InitializeSearch}(G_l)$\;
  \For{search iterations}{
    $dists \gets \DIS(candidates, q, l, X, M_{HS}, O_{HS}$)\;
    $TopM \gets UpdateTopM(TopM, candidates, dists)$\;
    $candidates \gets G_l[topM[\text{first unvisited node}]]$\;
  }
}
\ForEach{$(q, l)$ from ($Q_{LS}, L_{q})$ in parallel}{
  \BFS($\text{IVF-BFS-Index}, q, l$)
}
\KwRet Merge results and map to global IDs\;
\end{algorithm}

\paragraph{Search.} For queries with high-specificity labels, \name routes the query to the \hs partition and performs the graph-based search in the posting lists that have the query's labels. \revisionB{Algorithm~\ref{alg:dual_structured_search} demonstrates how \name utilizes its indexing structures during search. The key step $G_l \gets G_{HS}[O_{HS}[l] : O_{HS}[l] + S_{HS}[l] - 1]$ extracts precisely the portion of local virtual graph $G_l$ for label $l$, using both the \emph{Label Sizes} and \emph{Label Offsets} metadata} The actual search follows the top-M list and candidate list expansion procedure in CAGRA, where \name starts with random sampling (e.g., \name uses \emph{Label Sizes} to confine the range of random sampling) to get the initial candidate list and performs iterative graph traversal iterations, i.e., sort itopk, pick up next parents, compute distance for child node, expand the candidate list until the top-M list converges, i.e., the top-M list remains unchanged from the previous iteration. Along the search process, \name uses the same elemental technologies from CAGRA~\cite{cagra} to boost the GPU utilization for the graph traversal on GPUs, including (1) warp splitting that divides a warp into fine-grained thread groups (a.k.a, teams) to allow all threads within each warp to maximize the bandwidth utilization through 128-bit load instructions, (2) single warp-level bitonic sort to quickly sort the priority queue in the registers of a single warp without the shared memory footprint, and (3) the forgettable hash table management to keep the hash table in the fast shared memory for managing the visited node list.

Like CAGRA search, all of our search operations, such as top-\(k\) computations and internal list updates, are performed using \revisionB{local vertex IDs within $G_l$ with values in the range $[0, S_{HS}[l])$}. Only when computing distances, \name's \revisionB{\emph{Index Mapping} $M_{HS}[l]$ translates the local vertex IDs to corresponding global indices in the global vectors.} 


\subsubsection{Bottom-Level GPU-Friendly Brute Force Search for \ls}
\label{subsubsec:ivf-bfs}

While BFS for \ls is conceptually straightforward, requiring only target vectors and query vectors, performing BFS on GPUs presents challenges, especially when processing multiple queries in parallel (e.g., batch size $>$ 1), as each query in the batch must search different parts of the dataset based on its labels. Recent advancements in GPU-based KNN search introduced in the NVidia cuVS library~\cite{cuvs}, \chenghao{Missing citation} provide two efficient BFS implementations with bitmap filtering.  For these approaches, each row of the bitmap represents a different query's label and each column represents a data point in the dataset, with bits set to 1 indicating label matches between queries and data points. In the first method, a tiled brute force kNN search implementation is used to achieve high compute efficiency between the queries and the entire data points, followed by applying the bitmap as a post-processing filter to get the final results. However, this approach becomes computationally expensive with large datasets like YFCC, which has 10M data points, as it requires computing distances between each query and all data points regardless of whether they match the query's labels or not. The second approach transforms the kNN search into a sparse computation by converting the bitmap into the Compressed Sparse Row (CSR) format, which only stores the coordinates of the matching pairs. This enables masked matrix multiplication that computes distances only between each query vector and its label-matching data vectors. The implementation processes all queries together in a single masked matrix multiplication operation, followed by a k-selection step to find the nearest neighbors for each query. This approach requires both extra memory accesses to load the CSR mask and leads to scattered memory accesses from sparse computation patterns, which are less efficient than the contiguous accesses typical in dense operations. As a result, it is unclear how to achieve high performance filtered kNN on large-scale datasets for large batches.



\begin{figure}[t]
    \centering
    \includegraphics[width=\linewidth]{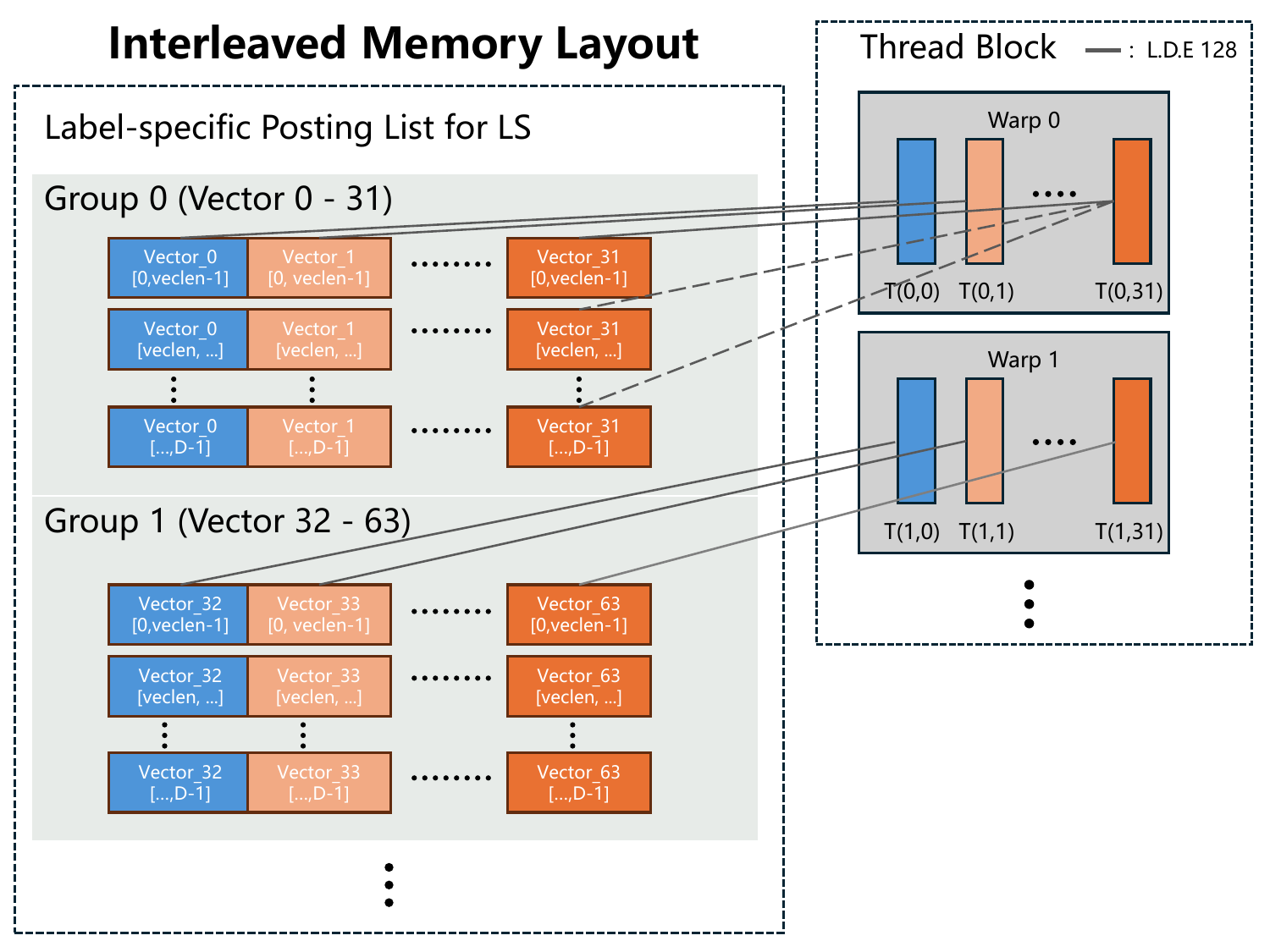}
    \caption{Interleaved memory layout for \name's IVF-BFS over \ls partition.}
    \label{fig:interleaved}
\end{figure}

Unlike these cuVS brute-force search methods, \name introduces an interleaved scan based IVF-BFS method for low-specificity data. This method is specifically designed for high-throughput distance computations on GPUs. The approach is built on top of cuVS IVF-Flat. During the index construction phase, \name employs an interleaved memory layout where vectors within each posting list of IVF are organized into blocks of interleaved components. As shown in Figure~\ref{fig:interleaved}, \name divide vectors into groups of 32 to match the GPU warp size. Within each group, instead of storing vectors sequentially ($(v_0[0:D], v_1[0:D], ...)$), \name stores them in an interleaved pattern where components are grouped based on $\text{veclen} = 16/\text{sizeof}(T)$, where $T$ is the data type. For example, with float32 vectors ($\text{sizeof}(T)=4$), $\text{veclen}=4$, so \name stores $(v_0[0:3], v_1[0:3], ..., v_{31}[0:3], v_0[4:7], v_1[4:7], ...)$. This organization enables efficient memory access patterns in two ways. First, when 32 threads in a warp process different vectors simultaneously, they access contiguous memory locations, allowing the GPU to coalesce these accesses into fewer memory transactions. Second, each thread can use wide load instructions (LD.E.128) to fetch 16 bytes at once, as the components for each vector are stored in chunks sized according to $\text{veclen}$. This dual-optimization of coalesced access across threads and vectorized loading within threads significantly improves utilization of GPU memory bandwidth during distance computations. In addition, this interleaved layout improves instruction-level parallelism and instruction pipeline efficiency by letting threads begin performance distance computations while vector coordinates continue to be fetched from memory.


For the search process, \name uses 
ivfflat\_interleaved\_scan kernel by cuVS which achieves high throughput through hierarchical parallelism and efficient distance computation. At the grid level, each CUDA block processes one query that searches inside its corresponding label-specific posting list from the IVF index. Within each block, the query vector is first loaded into shared memory for fast access by all threads. The block then processes vectors in its assigned posting list through coordinated warp execution. Each warp processes one group of 32 interleaved vectors from the memory layout described in Figure~\ref{fig:interleaved}. Within a warp, each thread computes the distance between one vector from the current group and the shared query vector.
Another important step during the brute force search is to filter out the top-k candidates. The kernel fuses an optimized block-select-k operation into the distance computation to identify nearest neighbors efficiently. This approach outperforms cuVS's brute force search implementations by avoiding extra distance computations through label-specific IVF posting lists and achieving better memory bandwidth utilization with optimized interleaved memory layout.
This hierarchical organization naturally aligns with the structure of small label-based clusters: blocks can independently process different query-label pairs, while warps provide efficient parallel distance computation and sorting within each cluster. Overall, this ensures complete coverage for exhaustive search without the overhead of graph traversal.

\revisionB{Same as IVF-Graph Redundancy-Bypassing , \ivfbfs maintains three key metadata components for each label, which are \emph{Label Sizes} $S_{LS}$, \emph{Label Offsets} $O_{LS}$ and \emph{Indexing Mapping} $M_{LS}$.Algorithm~\ref{alg:dual_structured_index_construction} and Algorithm~\ref{alg:dual_structured_search} also show the index construction and search process of \ivfbfs.} 
We note that, for low-specificity data points, \name introduces additional memory overhead compared to using base vectors directly. This overhead arises because vectors are stored within each label's IVF list in a continuous, interleaved layout to optimize GPU memory bandwidth utilization. Since a single data point may be associated with multiple low-specificity labels, its vector needs to be stored in multiple IVF lists. The total memory overhead can be expressed as $N \times D \times F$, where $F$ represents the average number of low-specificity labels per point. For example, in the YFCC dataset, about 5.95M points (59.5\% of the dataset) have at least one low-specificity label (e.g., labels associated with fewer than 1,000 points), with these points appearing on average in 3.54 low-specificity IVF lists (21.1M total entries across low-specificity IVF lists divided by 5.95M points), leading to 3.54x memory overhead compared to the original vector storage for LS partition.

\subsubsection{Persistent Kernel-based Search for Small Batch Queries}
\label{subsubsec:persistent-kernel}

Since all queries within a batch are independent, \name uses a single-kernel implementation for handling batched queries but use an individual CUDA thread-block (single-CTA) within a kernel to handle large batch sizes, similar as how CAGRA handles large batch sizes. Overall, when the batch size is large (e.g., $>$100), each query is mapped to a single thread block, and multiple of these blocks are executed in parallel. This approach effectively harnesses the massive parallelism of GPUs, maximizing their compute potential. However, when the batch size is small (e.g., $<$100), the GPU resource can be left underutilized with the single-CTA implementation. 

In practice, vector search services often must execute queries in small batches and maintain low latency. However, traditional GPU-based methods like CAGRA are optimized for large batches and suffer from high overhead when processing small batches due to repeated kernel launching overhead. One may wonder whether it is possible to hide the kernel launching overhead to some extent by launching a few kernels in parallel using streams, thereby overlapping the launch overhead with the execution of other kernels. However, this strategy also does not work very well in practice because of CPU thread synchronization overheads.
Modern GPUs require hundreds of thread blocks running in parallel to achieve maximum throughput. At this scale, even a few CUDA host API calls per-thread can become a bottleneck due to shared CUDA context locking.
To address this, \name introduces a persistent kernel approach that maintains a continuously running kernel on the GPU to process queries as they arrive, which not only eliminates the repeated kernel launch overhead but also avoid synchronization at all costs by using atomic-based control structures.

\begin{figure}[!ht]
    \centering
    \includegraphics[width=\linewidth]{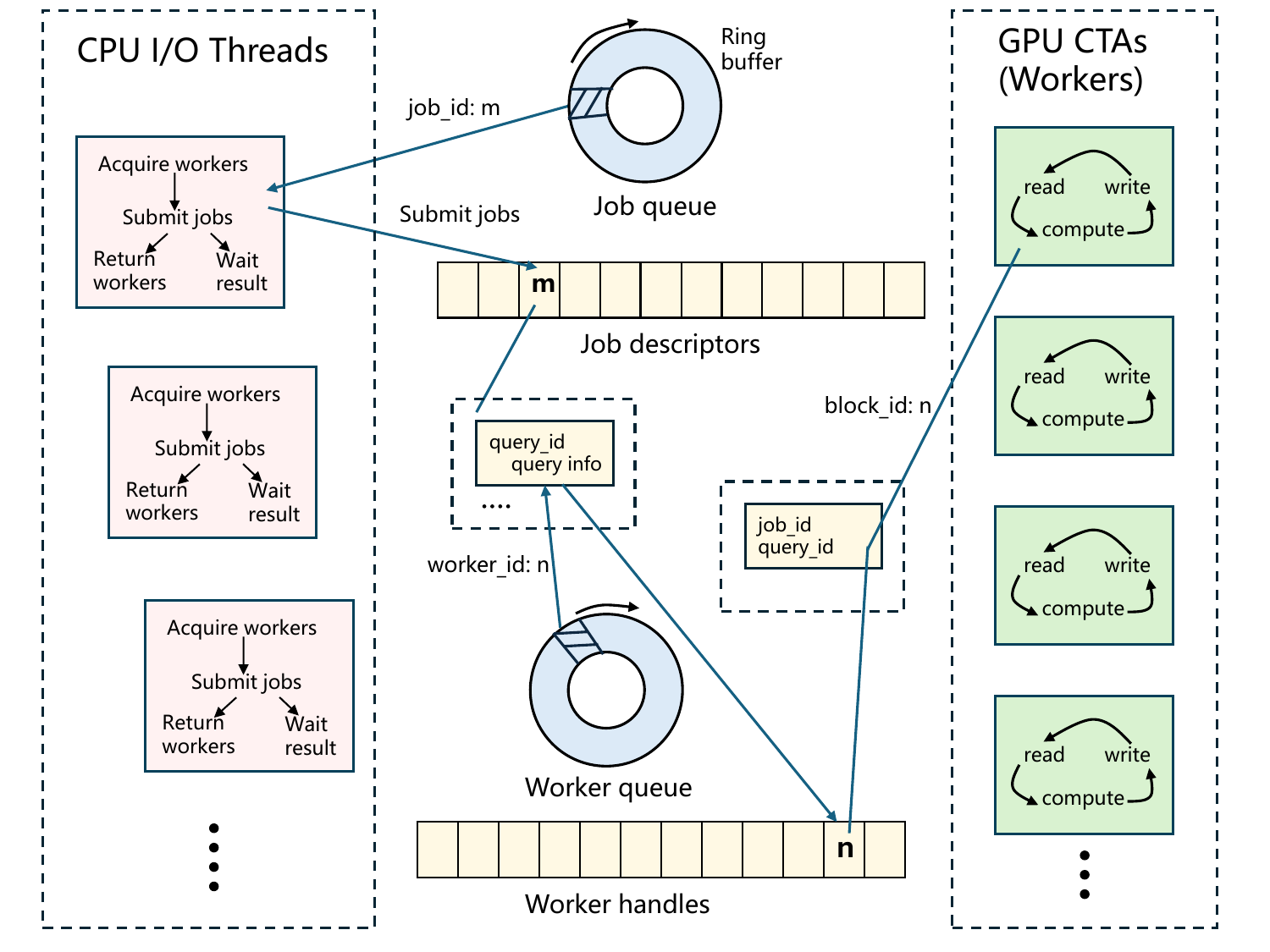}
    \caption{Persistent kernel design in \name.}
    \label{fig:persistent-kernel}
\end{figure}

\fref{fig:persistent-kernel} shows our persistent kernel design, which uses a job queue system with atomic ring buffers to manage incoming queries and a worker queue to track available GPU thread blocks. When a query arrives, it is assigned a job ID and mapped to an available worker. The persistent kernel continuously monitors for new work, with each thread block handling one query independently. 
This approach enables efficient parallel processing while maintaining high GPU utilization despite small batch sizes.


\subsubsection{Memory Consumption.} 

We express the memory consumption as the sum of bytes required to store both the \hs and \ls partition indices. Let \( F \) denote the average number of labels per point, with \( F_{HS} \) and \( F_{LS} \) representing the averages for the HS and LS partitions, respectively.
Without applying the redundancy-bypassing algorithm, the memory consumption of the \hs partition is \( N \times F_{HS} \times (D + R) \times b \), where \( R \) is the length of the out-degree, and \( b \) is the number of bytes of data type. After applying the redundancy-bypassing optimization, \name stores a single shared copy of the base data vectors while maintaining separate neighbor lists for each label-specific graph. This reduces memory usage significantly because \( D \) is often much larger than \( R \). The memory consumption of the \hs partition becomes \( N \times (D + F_{HS} \times R) \times b \).
Furthermore, the average cluster size per label is much smaller than \( N \) in practice. For example, in the YFCC dataset, this average label size reduces from 10M to 540. Consequently, we use a much smaller out-degree \( R'\), which shows no performance degradation based on empirical observation. Thus, the final memory consumption for the \hs partition is \( N \times (D + F_{HS} \times R') \times b \). Instead, we store only the base data in the LS partition, resulting in a memory consumption of \( N \times D \times F_{LS} \times b \). Thus the total memory consumption is \( N \times (D + F_{HS} \times R') \times b + N \times D \times F_{LS} \times b \). 
Assume we have 100 million points, with the following parameters: \( D = 128, R = 64, F = 3.17, F_{HS} = 3.0, F_{LS} = 0.17, b = 4, R' = 16\). Substituting the values, the memory consumption is \( 65.57 \, \text{GB} + 8.11 \, \text{GB} = 73.67 \, \text{GB} \), which is slightly higher than a single index's memory consumption of \( 71.53 \, \text{GB} \), calculated as \( N \times (D + R) \times b \). For metadata, the total memory consumption for local-global index mapping is \( N \times F \times b \), which is \( 1.18 \, \text{GB}. \) Regarding label metadata (\emph{Label Size} and \emph{Label Offset}, \(2\) integers for each label), since the number of labels is typically in the range of several thousands, its memory usage is negligible. For example, even in the YFCC dataset with 200K labels (which already represents a relatively large label set), the label metadata only accounts for \( 1.49 \, \text{MB} \). This value is sufficiently small that it can be ignored in most cases.



\subsection{Multi-Label Query Processing with Predicate Filtering and Early Stopping}
\label{subsec:multi-label}

\begin{figure}[t]
    \centering
    \includegraphics[width=\linewidth]{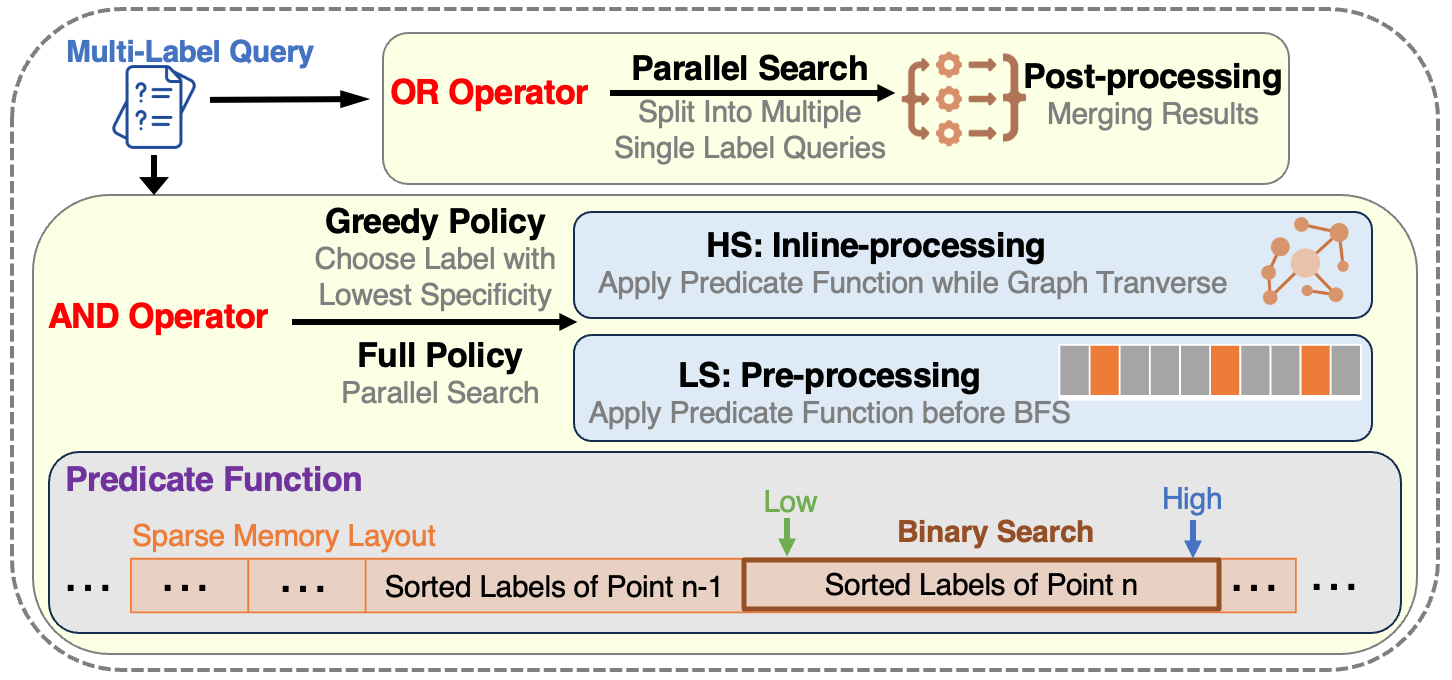}
    \caption{Multi-label query processing in \name.}
    \label{fig:enter-label}
\end{figure}

Prior work often assumes queries have a single label. However, in practice, one query is often associated with multiple labels, which introduces challenges. 
\revisionC{
    \name's label-centric indexing handles multi-label queries efficiently through GPU parallelism, where each thread block processes one query label. As shown in \fref{fig:enter-label}, \name implement different strategies for \emph{OR} operations (parallel processing with result merging) and \emph{AND} operations (either parallel search with filtering or selective greedy search, while both using predicate verification).
}

\subsubsection{Handling \emph{OR} Operations} 
For multi-label query connected by OR operations, we can simply convert it into single-label search problem, treating each label as a separate search request and launch multiple standard single-label searches \emph{in parallel}, using IVF-Graph search for high-specificity labels and IVF-BFS for low-specificity labels, followed by merging results to find the nearest neighbors that satisfy any of the query labels. 

\subsubsection{Handling \emph{AND} Operations} 
It becomes challenging to handle \emph{AND} operations efficiently because a point must satisfy all query labels simultaneously. 
For example, when looking for points that satisfy both label A \emph{AND} B, even if label A and B individually have many matching points, the set of points satisfying both constraints may be sparsely distributed in the single graph index. 
This forces the search to explore many irrelevant paths and points before finding valid results that satisfy all constraints. Our key insight to support efficient \emph{AND} Operations is that it is \emph{not necessary to check all paths} to identify candidates that satisfy A \emph{AND} B. Instead, we develop a predicate function that efficiently verifies whether a point contains all specified labels and uses this predicate function to obtain a \emph{selective early-stopping} search strategies for both \hs and \ls partitions to handle \emph{AND} queries effectively. 

\paragraph{Predicate function for efficient multi-label verification.} To efficiently verify whether a data point contains all specified labels, we use a compact data structure based on three arrays: a global label array, a label offset array, and a label size array. The global label array stores the label of all data points sequentially, 
keeping each data point's labels contiguous and sorted. The label offset array marks where each point's labels begin in the global array, and the label size array records how many labels each point has. We choose this structure over a matrix representation $N \times |\mathcal{L}|_{max}$(where $|\mathcal{L}|_{max}$ represents the max label size), 
because of highly variable label sizes. In the YFCC dataset, while the average is 10.8 labels per point, some points have up to 1,500 labels. 
A matrix approach would waste memory by allocating 1,500 columns for every point. Our array-based approach stores only the actual labels each point has, reducing memory fragmentation and enabling efficient GPU parallel processing through a compact, contiguous memory layout.

For each data point, we employ a \emph{binary search} algorithm to retrieve its portion of labels from the global label array. For instance, if a data point's offset is 1000 and it has 15 labels, we search within positions 1000 to 1014 of the global array. The search adjusts two pointers iteratively based on comparisons with the target label. If found, it returns true and the label's position; otherwise, it returns false, indicating the point fails the AND predicate.

For queries with multiple labels, we optimize the search process by first sorting the query labels. It then searches from the smallest and largest query labels to establish the valid range that must contain all other query labels. For example, suppose a point has labels [1,3,5,7,9,11,13,15] and the algorithm needs to verify query labels [5,9,11]. It first searches for the smallest label 5, finding it at index 2. If label 5 is not found, it immediately returns false. Otherwise, it searches for the largest label 11, finding it at index 5. If label 11 is not found, it returns false. After finding both boundary labels, it establishes that all remaining query labels must exist within range [3,4]. For the middle label 9, it only needs to search this smaller range. This approach maintains the worst-case complexity of $O(|\mathcal{L}_q|\log F)$, but typically performs better in practice as intermediate searches operate on a much smaller range. The function returns true only if all query labels are found.

\jingyi{Do we need to propose greedy policy and parallel policy for AND search?}\minjia{We should because we use those two in the eval section. Better to restructure the following section to make those two policies clear, but we don't have time. I added the name of those two policies. }
\jingyi{Agree, Better to restructure.}






\paragraph{\ivfgraph \emph{AND} search for \hs.} For queries with multiple labels connected by \emph{AND} operations in the high-specificity partition, we develop a \emph{greedy search policy} that combines IVF list selection with {inline-processing} during graph traversal. The key idea is to minimize the search space while maintaining high accuracy through two main components.
Our search space selection is straightforward: given a query with multiple labels connected by \emph{AND} operations, the search selects the IVF list corresponding to the label with the \emph{lowest} specificity. For example, if a query asks for points with labels A associated with 10,000 points \emph{AND} B associated with 1,000 points \emph{AND} C associated with 5,000 points, the search chooses label B's IVF list. This ensures we search within the smallest possible search space.
For each iteration of graph traversal, we employ the {inline-processing} approach that integrates label verification directly into the search process. After computing distances at each graph expansion iteration, we apply our predicate function to check whether candidate points contain all the rest higher-specificity labels from the query. Points that fail this verification step are removed from consideration before the next iteration begins. 
This filtering strategy ensures that graph traversal expands only from points that satisfy all label requirements.

This combined approach provides performance benefits through improved selectivity. By starting from the IVF list of the lowest-specificity label and using higher-specificity labels as filters during traversal, we increase the relative selectivity to points that satisfy all label constraints. For example, when searching for points with labels A \emph{AND} B where A is associated with 50,000 points and B with 5,000 points, searching in B's IVF list reduces the initial search space by 90\% compared to searching in A's list. Meanwhile, since a point in the dataset has a higher probability of having label A, filtering for label A during graph traversal becomes more effective. This means when we verify if a point has label A during traversal, we have a better chance of finding valid neighbors, making the search more efficient in exploring the relevant regions of the graph.


If very high accuracy is desired, we also provide a full \emph{parallel search policy} for handling AND queries in the HS partition. This approach converts \emph{AND} queries into single-label search problems, similar to what we do for \emph{OR} operators, but incorporates inline processing with our predicate function. Instead of starting from the IVF list with the lowest specificity, we launch parallel searches in each label's IVF list. Each search applies the predicate function during graph traversal to verify if points contain all other labels in the query. For instance, when searching for points with labels A \emph{AND} B, two parallel searches are launched: one in label A's IVF list filtering for label B, and another in label B's IVF list filtering for label A. Each search is executed by a dedicated CTA. After all CTAs complete their searches, the final results are obtained by merging the top-K candidates from all parallel searches. This method delivers higher QPS, particularly when the recall target is high (e.g., $>$90\%). 


\paragraph{\ivfbfs \emph{AND} search for \ls.}
For queries with multiple labels connected by \emph{AND} operations in the low-specificity partition, we select the IVF list corresponding to the label with the lowest specificity. This choice is sufficient since BFS will traverse all points in the selected list. Before performing a brute-force search within the selected IVF list, we apply our predicate function to verify whether each point satisfies all query labels. This pre-processing strategy effectively eliminates unnecessary distance computations for points that do not meet all label constraints.


\section{Evaluation}
\label{sec:evaluation}

In this section, we experimentally evaluate \name.


\subsection{Evaluation Methodology}

\noindent
\textbf{Datasets.} We use public datasets from prior work to evaluate our system, as listed below:
\begin{itemize}
    \item \textbf{YFCC.} \revisionB{
        The YFCC-10M dataset~\cite{yfcc}, utilized in the NeurIPS’23 Big-ANN Competition Filter Track~\cite{bigann}, contains 10 million images transformed into CLIP~\cite{clip} embeddings with 192 dimensions. Each is associated with labels extracted from multiple sources including image descriptions, camera model information, capture year, and location data. The dataset features 200,386 unique labels ($|\mathcal{L}|=200,386$)  following a highly skewed distribution -- a small number of labels appear frequently while most appear rarely. On average, each data point is associated with 10.8 labels.
    }
    \item \textbf{WIKI-ANN.} \revisionB{
        Wiki-ANN~\cite{wikiann} is a dataset introduced in \cite{wikiann-paper}, designed to evaluate AND predicates. Each query in the dataset is associated with two labels connected by an AND operator. The dataset contains 35 million Wikipedia passages transformed into embeddings with 768 dimensions. The set of all possible labels $\mathcal{L}$ consists of the 4,000 most frequent words in passages, where each passage's labels are those words from $\mathcal{L}$ that appear in its text. With the total embedding size exceeding 100GB, which surpasses single GPU memory capacity, we use a 1-million-point subset for our experiments, with each data point having an average of 22.5 labels.
    }
    \item \textbf{Semi-synthetic dataset.} Following the approach used in Filtered-DiskANN~\cite{filtered-diskann}, we employ Zipf's law to generate labels for the base dataset, effectively simulating the skewed distributions commonly observed in real-world applications. 
    \revisionB{
     For the base dataset, we use \textbf{SIFT-1M}(D=128), with a set of $|\mathcal{L}|=50$ unique labels, and an average of 3.17 labels assigned per data point. Additionally, to assess the scalability of our algorithm, we use the same method to generate $|\mathcal{L}|=2,500$ unique labels for the \textbf{DEEP-50M}(D=96) dataset, with an average of 5.88 labels per data point.
    }
\end{itemize}

\noindent
\textbf{Metrics.} We evaluate Filtered-ANNS search efficiency using recall and queries processed per second (QPS). The recall measures the fraction of the top-$K$ query results retrieved by the Filtered-ANNS that match exact nearest neighbors with query labels(measured by brute force search). In practice, we usually choose $K=10$ and take QPS at 90\% recall as an important metric. In addition, we also measure average latency (time to run a single batch) in latency mode (single-batch mode), and memory footprint consumption of our algorithm. 

\noindent
\textbf{Baselines.} We compare with the following baseline methods:

\begin{itemize}
    \item Filtered-DiskANN~\cite{filtered-diskann}: We include the two algorithms from Filtered-DiskANN: FilteredVamana and StichedVamana.  We use the parameters listed in the Filtered-DiskANN paper to build the index and search.
    \revisionB{
    \item FAISS~\cite{faiss}: The baseline from the NeurIPS'23 Competition BigANN Filter track.
    }
    \item IVF\textsuperscript{2}~\cite{ivf2}: The best-performing solution from the NeurIPS'23 Competition BigANN Filter track that uses CPU for filtered-ANNS. It is tailored for the YFCC dataset.
    \item CAGRA + Post-processing (CAGRA-Post)~\cite{cagra}: 
    We include a strong baseline by extending CAGRA with post-processing. We build a CAGRA graph for the entire dataset, query as usual, and employ post-processing to select only those results returned by the index that match the query filter. 
    \item CAGRA + Inline-processing (CAGRA-Inline)~\cite{cagra}: This offers a stronger baseline by extending the state-of-the-art GPU-based ANNS solution CAGRA with on-the-fly filtering, i.e., the graph traversal skips points in the graph that do not match the filters of the query.  
\end{itemize}

\noindent
\textbf{Testbeds} Experiments are conducted on nodes with NVIDIA A100 GPU (40GB) and AMD EPYC 7763 CPU (2.45 GHz, 128 cores, 256GB memory) and NVIDIA Grace Hopper GH200 Superchip (96GB).

\subsection{Main Result}

Our first experiment demonstrates the search performance across different datasets. We compare \name with the state-of-the-art CPU-based Filtered-DiskANN, IVF\textsuperscript{2}, and our GPU-based baselines CAGRA-Inline and CAGRA-Post. For Filtered-DiskANN, we build graphs with $L = 90$ and $R=96$ for FilteredVamana, $R_{small} = 32$, $L = 100$ and $R_{stitched} = 64$ for StitchedVamana; For IVF\textsuperscript{2} we use parameters suggested in \cite{big-ann-repo}. 
For CAGRA-Inline and CAGRA-Post we build a single CAGRA graph with R = 32 as suggested in ~\cite{cagra}. For \name, we set the specificity threshold \(T = 2000\), and we build the CAGRA graphs with R = 16 for the \hs group.

\begin{figure*}[!ht]
    \centering
    \begin{minipage}{\textwidth}
        \centering
        \includegraphics[width=0.75 \linewidth]{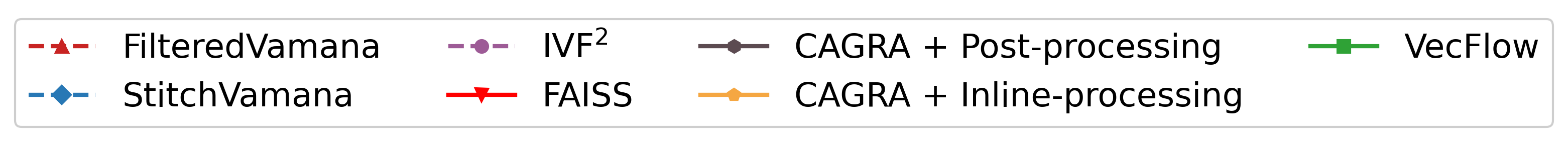}
    \end{minipage}
    \begin{minipage}{0.245\textwidth}
        \includegraphics[width=\linewidth]
        {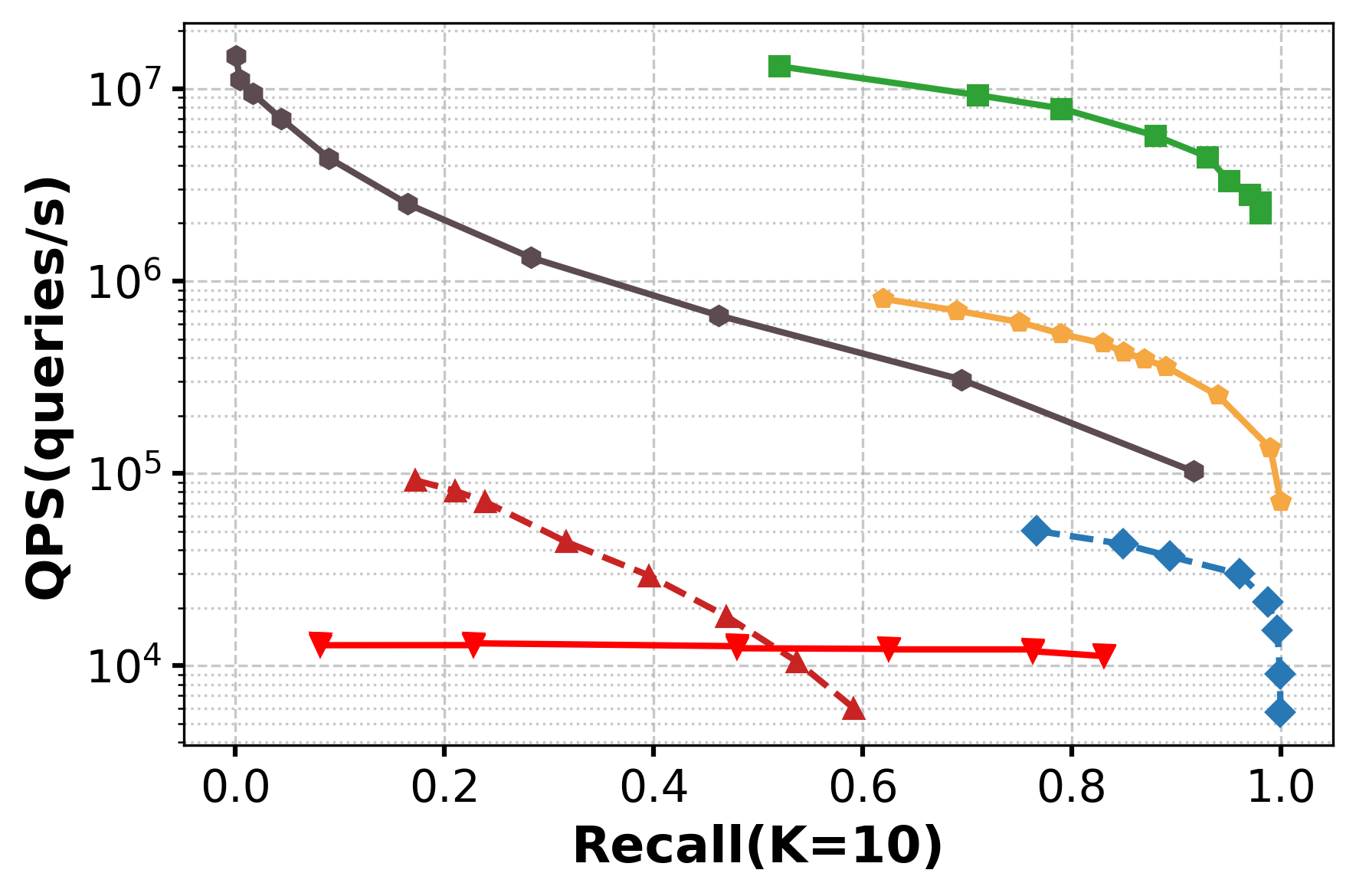}
        \caption*{(a) SIFT-1M}
    \end{minipage} 
    \begin{minipage}{0.245\textwidth}
        \includegraphics[width=\linewidth]
        {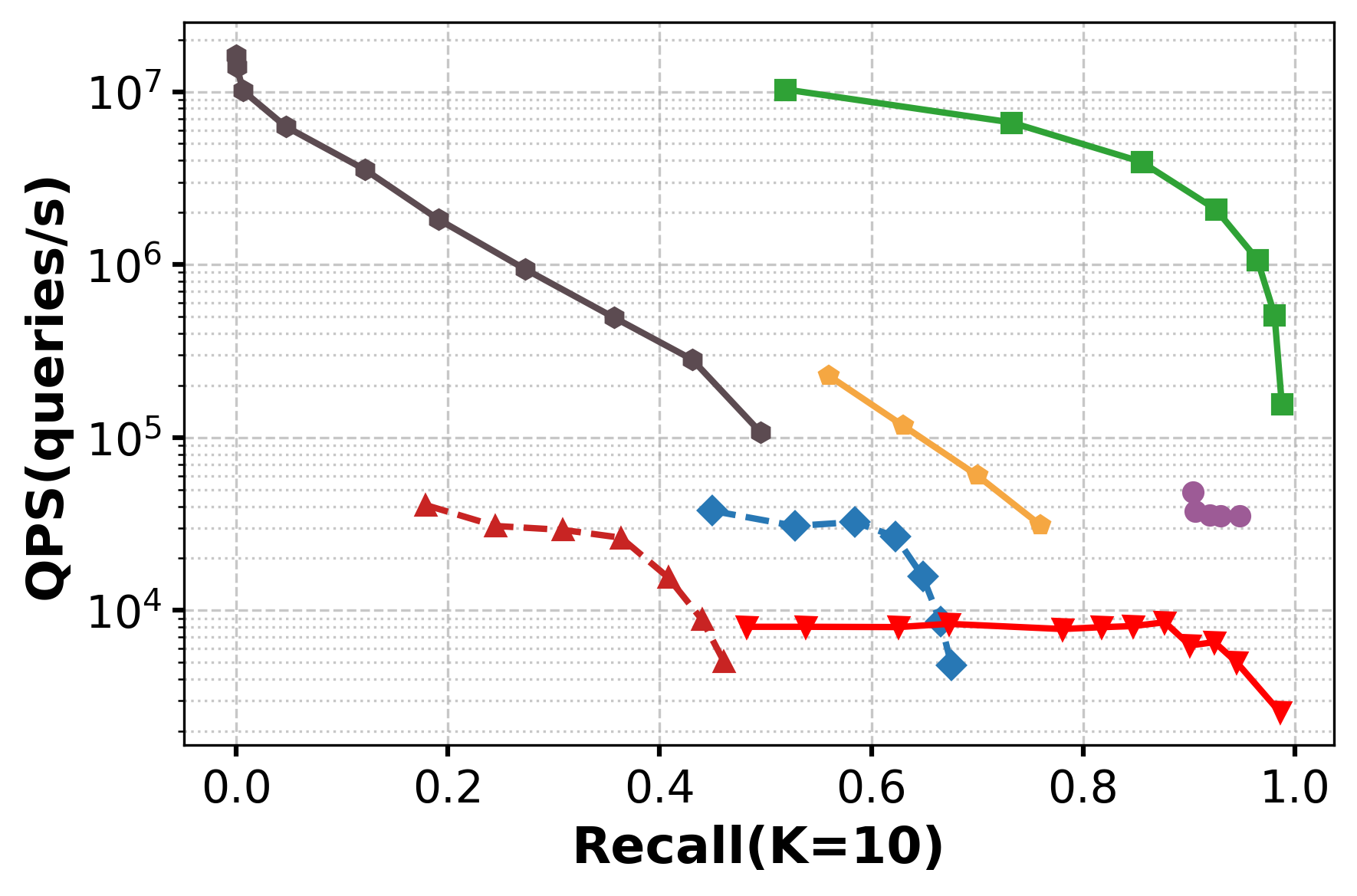}
        \caption*{(b) YFCC-10M}
    \end{minipage}
    \begin{minipage}{0.245\textwidth}
        \includegraphics[width=\linewidth]
        {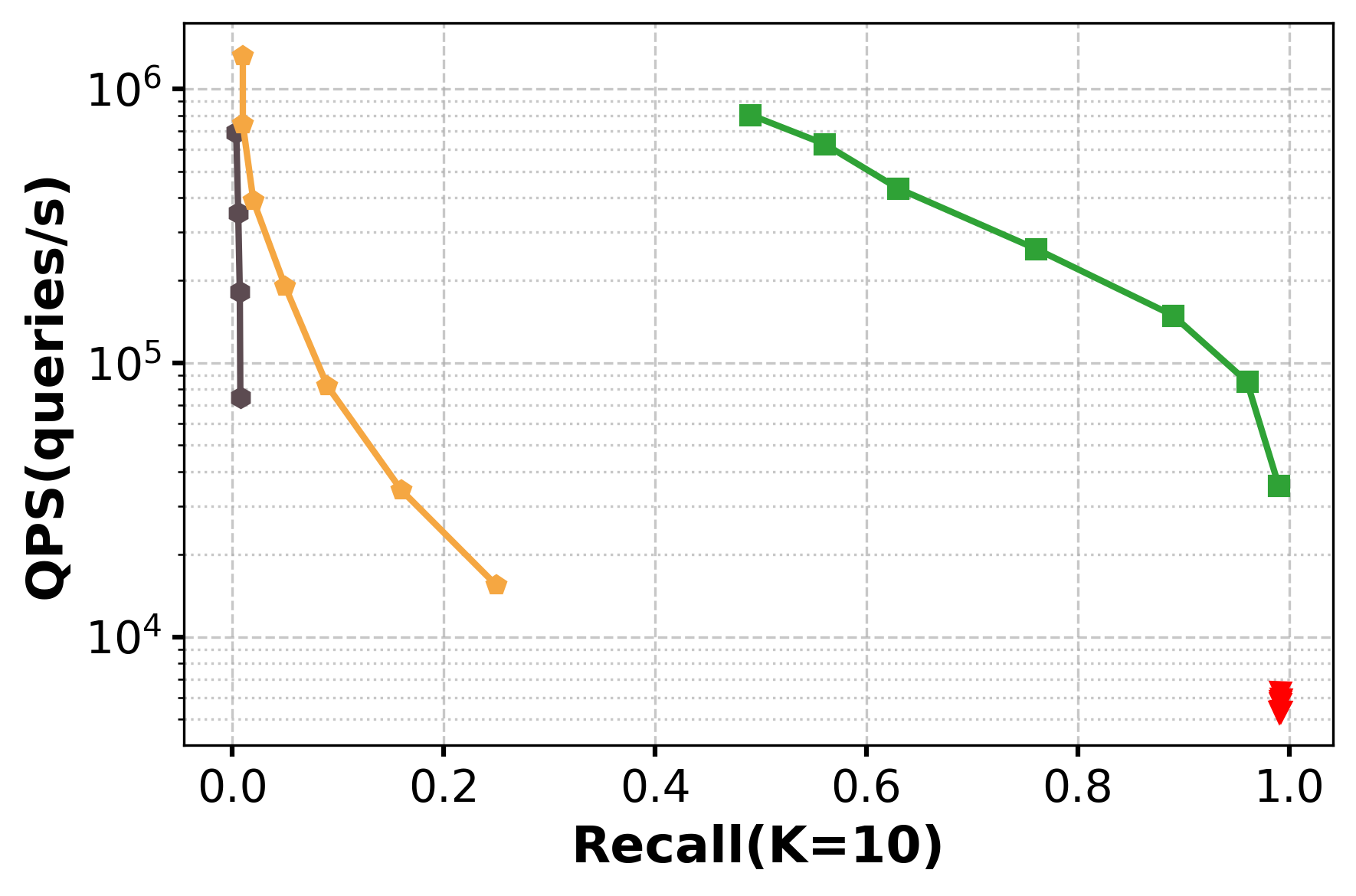}
        \caption*{(c) WIKI-1M}
    \end{minipage}
    \begin{minipage}{0.245\textwidth}
        \includegraphics[width=\linewidth]
        {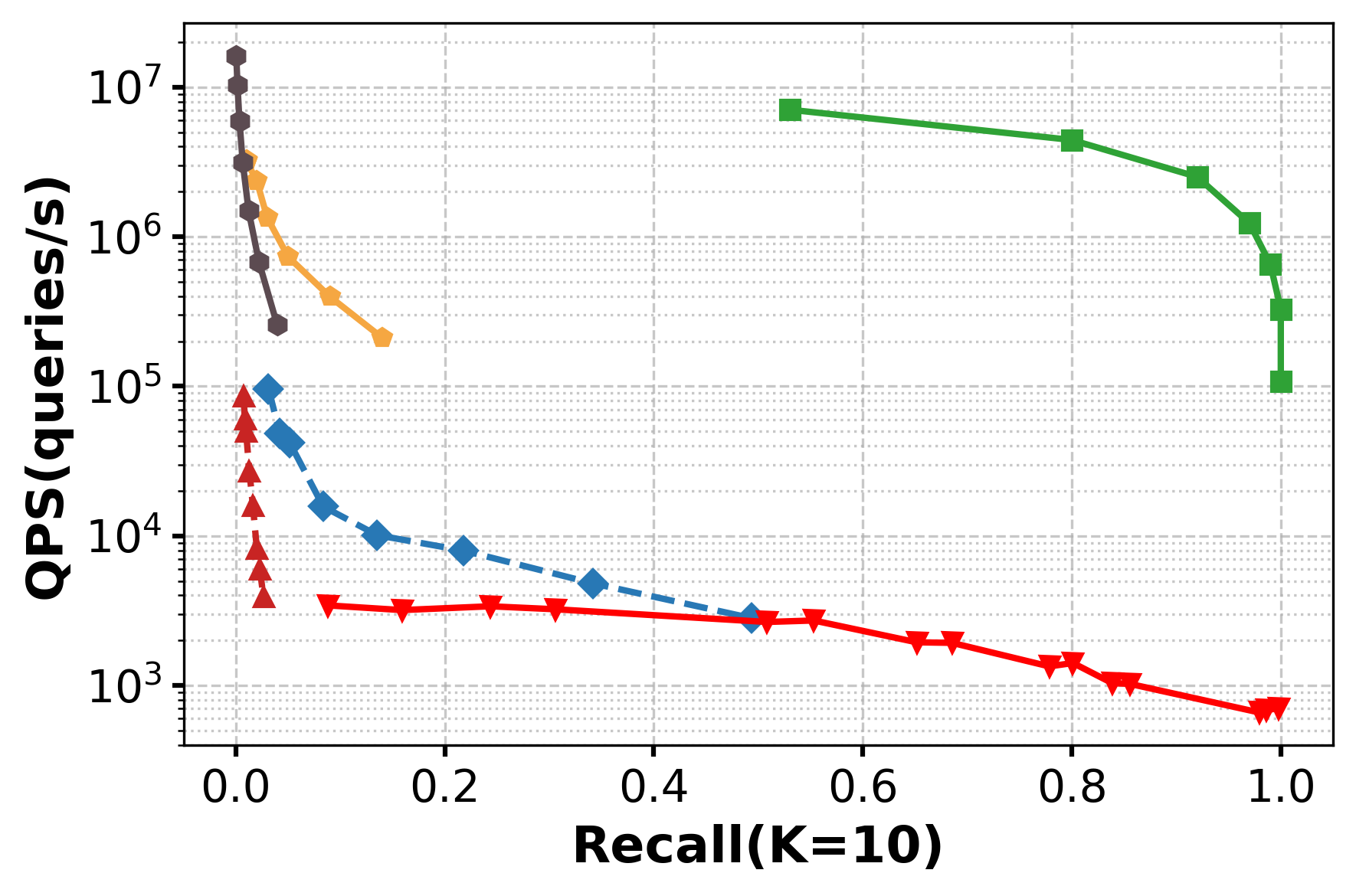}
        \caption*{(d) DEEP-50M}
    \end{minipage}
    \caption{\revisionB{The QPS vs. recall comparison results on (a) SIFT, (b) YFCC, (c) WIKI and (d) DEEP respectively.}}
    \label{fig:qps-vs-recall}
\end{figure*}

Our results, as shown in \fref{fig:qps-vs-recall} that follows the style of \cite{cagra,filtered-diskann,ivf2}, highlight \name's significant advantage in QPS-vs-recall across different datasets. We achieve these plots by varying search parameters for each algorithm to improve recall and reduce throughput: search width ($L_s$ from ~\cite{filtered-diskann}) for FilteredVamana and StitchedVamana, and $itopk$ for the CAGRA-based approaches and \name.
Overall, it is worth noting that \name achieves million-scale QPS at 90\% recall (K=10), which is one to two orders of magnitude higher than both CPU and GPU baselines.

On the semi-synthetic SIFT-1M dataset, FilteredVamana achieves a maximum recall of 60\% even with a large  $L_s$ value.
This suggests that traditional proximity graph with inline-filtering cannot effectively capture label-specific information within the graph's connections. StitchedVamana performs better than FilteredVamana, achieving 37K QPS at 90\% recall and capable of reaching 100\% recall. This is because the stitched graph retains some label-related connections, which helps improve filtered graph search. 
CAGRA with post-processing and CAGRA with inline-processing both outperform StitchedVamana, thanks to CAGRA's highly optimized search capability on GPUs. In comparison to those existing methods, \name achieves an impressive 5M QPS at 90\% recall, which is 135 times higher than StitchedVamana. \name is able to achieve high performance because the dataset has only 50 labels, all of which are high specificity (the least specific label cluster contains 14,000 data points). This causes each query to only search a small IVF list, resulting in significantly reduced distance computation. In addition, \name leverages CAGRA's high performance within each IVF, achieving high throughput and high recall by leveraging GPU-friendly graph traversal. This use case illustrates that \name performs very well for datasets without extreme label distributions.

Despite the more extreme filter labels for the YFCC-10M dataset, we observe similar trends with the performance of various methods. FilteredVamana struggles with lower recall levels even with large $L_s$, highlighting its limitations in capturing label-specific connections on a larger dataset. StitchedVamana also begins to struggle to achieve high recall, getting less than 70\% even with very large $L_s$. 
This is because YFCC contains over 200K labels, many of which exhibit low specificity. \revisionB{FAISS achieves high recall but with very low QPS because it needs to explore a larger portion of vectors.} For this dataset, IVF\textsuperscript{2} performs the best among the CPU baselines, achieving 48K QPS at 90\% recall. This is because it employs a label-centric IVF method similar to ours, which is particularly well-suited for datasets like YFCC that have a large number of labels with low specificity.
CAGRA + Post-processing and CAGRA + Inline-processing outperform StitchedVamana again, leveraging GPU efficiency for rapid search, but are unable to achieve more than 80\% recall. Finally, \name continues to outperform all other methods, achieving a remarkable QPS of 2.6M at 90\% recall, demonstrating its scalability and effectiveness on a real dataset as large as YFCC-10M and as complex as 200K labels. 

The WIKI-1M dataset is particularly challenging due to the multi-label search queries. FilteredVamana and StitchedVamana are not included in the results because they do not support multi-label filter searches. 
\revisionB{FAISS achieves high recall by performing BFS for almost every query due to the extremely small intersection between two labels. However, the time taken to find the intersection for each query is too slow.}
Both CAGRA Post-processing and Inline-processing methods achieve nearly 0 recall due to the extremely small intersection between two labels, making it unlikely to find points with the same label as the query. In contrast, \name still delivers high performance, achieving 150K QPS at 90\% recall. This is attributed to our highly efficient predicate function and optimized search policy for multi-label queries described in \sref{subsec:multi-label}. 
Finally, our evaluation on DEEP-50M, which is the largest dataset that can accommodate on a NVidia A100 40GB GPU, reveals that \name achieves around 3M QPS at 90\% recall, whereas other methods struggle to surpass 50\% recall and many fail to even reach 20\% recall, which is consistent with our previous analysis over SIFT-1M.  


\subsection{Analysis Results}

Next we will justify each part of our design by conducting related performance evaluations. We also evaluate \name's effectiveness over other type of GPUs.  

\begin{figure}[!ht]
    \centering
    \includegraphics[width=\linewidth]{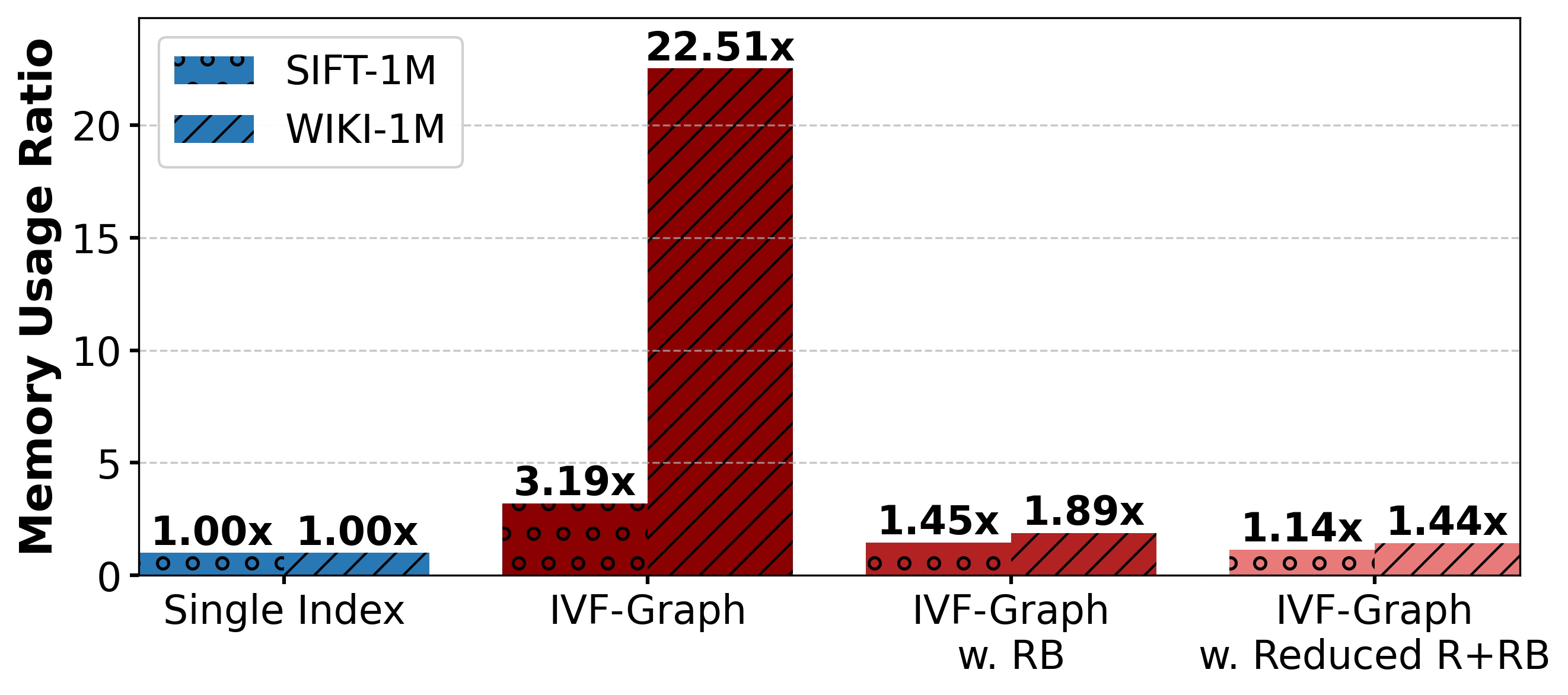}
    \caption{\revisionB{Memory efficiency of redundancy-bypassing.}}
    \label{fig:rb_memory}
\minjia{In that case, can you change the x-axis labels to Single index, IVF-Graph, IVF-Graph w. Reduced R, IVF-Graph w. Reduced R + RB, IVF-Graph + RB?}
\jingyi{Ok.}
\end{figure}

\paragraph{How does redundancy-bypassing \ivfgraph bring benefits?} In \sref{subsubsec:ivf-graph} we introduce the \ivfgraph with redundancy-bypassing (RB) optimization. We evaluate its effectiveness over SIFT-1M and WIKI-1M datasets by comparing (i) single-index graph-based approach using CAGRA (ii) IVF-Graph without RB, and (iii) IVF-Graph with RB, \revisionB{in the same graph degree of 32}, where "without RB" means without additional layers of indirection but with
data vectors copied for each IVF-Graph. In addition, we also report IVF-Graph with reduced degree. \fref{fig:rb_memory} shows that without RB, the memory consumption of IVF-Graph increases dramatically, especially when each data point is associated with a large number of labels. 
In contrast, the RB optimization drastically reducing index size to \revisionB{\(1.45\times\) and \(1.89\times\)} for the single-index baseline for SIFT-1M and WIKI-1M, respectively. This is achieved through our local virtual graph plus local-global mapping that enables sharing of the embedding vectors across IVF lists, effectively eliminating redundancy.

\begin{figure}[!ht]
    \centering
    \includegraphics[width=0.7\linewidth]{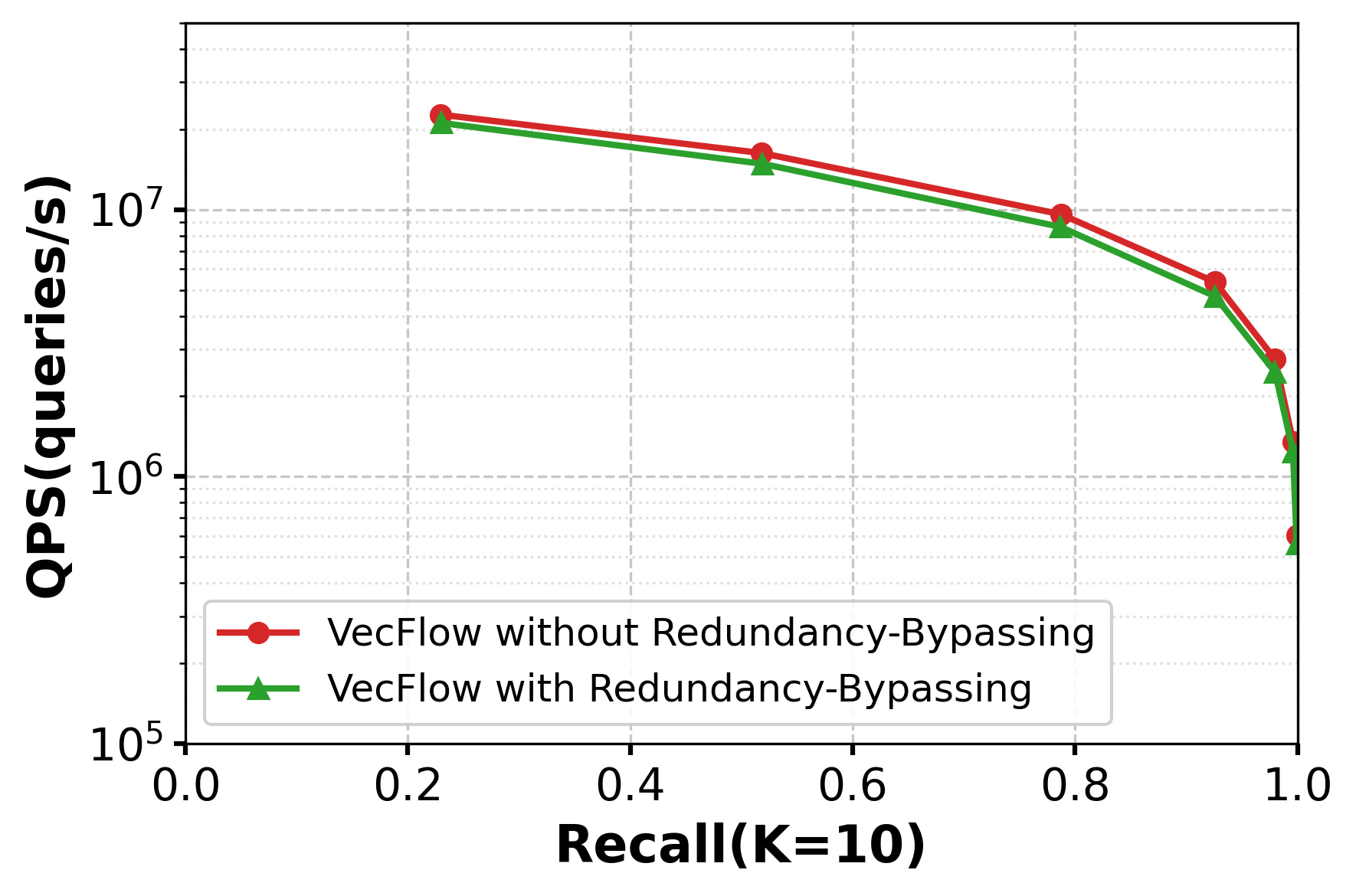}
    \caption{\revisionA{Comparison results between \name with Redundancy-bypassing and \name without Redundancy-bypassing.}}
    \label{fig:unfilter}
\end{figure}

\revisionA{
\paragraph{What is the speed-memory trade-off of redundancy-bypassing?}
We study \name with and without RB on the SIFT dataset for filtered search. \fref{fig:unfilter} shows that at recall 0.925, \name with RB achieves 4.7M QPS while \name without RB achieves 5.3M QPS, indicating that the additional layer of indirection causes approximately 11.6\% performance loss. We consider this a desirable trade-off given that \name with RB uses only 45\% of the memory required by \name without RB as shown in \fref{fig:unfilter}.
}

\begin{figure}
    \centering
    \includegraphics[width=0.7\linewidth]{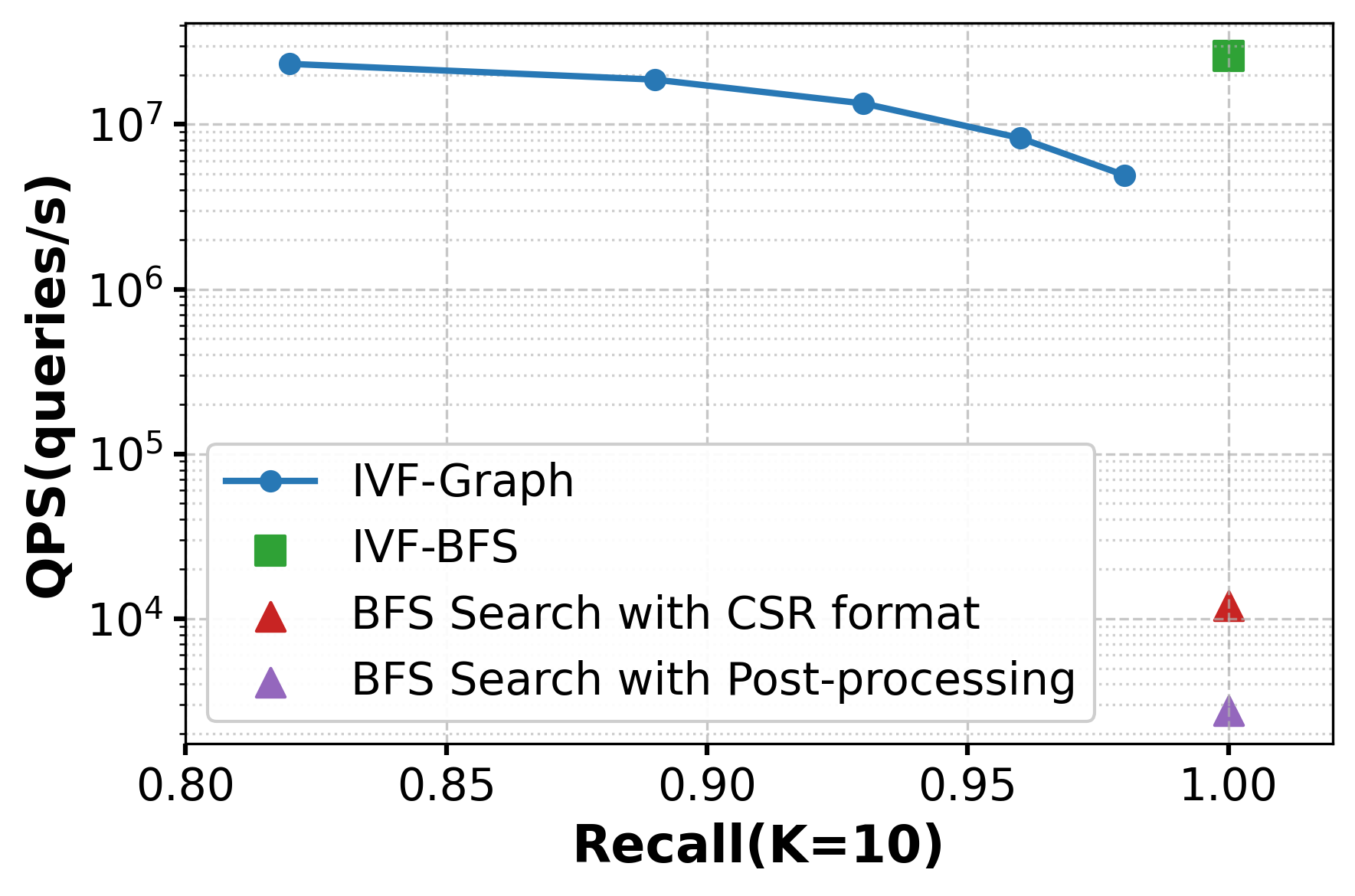}
    \caption{IVF-BFS performance analysis on YFCC dataset(T = 2000).}
    \label{fig:bfs-ablation}
\end{figure}

\paragraph{Is it necessary to perform IVF-BFS for search in \ls?} 
\fref{fig:bfs-ablation} presents the performance comparison between \ivfbfs and two baseline methods on the YFCC dataset for clusters with sizes less than 2000. The results show that \ivfbfs achieves 26M QPS, which is 2167 times and 9455 times faster compared to baseline BFS search with CSR format and BFS search with postprocessing, respectively. Additionally, it is 1.32 times faster than IVF-Graph at 90\% recall. This performance advantage comes from our interleaved scan-based kernel optimization, which makes IVF-BFS suitable for accelerating search in \ls.

\begin{figure}[!ht]
    \centering
    \begin{minipage}{0.495\linewidth}
        \centering
        \includegraphics[width=\linewidth]{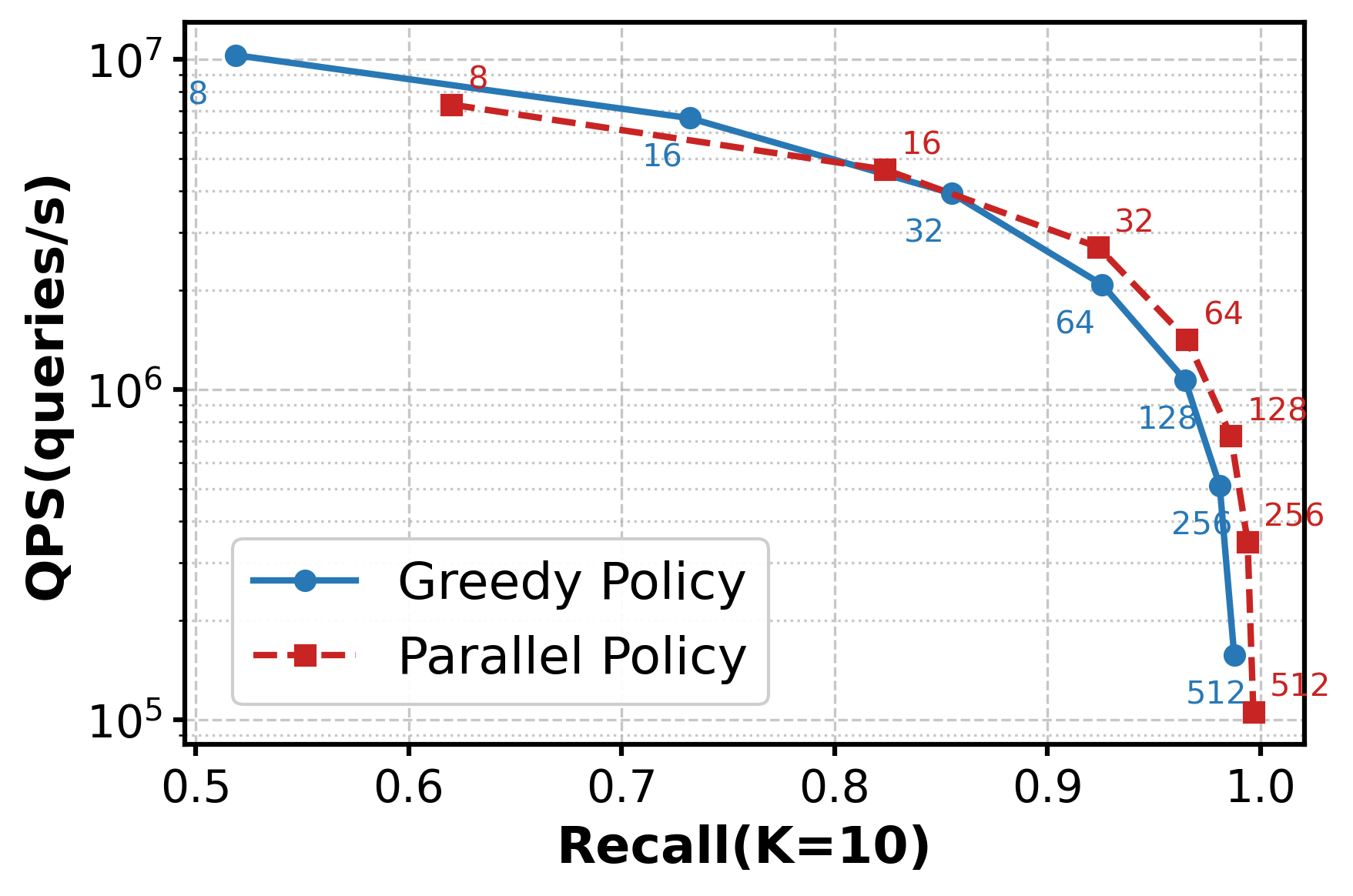}
        \caption*{(a) YFCC}
    \end{minipage}
    \begin{minipage}{0.495\linewidth}
        \centering
        \includegraphics[width=\linewidth]{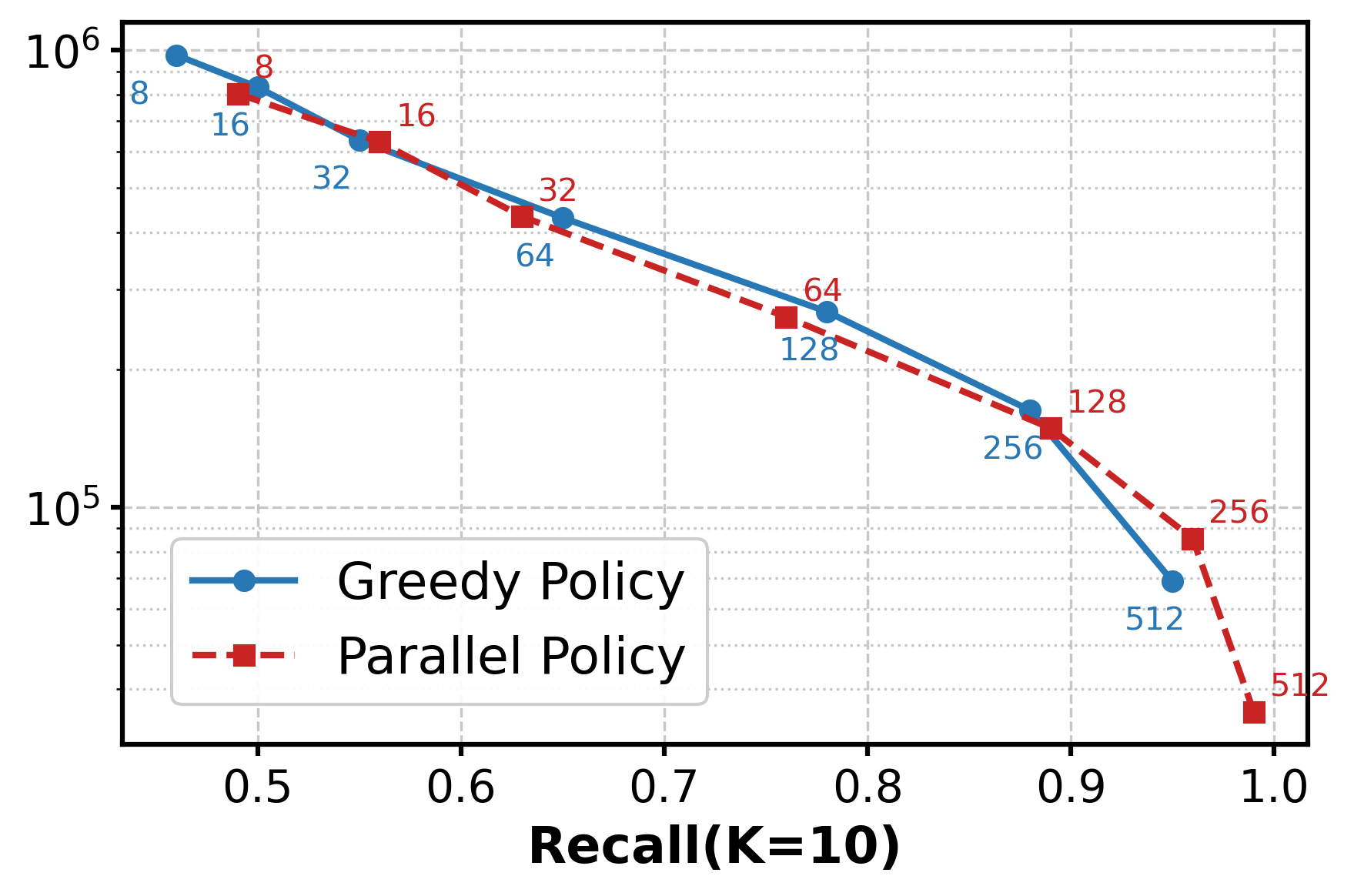}
        \caption*{(b) WIKI}
    \end{minipage}
    \caption{The comparison of multi-label search policy in \name. The \textit{itopk} size used for each point is labeled next to the corresponding data point.}
    \label{fig:mutil-label}
\end{figure}

\paragraph{How does \name's multi-label search policy perform?}
\fref{fig:mutil-label} illustrates the comparison between the two search strategies we proposed for the AND query. Here, we discuss the advantages and disadvantages of both approaches. The results show that the parallel strategy outperforms the greedy policy, especially in the high-recall region. For instance, on the wiki dataset, the greedy policy achieves a maximum recall of only 94\% even with the largest \textit{itopk}. However, with the same \textit{itopk} size, greedy policy searches only the cluster with the lowest specificity label, resulting in reduced computational costs while achieving faster search speeds. As shown in the figure, with the same \textit{itopk}, the QPS of the greedy policy is significantly higher than that of the parallel strategy, with minimal loss in accuracy.
\minjia{The last sentence needs a double-check. The figure does not have itopk information in it. Therefore, it is confusing to say "As shown in the figure, with the same \textit{itopk}". Also, it does not seem like parallel policy "consistently" outperforms greedy policy. I removed "consistently" because greedy outperforms parallel at the low recall regime. Please double check the statement.}
\jingyi{I add itopk description in this figure's caption. removing consistently is correct.}

\begin{figure}[!ht]
    \centering
    \begin{minipage}{0.48\linewidth}
    \centering
    \includegraphics[width=\linewidth]{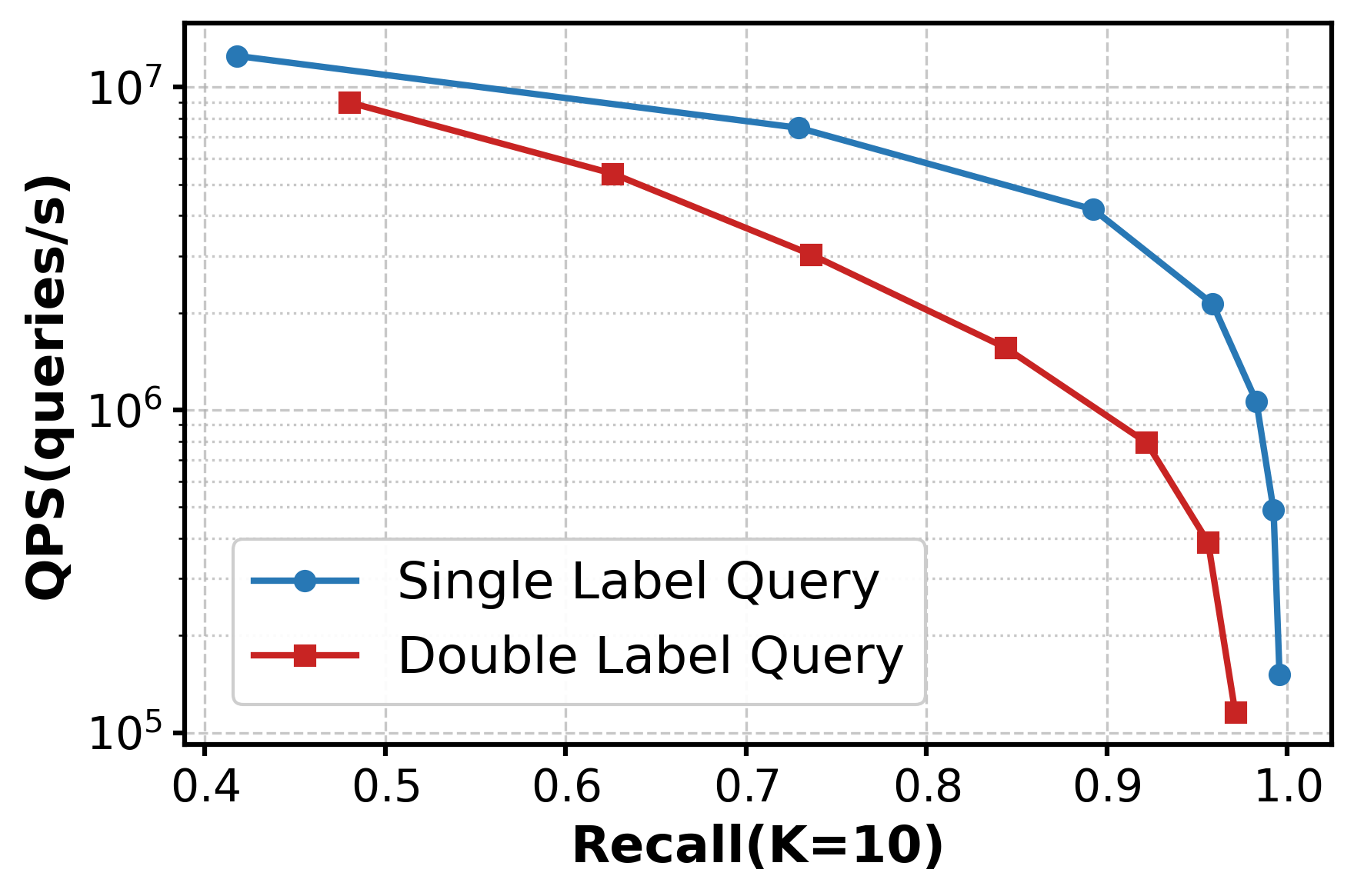}
    \caption{Performance of single label and double label query on YFCC dataset.}
    \label{fig:single-double-query}
    \end{minipage}
    ~~~~
    \begin{minipage}{0.47\linewidth}
    \centering
    \includegraphics[width=\linewidth]{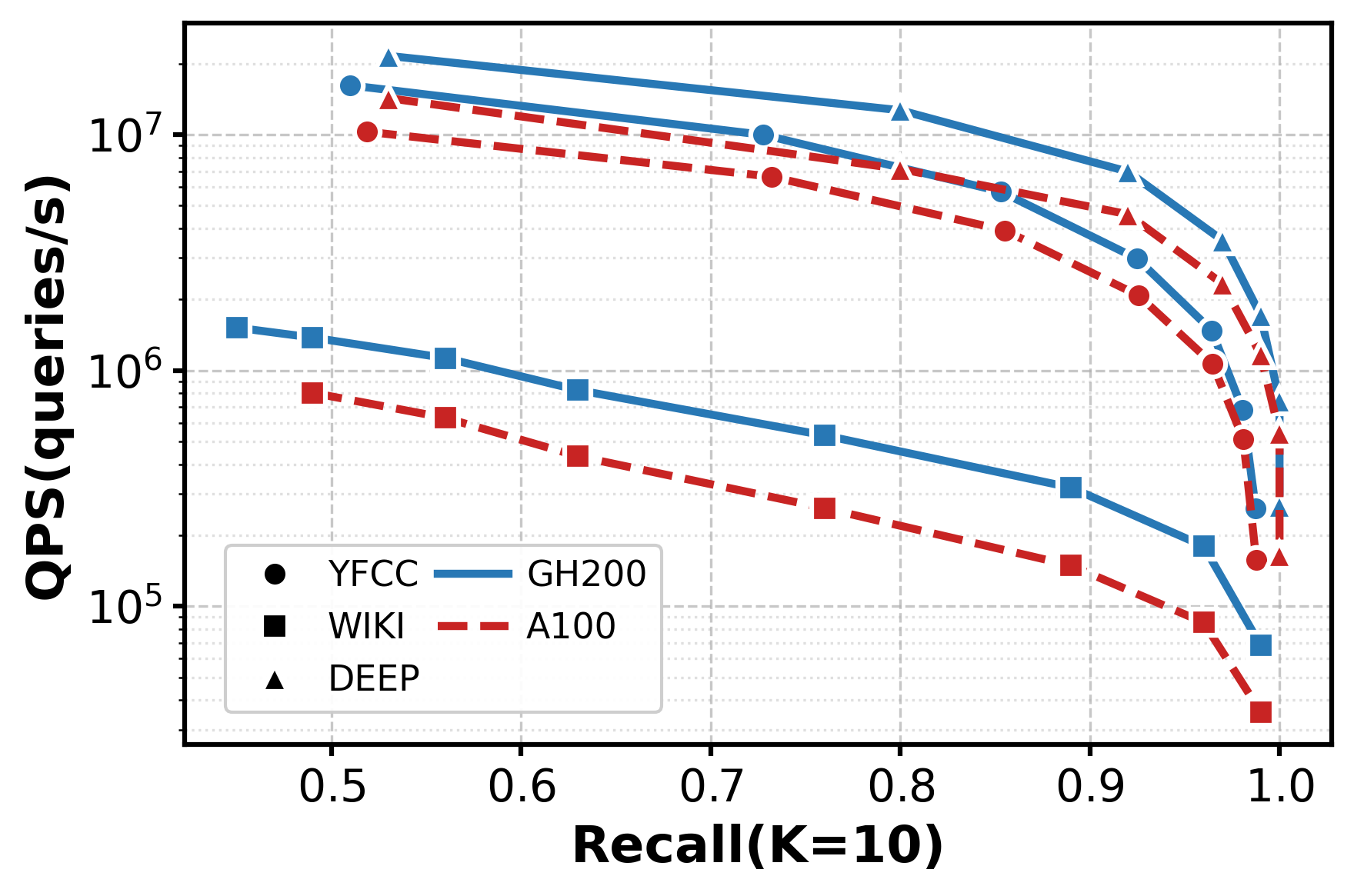}
    \caption{Performance of \name on NVidia GH200 GPU.}
    \label{fig:h200}
    \end{minipage}
\end{figure}

We also evaluate single-label and double-label queries separately on the YFCC dataset. As shown in \fref{fig:single-double-query}, both types of queries can achieve high recall of \(>97\%\). Regarding QPS, single-label queries are 4 times faster than double-label queries at 90\% recall, as the inline-processing slows down the search for double-label queries.

\begin{figure}[!ht]
    \centering
    \begin{minipage}{0.495\linewidth}
        \centering
        \includegraphics[width=\linewidth]{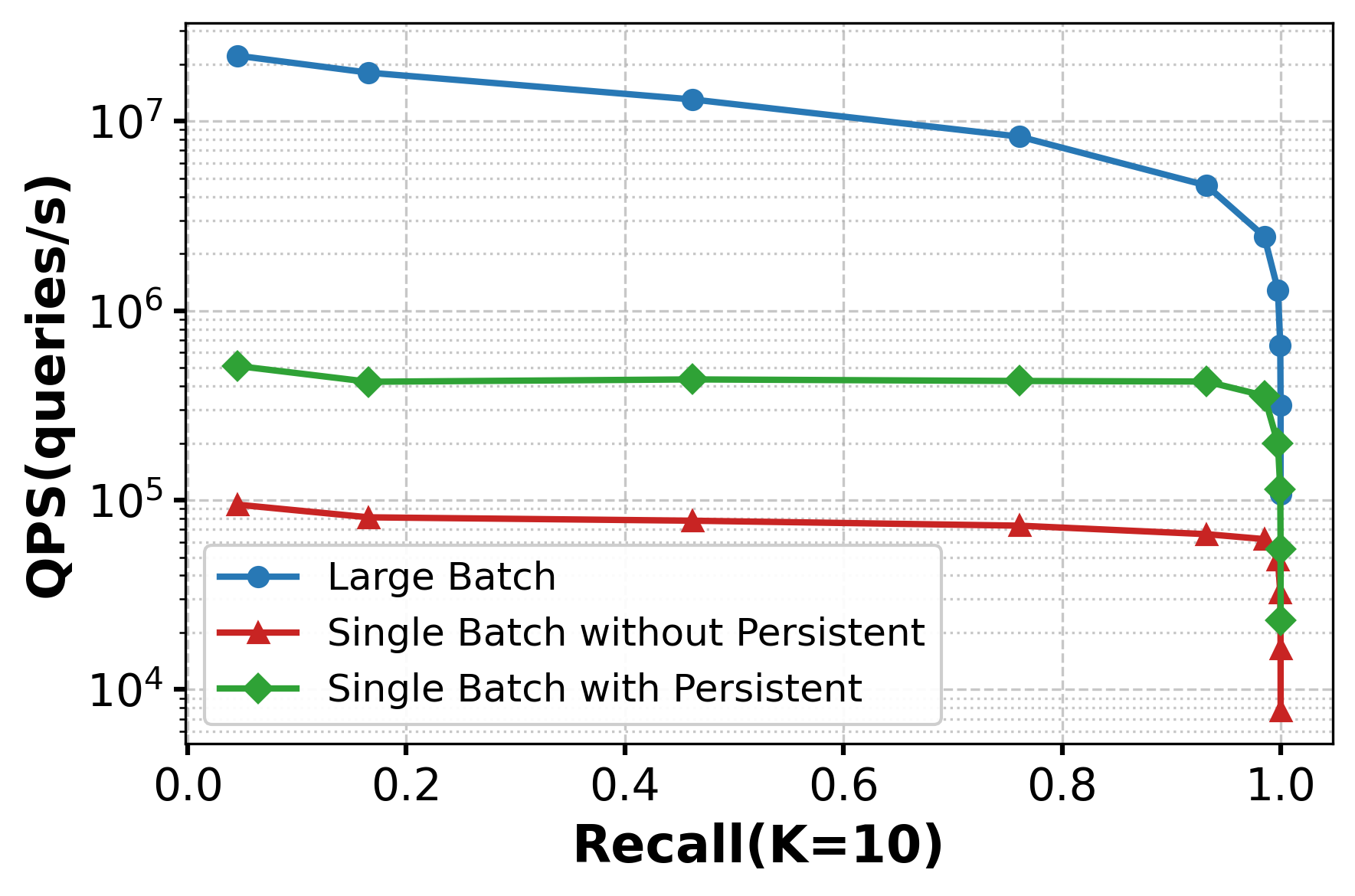}
    \end{minipage}
    \begin{minipage}{0.495\linewidth}
        \centering
        \includegraphics[width=\linewidth]{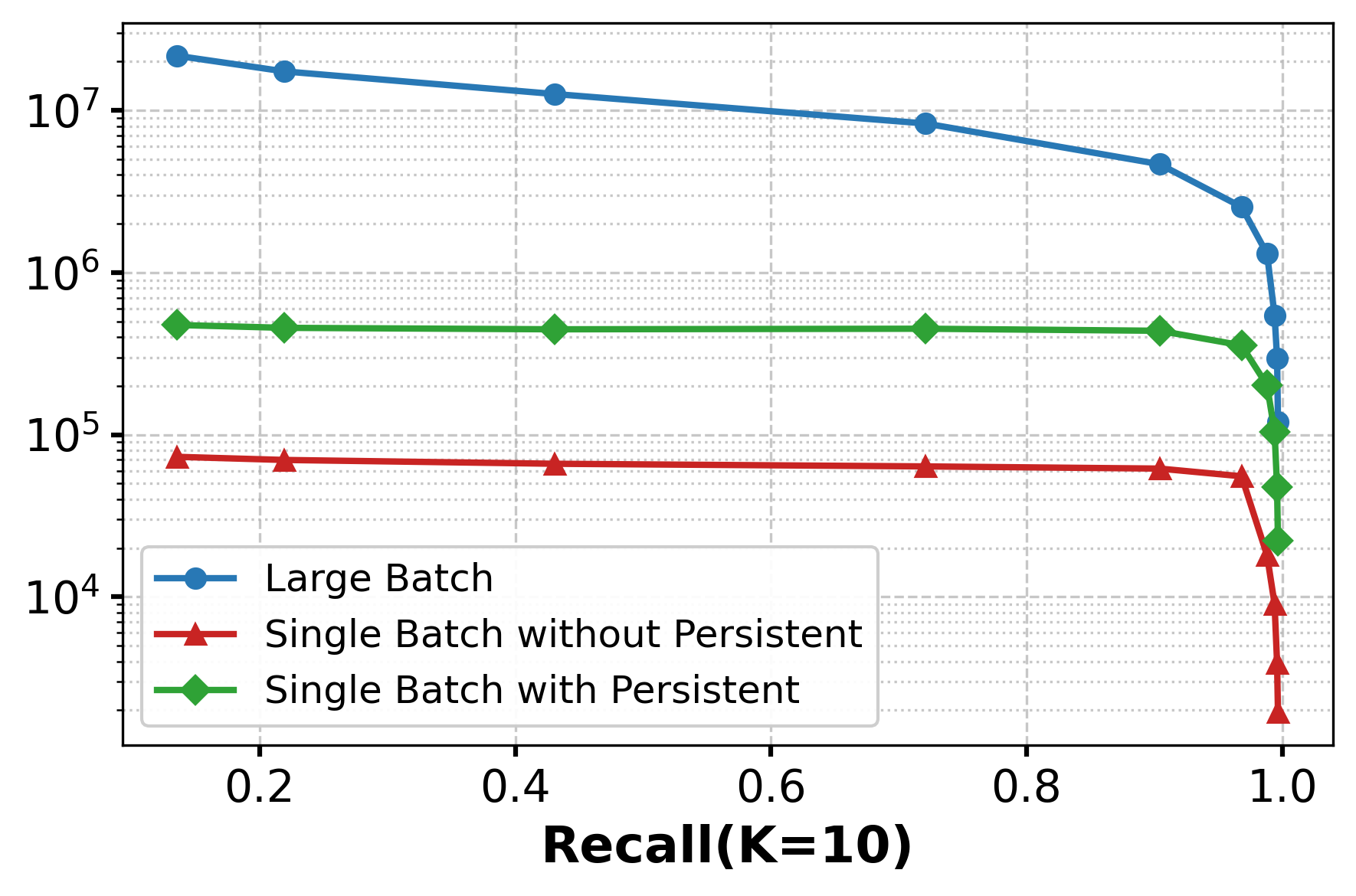}
    \end{minipage}
    \begin{minipage}{0.495\linewidth}
        \centering
        \includegraphics[width=\linewidth]{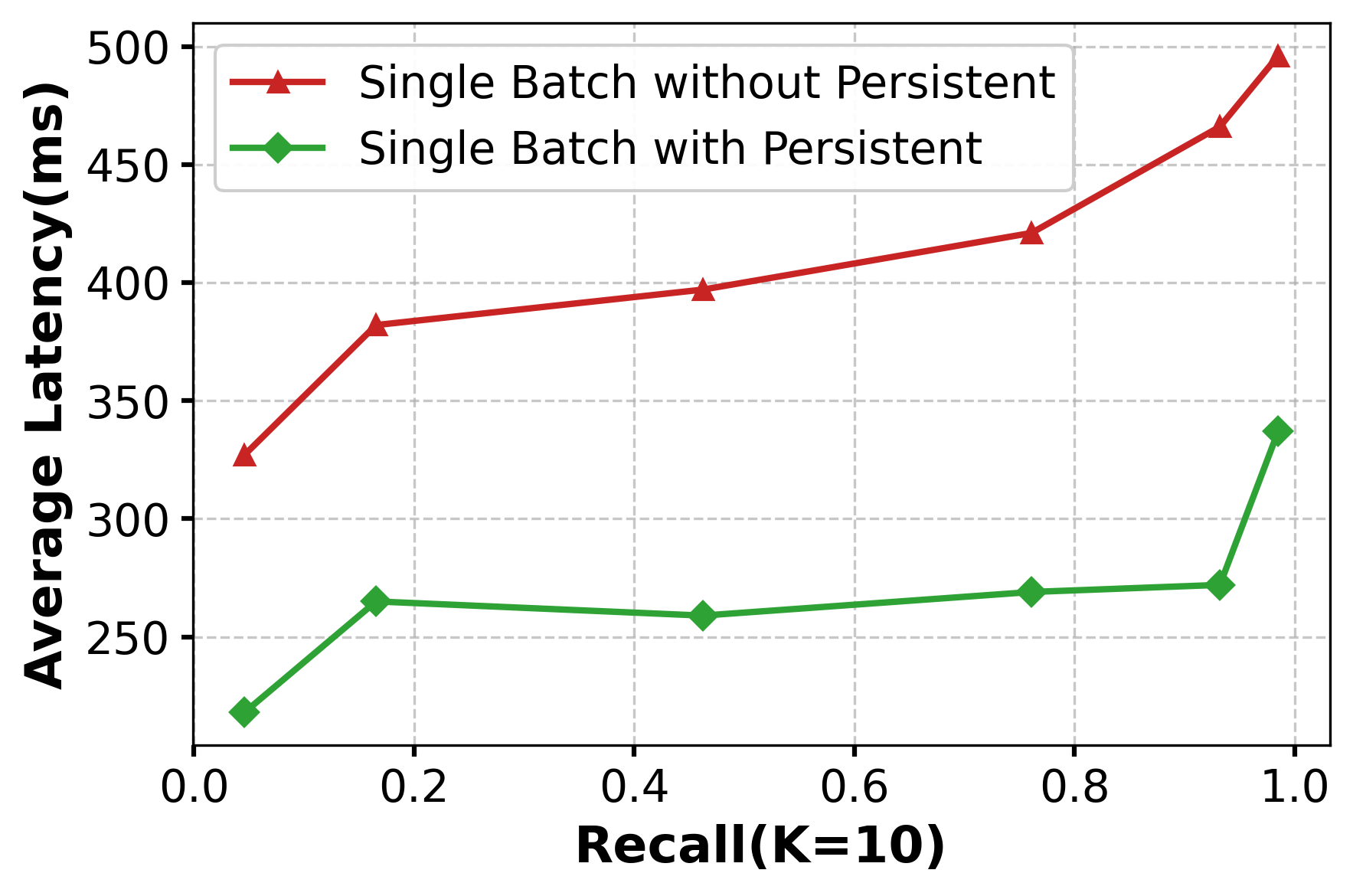}
        \caption*{(a) SIFT-1M}
    \end{minipage}
    \begin{minipage}{0.495\linewidth}
        \centering
        \includegraphics[width=\linewidth]{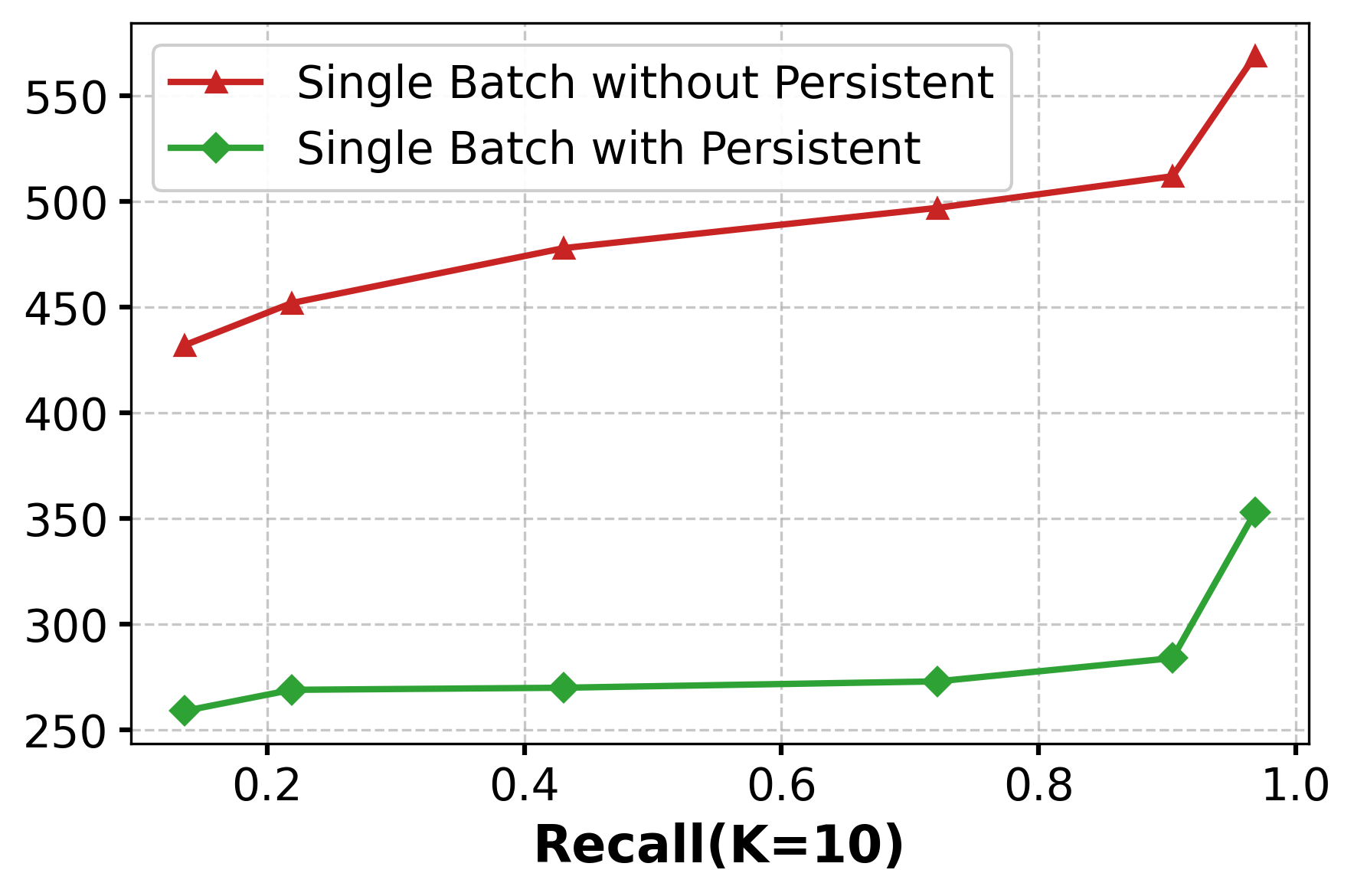}
        \caption*{(b) YFCC-10M}
    \end{minipage}
    \caption{Persistent kernel results on SIFT and YFCC datasets.}
    \label{fig:persistent-results}
\end{figure}

\paragraph{How does persistent kernel improve the performance of small batch queries?}
We test \name on single-batch mode. For QPS, the persistent kernel achieves an average speedup of 5.68x and a maximum speedup of 6.39x on the SIFT dataset than the baseline, as well as an average speedup of 6.72x and a maximum speedup of 7.08x on the YFCC dataset.
For the average latency per query, the persistent kernel achieves an average speedup of 1.54x and a maximum speedup of 1.71x on the SIFT dataset, and an average speedup of 1.73x and a maximum speedup of 1.82x on the YFCC dataset, as shown in Figure \ref{fig:persistent-results}. Persistent kernel-based search improves the QPS when batch size is small because it uses a single-kernel to process incoming queries, without launching additional kernels, largely reducing the kernel launching overhead. In addition, by continuously accepting new queries to process, it effectively increases the GPU utilization, achieving high QPS. 

\begin{figure}[!ht]
    \centering
    \includegraphics[width=0.7\linewidth]{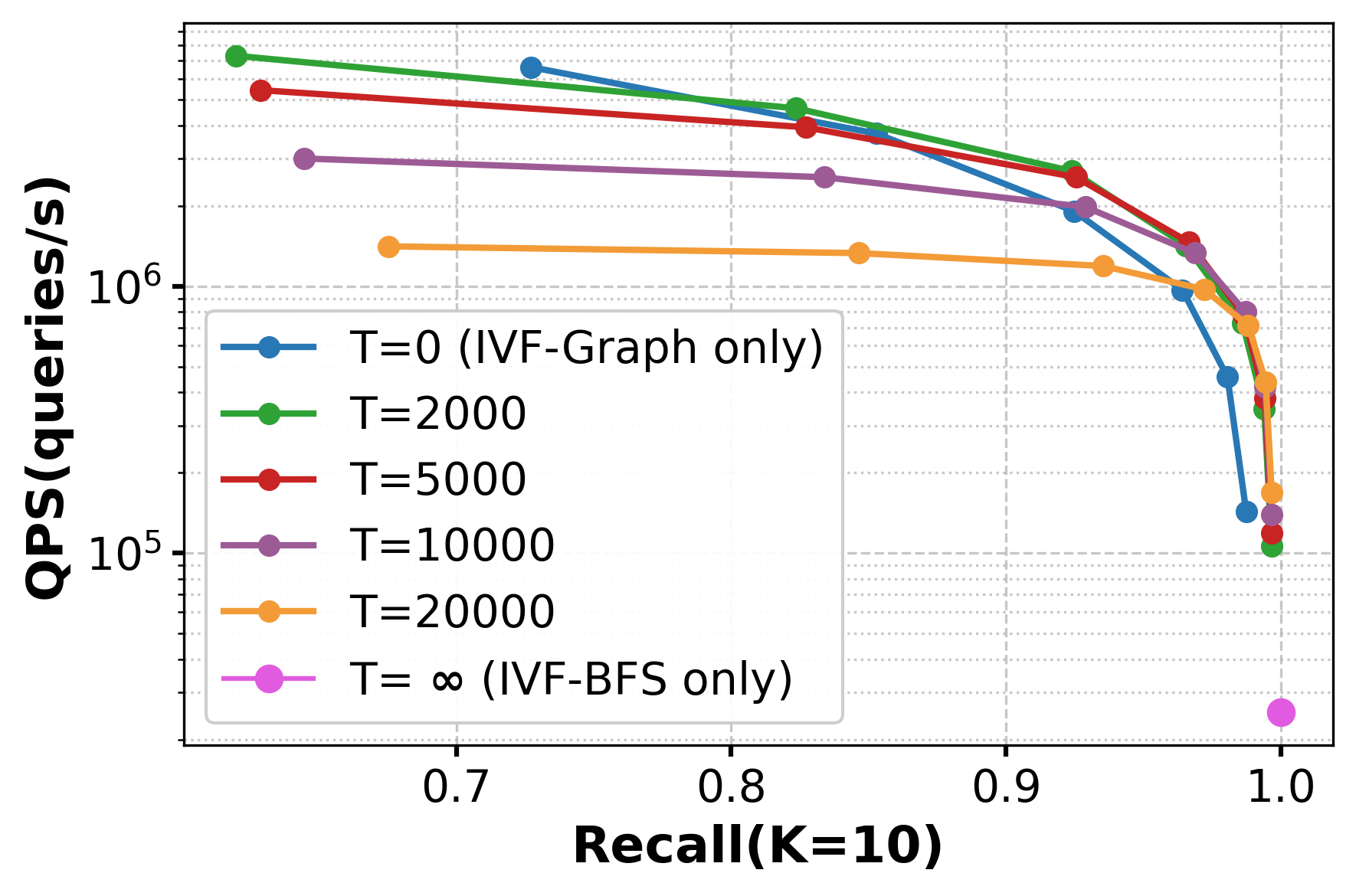}
    \caption{\name varying the specificity threshold $T$ on YFCC dataset.}
    \label{fig:threshold}
\end{figure}

\paragraph{How does the choice of specificity threshold T affect the search performance?}
To analyze the impact of the specificity threshold \( T \) on our algorithm, we vary \( T \) and observe its influence on performance, as shown in \fref{fig:threshold}. The parameter \( T \) controls the transition between IVF-BFS-only search (\( T=\infty \)) and IVF-Graph-only search (\( T=0 \)). At \( T=0 \), the graph search results in worse performance compared to BFS due to its constant overhead, as analyzed in \sref{subsubsec:decoupling}.
At moderate thresholds, such as \( T=2000 \) or \( T=5000 \), the algorithm achieves a balanced trade-off, providing both competitive QPS and high recall. As \( T \) increases further, BFS becomes less efficient due to its exhaustive nature. This demonstrates that the threshold \( T \) is a key parameter for tuning to get the best performance.

\paragraph{How does \name perform on different GPUs?} 
\fref{fig:h200} shows the comparison of \name over different types of GPUs. On GH200, \name achieves 8.3M, 4M, and 300K QPS on the DEEP, YFCC, and WIKI datasets, respectively, which are 1.66x, 1.6x, and 2x faster than the results on A100. This is expected because H200 GPU has a memory bandwidth of 4.8TB/s, which is 2.4x higher than A100. 


\revisionA{
    \paragraph{How does \name perform on real-world production workloads?}
    We evaluated VecFlow on a real-world recommendation dataset containing 4 million vectors (D=64) with over 10,000 labels showing a long-tailed distribution. Using a specificity threshold of 2000, VecFlow processed 110 high-specificity labels through IVF-Graph and 10,000 low-specificity labels with IVF-BFS. At 92\% recall, VecFlow achieves 6.51 million QPS, while baseline approaches struggle with accuracy -- CAGRA with post-processing reaches only 53\% recall even at reduced throughput (175K QPS), and CAGRA with inline filtering achieves only 55\% recall at 135K QPS. This demonstrates VecFlow's ability to maintain both high throughput and high accuracy in real production environments, where filtered-ANNS is critical for personalized recommendations.
}

\begin{figure}[!ht]
    \centering
    \includegraphics[width=0.8\linewidth]{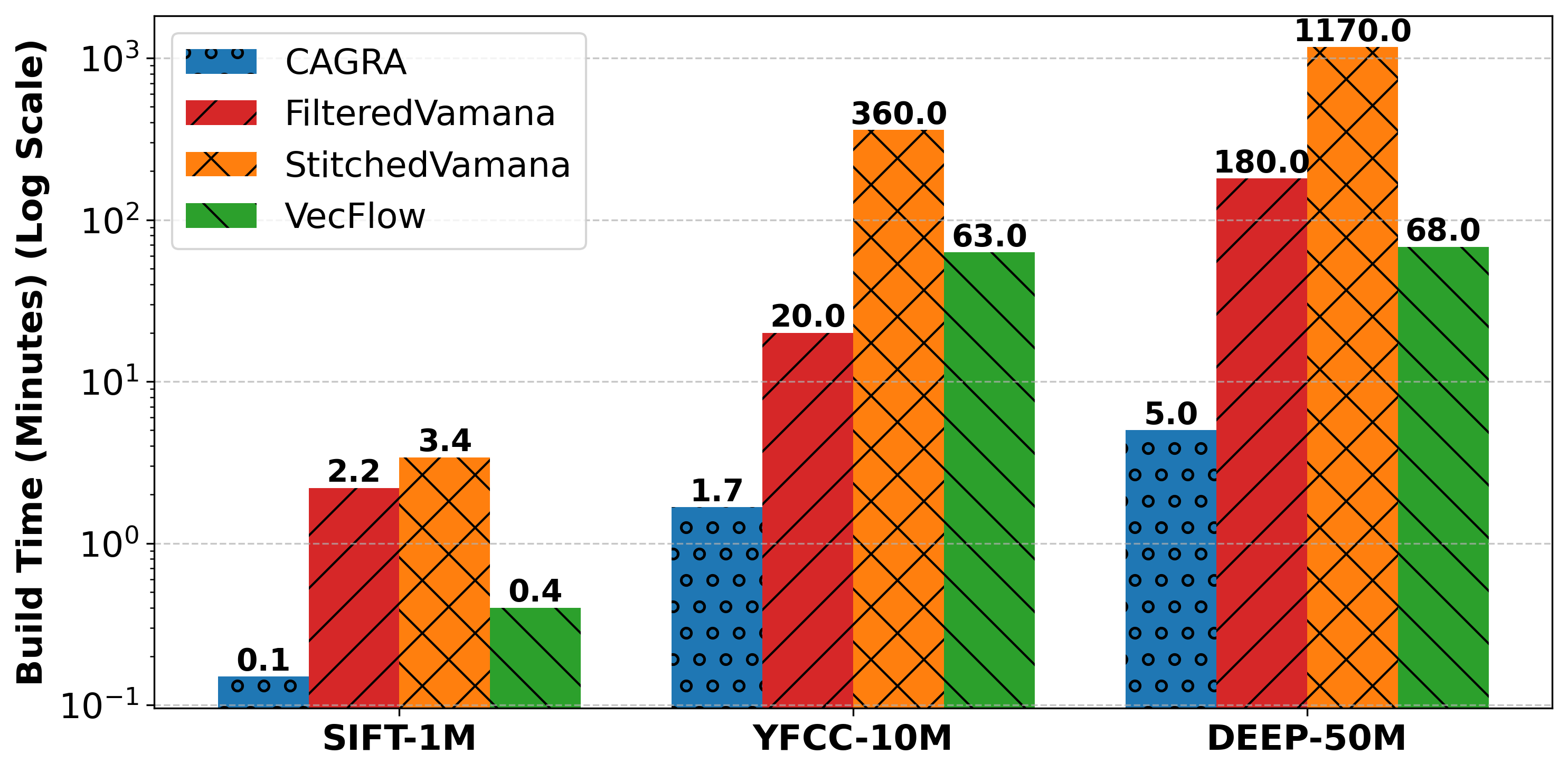}
    \caption{Index construction time comparison.}
    \label{fig:build-time}
\end{figure}

\paragraph{Does \name add high index construction time?}
\fref{fig:build-time} compares the index construction times for CAGRA, FilteredVamana, StitchedVamana, and \name. 
CAGRA demonstrates exceptional index build speed, primarily due to its efficient GPU-based implementation. In contrast, FilteredVamana and StitchedVamana, which rely on CPU computation, show considerably slower performance. StitchedVamana, in particular, is further hindered by the additional overhead of constructing and stitching separate indices.
\revisionB{For our method, \name first collects the data IDs associated with each label. For \ivfgraph, \name aggregates the vectors belonging to each label  $X_l = \{X_i | i \in C_l\}$ with $C_l$ which contains data point indices and use multiple CUDA streams to construct different label graphs concurrently while sharing the same memory pool. After construction, \name releases the vectors per label since \name's final index shares only one global vector storage. For IVF-BFS, \name has a highly optimized kernel that arranges the data in an interleaved format without requiring computational operations like ANN-based graph indexing, making this component's processing time negligible due to GPU's high memory bandwidth.} This approach enables \name to achieve significantly faster build times than both FilteredVamana and StitchedVamana, except on the YFCC-10M dataset with 200K labels, where building separate graph indices takes longer than FilteredVamana's single-index approach. \revisionC{However, \name still lags behind CAGRA's single-index construction speed. Enhancing its index construction efficiency remains an important area for future research, particularly for large-scale applications requiring frequent index updates.}

\section{Conclusion}

Modern AI-based applications require ANNS with filtered search. However, indexing and search algorithms with filters on GPUs must be re-designed to achieve high performance potential. We present a new GPU-based indexing and search algorithm called \name, which applies a hierarchical and divide-and-conquer design of filtered-ANNS for GPUs. Furthermore, \name designs redundancy-bypassing IVF-Graph and interleaved-scan based IVF-BFS to achieve high compute and memory efficiency while attaining high accuracy. Evaluation on both public and semi-synthetic datasets show that \name outperforms existing solutions by two orders of magnitude in QPS and establishes the new state-of-the-art for ANNS with filters.

\minjia{TODO: Double check all citations to make sure (1) no duplicated citations, (2) Fix the format to make the style consistent.}

\section*{Acknowledgments}

We sincerely appreciate the insightful feedback from the anonymous reviewers. This research was supported by the National Science Foundation (NSF) under Grant No. 2441601. The work utilized the DeltaAI system at the National Center for Supercomputing Applications (NCSA) through allocation CIS240055 from the Advanced Cyberinfrastructure Coordination Ecosystem: Services \& Support (ACCESS) program, which is supported by National Science Foundation grants \#2138259, \#2138286, \#2138307, \#2137603, and \#2138296. The Delta advanced computing resource is a collaborative effort between the University of Illinois Urbana-Champaign and NCSA, supported by the NSF (award OAC 2005572) and the State of Illinois. This work also utilized the Illinois Campus Cluster and NCSA NFI Hydro cluster, both supported by the University of Illinois Urbana-Champaign and the University of Illinois System.




\clearpage

\bibliographystyle{ACM-Reference-Format}
\bibliography{reference.bib}


\begin{thebibliography}{69}


\ifx \showCODEN    \undefined \def \showCODEN     #1{\unskip}     \fi
\ifx \showDOI      \undefined \def \showDOI       #1{#1}\fi
\ifx \showISBNx    \undefined \def \showISBNx     #1{\unskip}     \fi
\ifx \showISBNxiii \undefined \def \showISBNxiii  #1{\unskip}     \fi
\ifx \showISSN     \undefined \def \showISSN      #1{\unskip}     \fi
\ifx \showLCCN     \undefined \def \showLCCN      #1{\unskip}     \fi
\ifx \shownote     \undefined \def \shownote      #1{#1}          \fi
\ifx \showarticletitle \undefined \def \showarticletitle #1{#1}   \fi
\ifx \showURL      \undefined \def \showURL       {\relax}        \fi
\providecommand\bibfield[2]{#2}
\providecommand\bibinfo[2]{#2}
\providecommand\natexlab[1]{#1}
\providecommand\showeprint[2][]{arXiv:#2}

\bibitem[pgv(2025)]%
        {pgvector}
 \bibinfo{year}{Accessed: 04-13-2025}\natexlab{}.
\newblock \bibinfo{title}{{pgvector: Open-source vector similarity search for Postgres}}.
\newblock \bibinfo{howpublished}{\url{https://github.com/pgvector/pgvector}}.
\newblock


\bibitem[AI(2025)]%
        {cuvs}
\bibfield{author}{\bibinfo{person}{RAPIDS AI}.} \bibinfo{year}{2025}\natexlab{}.
\newblock \bibinfo{title}{cuVS}.
\newblock \bibinfo{howpublished}{\url{https://github.com/rapidsai/cuvs}}.
\newblock
\newblock
\shownote{Accessed: 2025-01-18}.


\bibitem[Aliannejadi et~al\mbox{.}(2018)]%
        {mobile-search}
\bibfield{author}{\bibinfo{person}{Mohammad Aliannejadi}, \bibinfo{person}{Hamed Zamani}, \bibinfo{person}{Fabio Crestani}, {and} \bibinfo{person}{W.~Bruce Croft}.} \bibinfo{year}{2018}\natexlab{}.
\newblock \showarticletitle{{Target Apps Selection: Towards a Unified Search Framework for Mobile Devices}}. In \bibinfo{booktitle}{\emph{{SIGIR} 2018}}. \bibinfo{pages}{215--224}.
\newblock


\bibitem[Andoni et~al\mbox{.}(2018)]%
        {ann-survey}
\bibfield{author}{\bibinfo{person}{Alexandr Andoni}, \bibinfo{person}{Piotr Indyk}, {and} \bibinfo{person}{Ilya Razenshteyn}.} \bibinfo{year}{2018}\natexlab{}.
\newblock \showarticletitle{{Approximate Nearest Neighbor Search in High Dimensions}}.
\newblock \bibinfo{journal}{\emph{arXiv preprint arXiv:1806.09823}} (\bibinfo{year}{2018}).
\newblock


\bibitem[Beckmann et~al\mbox{.}(1990)]%
        {r-star-tree}
\bibfield{author}{\bibinfo{person}{Norbert Beckmann}, \bibinfo{person}{Hans{-}Peter Kriegel}, \bibinfo{person}{Ralf Schneider}, {and} \bibinfo{person}{Bernhard Seeger}.} \bibinfo{year}{1990}\natexlab{}.
\newblock \showarticletitle{{The R*-Tree: An Efficient and Robust Access Method for Points and Rectangles}}. In \bibinfo{booktitle}{\emph{{SIGMOD} 1990}}. \bibinfo{pages}{322--331}.
\newblock


\bibitem[{Ben Landrum and Magdalen Dobson Manohar and Mazin Karjikar and Laxman Dhulipala }(2024)]%
        {ivf2}
\bibfield{author}{\bibinfo{person}{{Ben Landrum and Magdalen Dobson Manohar and Mazin Karjikar and Laxman Dhulipala }}.} \bibinfo{year}{2024}\natexlab{}.
\newblock \bibinfo{title}{{IVF2: Fusing Classic and Spatial Inverted Indices for Fast Filtered ANNS}}.
\newblock \bibinfo{howpublished}{\url{https://big-ann-benchmarks.com/neurips23_slides/IVF_2_filter_Ben.pdf}}.
\newblock


\bibitem[Bentley(1975)]%
        {kd-tree}
\bibfield{author}{\bibinfo{person}{Jon~Louis Bentley}.} \bibinfo{year}{1975}\natexlab{}.
\newblock \showarticletitle{{Multidimensional Binary Search Trees Used for Associative Searching}}.
\newblock \bibinfo{journal}{\emph{Commun. ACM}} \bibinfo{volume}{18}, \bibinfo{number}{9} (\bibinfo{date}{Sept.} \bibinfo{year}{1975}), \bibinfo{pages}{509--517}.
\newblock
\showISSN{0001-0782}


\bibitem[Bernstein et~al\mbox{.}({[n.\,d.]})]%
        {wikiann-paper}
\bibfield{author}{\bibinfo{person}{Philip~A Bernstein}, \bibinfo{person}{Siddharth Gollapudi}, \bibinfo{person}{Suryansh Gupta}, \bibinfo{person}{Ravishankar Krishnaswamy}, \bibinfo{person}{Sepideh Mahabadi}, \bibinfo{person}{Sandeep Silwal}, \bibinfo{person}{Gopal~R Srinivasa}, \bibinfo{person}{Varun Suriyanarayana}, \bibinfo{person}{Jakub Tarnawski}, \bibinfo{person}{Haiyang Xu}, {et~al\mbox{.}}} \bibinfo{year}{[n.\,d.]}\natexlab{}.
\newblock \showarticletitle{Graph-based algorithms for nearest neighbor search with multiple filters}.
\newblock  (\bibinfo{year}{[n.\,d.]}).
\newblock


\bibitem[Big-ANN({[n.\,d.]})]%
        {bigann}
\bibfield{author}{\bibinfo{person}{Big-ANN}.} \bibinfo{year}{[n.\,d.]}\natexlab{}.
\newblock \bibinfo{title}{NeurIPS'23 Competition Track: Big-ANN}.
\newblock \bibinfo{howpublished}{\url{https://big-ann-benchmarks.com/neurips23.html}}.
\newblock
\newblock
\shownote{Accessed: 2024}.


\bibitem[Chen et~al\mbox{.}(2024)]%
        {singlestore-v}
\bibfield{author}{\bibinfo{person}{Cheng Chen}, \bibinfo{person}{Chenzhe Jin}, \bibinfo{person}{Yunan Zhang}, \bibinfo{person}{Sasha Podolsky}, \bibinfo{person}{Chun Wu}, \bibinfo{person}{Szu-Po Wang}, \bibinfo{person}{Eric Hanson}, \bibinfo{person}{Zhou Sun}, \bibinfo{person}{Robert Walzer}, {and} \bibinfo{person}{Jianguo Wang}.} \bibinfo{year}{2024}\natexlab{}.
\newblock \showarticletitle{SingleStore-V: An Integrated Vector Database System in SingleStore}.
\newblock \bibinfo{journal}{\emph{Proc. VLDB Endow.}} \bibinfo{volume}{17}, \bibinfo{number}{12} (\bibinfo{date}{Aug.} \bibinfo{year}{2024}), \bibinfo{pages}{3772–3785}.
\newblock
\showISSN{2150-8097}
\urldef\tempurl%
\url{https://doi.org/10.14778/3685800.3685805}
\showDOI{\tempurl}


\bibitem[Chen et~al\mbox{.}(2018)]%
        {sptag}
\bibfield{author}{\bibinfo{person}{Qi Chen}, \bibinfo{person}{Haidong Wang}, \bibinfo{person}{Mingqin Li}, \bibinfo{person}{Gang Ren}, \bibinfo{person}{Scarlett Li}, \bibinfo{person}{Jeffery Zhu}, \bibinfo{person}{Jason Li}, \bibinfo{person}{Chuanjie Liu}, \bibinfo{person}{Lintao Zhang}, {and} \bibinfo{person}{Jingdong Wang}.} \bibinfo{year}{2018}\natexlab{}.
\newblock \bibinfo{booktitle}{\emph{{SPTAG: A library for fast approximate nearest neighbor search}}}.
\newblock
\urldef\tempurl%
\url{https://github.com/Microsoft/SPTAG}
\showURL{%
\tempurl}


\bibitem[Chen et~al\mbox{.}(2019)]%
        {vlq-adc}
\bibfield{author}{\bibinfo{person}{Wei Chen}, \bibinfo{person}{Jincai Chen}, \bibinfo{person}{Fuhao Zou}, \bibinfo{person}{Yuan{-}Fang Li}, \bibinfo{person}{Ping Lu}, \bibinfo{person}{Qiang Wang}, {and} \bibinfo{person}{Wei Zhao}.} \bibinfo{year}{2019}\natexlab{}.
\newblock \showarticletitle{Vector and line quantization for billion-scale similarity search on GPUs}.
\newblock \bibinfo{journal}{\emph{Future Gener. Comput. Syst.}}  \bibinfo{volume}{99} (\bibinfo{year}{2019}), \bibinfo{pages}{295--307}.
\newblock


\bibitem[Cormode and Muthukrishnan(2005)]%
        {sketches}
\bibfield{author}{\bibinfo{person}{Graham Cormode} {and} \bibinfo{person}{S. Muthukrishnan}.} \bibinfo{year}{2005}\natexlab{}.
\newblock \showarticletitle{An improved data stream summary: the count-min sketch and its applications}.
\newblock \bibinfo{journal}{\emph{Journal of Algorithms}} \bibinfo{volume}{55}, \bibinfo{number}{1} (\bibinfo{year}{2005}), \bibinfo{pages}{58--75}.
\newblock
\showISSN{0196-6774}
\urldef\tempurl%
\url{https://doi.org/10.1016/j.jalgor.2003.12.001}
\showDOI{\tempurl}


\bibitem[Dehghani et~al\mbox{.}(2017)]%
        {nrm-weak-supervision}
\bibfield{author}{\bibinfo{person}{Mostafa Dehghani}, \bibinfo{person}{Hamed Zamani}, \bibinfo{person}{Aliaksei Severyn}, \bibinfo{person}{Jaap Kamps}, {and} \bibinfo{person}{W.~Bruce Croft}.} \bibinfo{year}{2017}\natexlab{}.
\newblock \showarticletitle{{Neural Ranking Models with Weak Supervision}}. In \bibinfo{booktitle}{\emph{SIGIR 2017}}. \bibinfo{pages}{65--74}.
\newblock


\bibitem[Dong et~al\mbox{.}(2011)]%
        {nn-descent}
\bibfield{author}{\bibinfo{person}{Wei Dong}, \bibinfo{person}{Charikar Moses}, {and} \bibinfo{person}{Kai Li}.} \bibinfo{year}{2011}\natexlab{}.
\newblock \showarticletitle{Efficient k-nearest neighbor graph construction for generic similarity measures}. In \bibinfo{booktitle}{\emph{Proceedings of the 20th International Conference on World Wide Web}} (Hyderabad, India) \emph{(\bibinfo{series}{WWW '11})}. \bibinfo{publisher}{Association for Computing Machinery}, \bibinfo{address}{New York, NY, USA}, \bibinfo{pages}{577–586}.
\newblock
\showISBNx{9781450306324}
\urldef\tempurl%
\url{https://doi.org/10.1145/1963405.1963487}
\showDOI{\tempurl}


\bibitem[Douze et~al\mbox{.}(2024)]%
        {faiss}
\bibfield{author}{\bibinfo{person}{Matthijs Douze}, \bibinfo{person}{Alexandr Guzhva}, \bibinfo{person}{Chengqi Deng}, \bibinfo{person}{Jeff Johnson}, \bibinfo{person}{Gergely Szilvasy}, \bibinfo{person}{Pierre-Emmanuel Mazar{\'e}}, \bibinfo{person}{Maria Lomeli}, \bibinfo{person}{Lucas Hosseini}, {and} \bibinfo{person}{Herv{\'e} J{\'e}gou}.} \bibinfo{year}{2024}\natexlab{}.
\newblock \showarticletitle{The Faiss library}.
\newblock \bibinfo{journal}{\emph{arXiv preprint arXiv:2401.08281}} (\bibinfo{year}{2024}).
\newblock
\urldef\tempurl%
\url{https://doi.org/10.48550/arXiv.2401.08281}
\showDOI{\tempurl}


\bibitem[Fang et~al\mbox{.}(2020)]%
        {fang2020memory}
\bibfield{author}{\bibinfo{person}{Jian Fang}, \bibinfo{person}{Yvo~TB Mulder}, \bibinfo{person}{Jan Hidders}, \bibinfo{person}{Jinho Lee}, {and} \bibinfo{person}{H~Peter Hofstee}.} \bibinfo{year}{2020}\natexlab{}.
\newblock \showarticletitle{In-memory database acceleration on FPGAs: a survey}.
\newblock \bibinfo{journal}{\emph{The VLDB Journal}}  \bibinfo{volume}{29} (\bibinfo{year}{2020}), \bibinfo{pages}{33--59}.
\newblock


\bibitem[Fu et~al\mbox{.}(2019)]%
        {spread-out-graph}
\bibfield{author}{\bibinfo{person}{Cong Fu}, \bibinfo{person}{Chao Xiang}, \bibinfo{person}{Changxu Wang}, {and} \bibinfo{person}{Deng Cai}.} \bibinfo{year}{2019}\natexlab{}.
\newblock \showarticletitle{{Fast Approximate Nearest Neighbor Search with the Navigating Spreading-out Graph}}. In \bibinfo{booktitle}{\emph{VLDB'19}}.
\newblock


\bibitem[Ganguly et~al\mbox{.}(1996)]%
        {bi-focal}
\bibfield{author}{\bibinfo{person}{Sumit Ganguly}, \bibinfo{person}{Phillip~B. Gibbons}, \bibinfo{person}{Yossi Matias}, {and} \bibinfo{person}{Avi Silberschatz}.} \bibinfo{year}{1996}\natexlab{}.
\newblock \showarticletitle{Bifocal sampling for skew-resistant join size estimation}.
\newblock \bibinfo{journal}{\emph{SIGMOD Rec.}} \bibinfo{volume}{25}, \bibinfo{number}{2} (\bibinfo{date}{June} \bibinfo{year}{1996}), \bibinfo{pages}{271–281}.
\newblock
\showISSN{0163-5808}
\urldef\tempurl%
\url{https://doi.org/10.1145/235968.233340}
\showDOI{\tempurl}


\bibitem[Ge et~al\mbox{.}(2013)]%
        {opq}
\bibfield{author}{\bibinfo{person}{Tiezheng Ge}, \bibinfo{person}{Kaiming He}, \bibinfo{person}{Qifa Ke}, {and} \bibinfo{person}{Jian Sun}.} \bibinfo{year}{2013}\natexlab{}.
\newblock \showarticletitle{{Optimized Product Quantization for Approximate Nearest Neighbor Search}}. In \bibinfo{booktitle}{\emph{CVPR 2013}}.
\newblock


\bibitem[Gionis et~al\mbox{.}(1999)]%
        {lsh}
\bibfield{author}{\bibinfo{person}{Aristides Gionis}, \bibinfo{person}{Piotr Indyk}, {and} \bibinfo{person}{Rajeev Motwani}.} \bibinfo{year}{1999}\natexlab{}.
\newblock \showarticletitle{{Similarity Search in High Dimensions via Hashing}}. In \bibinfo{booktitle}{\emph{VLDB'99}}. \bibinfo{pages}{518--529}.
\newblock


\bibitem[Gollapudi et~al\mbox{.}(2023)]%
        {filtered-diskann}
\bibfield{author}{\bibinfo{person}{Siddharth Gollapudi}, \bibinfo{person}{Neel Karia}, \bibinfo{person}{Varun Sivashankar}, \bibinfo{person}{Ravishankar Krishnaswamy}, \bibinfo{person}{Nikit Begwani}, \bibinfo{person}{Swapnil Raz}, \bibinfo{person}{Yiyong Lin}, \bibinfo{person}{Yin Zhang}, \bibinfo{person}{Neelam Mahapatro}, \bibinfo{person}{Premkumar Srinivasan}, {et~al\mbox{.}}} \bibinfo{year}{2023}\natexlab{}.
\newblock \showarticletitle{Filtered-diskann: Graph algorithms for approximate nearest neighbor search with filters}. In \bibinfo{booktitle}{\emph{Proceedings of the ACM Web Conference 2023}}. \bibinfo{pages}{3406--3416}.
\newblock


\bibitem[Google(2022)]%
        {google-multisearch}
\bibfield{author}{\bibinfo{person}{Google}.} \bibinfo{year}{2022}\natexlab{}.
\newblock \bibinfo{title}{Go beyond the search box: Introducing multisearch}.
\newblock \bibinfo{howpublished}{\url{https://blog.google/products/search/multisearch/}}.
\newblock
\newblock
\shownote{Accessed: 2025}.


\bibitem[Groh et~al\mbox{.}(2023)]%
        {ggnn}
\bibfield{author}{\bibinfo{person}{Fabian Groh}, \bibinfo{person}{Lukas Ruppert}, \bibinfo{person}{Patrick Wieschollek}, {and} \bibinfo{person}{Hendrik P.~A. Lensch}.} \bibinfo{year}{2023}\natexlab{}.
\newblock \showarticletitle{GGNN: Graph-Based GPU Nearest Neighbor Search}.
\newblock \bibinfo{journal}{\emph{IEEE Transactions on Big Data}} \bibinfo{volume}{9}, \bibinfo{number}{1} (\bibinfo{year}{2023}), \bibinfo{pages}{267--279}.
\newblock
\urldef\tempurl%
\url{https://doi.org/10.1109/TBDATA.2022.3161156}
\showDOI{\tempurl}


\bibitem[Guo et~al\mbox{.}(2016)]%
        {dl-matching-model}
\bibfield{author}{\bibinfo{person}{Jiafeng Guo}, \bibinfo{person}{Yixing Fan}, \bibinfo{person}{Qingyao Ai}, {and} \bibinfo{person}{W.~Bruce Croft}.} \bibinfo{year}{2016}\natexlab{}.
\newblock \showarticletitle{{A Deep Relevance Matching Model for Ad-hoc Retrieval}}. In \bibinfo{booktitle}{\emph{{CIKM} 2016}}. \bibinfo{pages}{55--64}.
\newblock


\bibitem[Gupta(2021)]%
        {gupta2021introduction}
\bibfield{author}{\bibinfo{person}{Neha Gupta}.} \bibinfo{year}{2021}\natexlab{}.
\newblock \showarticletitle{Introduction to hardware accelerator systems for artificial intelligence and machine learning}.
\newblock In \bibinfo{booktitle}{\emph{Advances in Computers}}. Vol.~\bibinfo{volume}{122}. \bibinfo{publisher}{Elsevier}, \bibinfo{pages}{1--21}.
\newblock


\bibitem[Huang et~al\mbox{.}(2013)]%
        {dssm}
\bibfield{author}{\bibinfo{person}{Po-Sen Huang}, \bibinfo{person}{Xiaodong He}, \bibinfo{person}{Jianfeng Gao}, \bibinfo{person}{Li Deng}, \bibinfo{person}{Alex Acero}, {and} \bibinfo{person}{Larry Heck}.} \bibinfo{year}{2013}\natexlab{}.
\newblock \showarticletitle{{Learning deep structured semantic models for web search using clickthrough data}}. In \bibinfo{booktitle}{\emph{CIKM '13}}. \bibinfo{pages}{2333--2338}.
\newblock


\bibitem[{HuggingFace}(2025)]%
        {wikiann}
\bibfield{author}{\bibinfo{person}{{HuggingFace}}.} \bibinfo{year}{2025}\natexlab{}.
\newblock \bibinfo{title}{{WikiANN dataset}}.
\newblock \bibinfo{howpublished}{\url{https://huggingface.co/2024annonymous/wiki-ann}}.
\newblock


\bibitem[Jayaram~Subramanya et~al\mbox{.}(2019)]%
        {diskann}
\bibfield{author}{\bibinfo{person}{Suhas Jayaram~Subramanya}, \bibinfo{person}{Fnu Devvrit}, \bibinfo{person}{Harsha~Vardhan Simhadri}, \bibinfo{person}{Ravishankar Krishnawamy}, {and} \bibinfo{person}{Rohan Kadekodi}.} \bibinfo{year}{2019}\natexlab{}.
\newblock \showarticletitle{Diskann: Fast accurate billion-point nearest neighbor search on a single node}.
\newblock \bibinfo{journal}{\emph{Advances in Neural Information Processing Systems}}  \bibinfo{volume}{32} (\bibinfo{year}{2019}).
\newblock


\bibitem[Jegou et~al\mbox{.}(2011)]%
        {product-quantization}
\bibfield{author}{\bibinfo{person}{Herve Jegou}, \bibinfo{person}{Matthijs Douze}, {and} \bibinfo{person}{Cordelia Schmid}.} \bibinfo{year}{2011}\natexlab{}.
\newblock In \bibinfo{booktitle}{{Product Quantization for Nearest Neighbor Search}}.
\newblock \bibinfo{journal}{\emph{TPAMI 2011}}.
\newblock


\bibitem[Johnson et~al\mbox{.}(2017)]%
        {billion-scale-search-on-gpus}
\bibfield{author}{\bibinfo{person}{Jeff Johnson}, \bibinfo{person}{Matthijs Douze}, {and} \bibinfo{person}{Herv{\'{e}} J{\'{e}}gou}.} \bibinfo{year}{2017}\natexlab{}.
\newblock \showarticletitle{{Billion-scale similarity search with GPUs}}.
\newblock \bibinfo{journal}{\emph{CoRR}}  \bibinfo{volume}{abs/1702.08734} (\bibinfo{year}{2017}).
\newblock
\showeprint[arxiv]{1702.08734}
\urldef\tempurl%
\url{http://arxiv.org/abs/1702.08734}
\showURL{%
\tempurl}


\bibitem[Kalantidis and Avrithis(2014)]%
        {lopq}
\bibfield{author}{\bibinfo{person}{Yannis Kalantidis} {and} \bibinfo{person}{Yannis~S. Avrithis}.} \bibinfo{year}{2014}\natexlab{}.
\newblock \showarticletitle{{Locally Optimized Product Quantization for Approximate Nearest Neighbor Search}}. In \bibinfo{booktitle}{\emph{{CVPR} 2014}}. \bibinfo{pages}{2329--2336}.
\newblock


\bibitem[Kanade et~al\mbox{.}(2020)]%
        {code-embedding}
\bibfield{author}{\bibinfo{person}{Aditya Kanade}, \bibinfo{person}{Petros Maniatis}, \bibinfo{person}{Gogul Balakrishnan}, {and} \bibinfo{person}{Kensen Shi}.} \bibinfo{year}{2020}\natexlab{}.
\newblock \showarticletitle{Learning and Evaluating Contextual Embedding of Source Code}. In \bibinfo{booktitle}{\emph{Proceedings of the 37th International Conference on Machine Learning, {ICML} 2020, 13-18 July 2020, Virtual Event}} \emph{(\bibinfo{series}{Proceedings of Machine Learning Research}, Vol.~\bibinfo{volume}{119})}. \bibinfo{publisher}{{PMLR}}, \bibinfo{pages}{5110--5121}.
\newblock


\bibitem[Lempitsky(2012)]%
        {inverted-multi-index}
\bibfield{author}{\bibinfo{person}{Victor Lempitsky}.} \bibinfo{year}{2012}\natexlab{}.
\newblock \showarticletitle{The Inverted Multi-index}. In \bibinfo{booktitle}{\emph{CVPR '12}}. \bibinfo{pages}{3069--3076}.
\newblock


\bibitem[Lewis et~al\mbox{.}(2020)]%
        {rag}
\bibfield{author}{\bibinfo{person}{Patrick Lewis}, \bibinfo{person}{Ethan Perez}, \bibinfo{person}{Aleksandra Piktus}, \bibinfo{person}{Fabio Petroni}, \bibinfo{person}{Vladimir Karpukhin}, \bibinfo{person}{Naman Goyal}, \bibinfo{person}{Heinrich K\"{u}ttler}, \bibinfo{person}{Mike Lewis}, \bibinfo{person}{Wen-tau Yih}, \bibinfo{person}{Tim Rockt\"{a}schel}, \bibinfo{person}{Sebastian Riedel}, {and} \bibinfo{person}{Douwe Kiela}.} \bibinfo{year}{2020}\natexlab{}.
\newblock \showarticletitle{Retrieval-augmented generation for knowledge-intensive NLP tasks}. In \bibinfo{booktitle}{\emph{Proceedings of the 34th International Conference on Neural Information Processing Systems}} (Vancouver, BC, Canada) \emph{(\bibinfo{series}{NIPS '20})}. \bibinfo{publisher}{Curran Associates Inc.}, \bibinfo{address}{Red Hook, NY, USA}, Article \bibinfo{articleno}{793}, \bibinfo{numpages}{16}~pages.
\newblock
\showISBNx{9781713829546}


\bibitem[Li et~al\mbox{.}(2022)]%
        {blip}
\bibfield{author}{\bibinfo{person}{Junnan Li}, \bibinfo{person}{Dongxu Li}, \bibinfo{person}{Caiming Xiong}, {and} \bibinfo{person}{Steven C.~H. Hoi}.} \bibinfo{year}{2022}\natexlab{}.
\newblock \showarticletitle{{BLIP:} Bootstrapping Language-Image Pre-training for Unified Vision-Language Understanding and Generation}. In \bibinfo{booktitle}{\emph{International Conference on Machine Learning, {ICML} 2022, 17-23 July 2022, Baltimore, Maryland, {USA}}} \emph{(\bibinfo{series}{Proceedings of Machine Learning Research}, Vol.~\bibinfo{volume}{162})}. \bibinfo{publisher}{{PMLR}}, \bibinfo{pages}{12888--12900}.
\newblock


\bibitem[Li et~al\mbox{.}(2020)]%
        {li2020approximate}
\bibfield{author}{\bibinfo{person}{Wen Li}, \bibinfo{person}{Ying Zhang}, \bibinfo{person}{Yifang Sun}, \bibinfo{person}{Wei Wang}, \bibinfo{person}{Mingjie Li}, \bibinfo{person}{Wenjie Zhang}, {and} \bibinfo{person}{Xuemin Lin}.} \bibinfo{year}{2020}\natexlab{}.
\newblock \showarticletitle{Approximate Nearest Neighbor Search on High Dimensional Data -- Experiments, Analyses, and Improvement}.
\newblock \bibinfo{journal}{\emph{IEEE Transactions on Knowledge and Data Engineering}} \bibinfo{volume}{32}, \bibinfo{number}{8} (\bibinfo{year}{2020}), \bibinfo{pages}{1475--1488}.
\newblock
\urldef\tempurl%
\url{https://doi.org/10.1109/TKDE.2019.2909204}
\showDOI{\tempurl}


\bibitem[Liu et~al\mbox{.}(2019)]%
        {roberta}
\bibfield{author}{\bibinfo{person}{Yinhan Liu}, \bibinfo{person}{Myle Ott}, \bibinfo{person}{Naman Goyal}, \bibinfo{person}{Jingfei Du}, \bibinfo{person}{Mandar Joshi}, \bibinfo{person}{Danqi Chen}, \bibinfo{person}{Omer Levy}, \bibinfo{person}{Mike Lewis}, \bibinfo{person}{Luke Zettlemoyer}, {and} \bibinfo{person}{Veselin Stoyanov}.} \bibinfo{year}{2019}\natexlab{}.
\newblock \showarticletitle{RoBERTa: {A} Robustly Optimized {BERT} Pretraining Approach}.
\newblock \bibinfo{journal}{\emph{CoRR}}  \bibinfo{volume}{abs/1907.11692} (\bibinfo{year}{2019}).
\newblock


\bibitem[LongChain({[n.\,d.]})]%
        {long-chain}
\bibfield{author}{\bibinfo{person}{LongChain}.} \bibinfo{year}{[n.\,d.]}\natexlab{}.
\newblock \bibinfo{title}{LongChain: Build context-aware reasoning applications}.
\newblock \bibinfo{howpublished}{\url{https://github.com/langchain-ai/langchain}}.
\newblock
\newblock
\shownote{Accessed: 2025}.


\bibitem[Malkov and Yashunin(2016)]%
        {hnsw}
\bibfield{author}{\bibinfo{person}{Yury~A. Malkov} {and} \bibinfo{person}{D.~A. Yashunin}.} \bibinfo{year}{2016}\natexlab{}.
\newblock \showarticletitle{{Efficient and robust approximate nearest neighbor search using Hierarchical Navigable Small World graphs}}.
\newblock \bibinfo{journal}{\emph{CoRR}}  \bibinfo{volume}{arXiv preprint abs/1603.09320} (\bibinfo{year}{2016}).
\newblock


\bibitem[Mikolov et~al\mbox{.}(2013)]%
        {word2vec}
\bibfield{author}{\bibinfo{person}{Tomas Mikolov}, \bibinfo{person}{Ilya Sutskever}, \bibinfo{person}{Kai Chen}, \bibinfo{person}{Gregory~S. Corrado}, {and} \bibinfo{person}{Jeffrey Dean}.} \bibinfo{year}{2013}\natexlab{}.
\newblock \showarticletitle{Distributed Representations of Words and Phrases and their Compositionality}. In \bibinfo{booktitle}{\emph{Advances in Neural Information Processing Systems 26: 27th Annual Conference on Neural Information Processing Systems 2013}}. \bibinfo{pages}{3111--3119}.
\newblock


\bibitem[Milvus-io(2022)]%
        {milvus}
\bibfield{author}{\bibinfo{person}{Milvus-io}.} \bibinfo{year}{2022}\natexlab{}.
\newblock \bibinfo{title}{Milvus-docs: Conduct a hybrid search}.
\newblock \bibinfo{howpublished}{\url{https://github.com/milvus-io/milvus-docs/blob/v2.1.x/site/en/userGuide/search/hybridsearch.md}}.
\newblock
\newblock
\shownote{Accessed: 2025}.


\bibitem[Mitra et~al\mbox{.}(2017)]%
        {learn-to-match}
\bibfield{author}{\bibinfo{person}{Bhaskar Mitra}, \bibinfo{person}{Fernando Diaz}, {and} \bibinfo{person}{Nick Craswell}.} \bibinfo{year}{2017}\natexlab{}.
\newblock \showarticletitle{{Learning to Match using Local and Distributed Representations of Text for Web Search}}. In \bibinfo{booktitle}{\emph{{WWW} 2017}}.
\newblock


\bibitem[Muja and Lowe(2014)]%
        {flann}
\bibfield{author}{\bibinfo{person}{Marius Muja} {and} \bibinfo{person}{David~G. Lowe}.} \bibinfo{year}{2014}\natexlab{}.
\newblock \showarticletitle{{Scalable Nearest Neighbor Algorithms for High Dimensional Data}}.
\newblock \bibinfo{journal}{\emph{TPAMI 2014}} \bibinfo{volume}{36}, \bibinfo{number}{11} (\bibinfo{year}{2014}), \bibinfo{pages}{2227--2240}.
\newblock


\bibitem[Norouzi and Fleet(2013)]%
        {cartesian-kmeans}
\bibfield{author}{\bibinfo{person}{Mohammad Norouzi} {and} \bibinfo{person}{David~J. Fleet}.} \bibinfo{year}{2013}\natexlab{}.
\newblock \showarticletitle{{Cartesian K-Means}}. In \bibinfo{booktitle}{\emph{CVPR 2013}}.
\newblock


\bibitem[Ootomo et~al\mbox{.}(2023)]%
        {cagra}
\bibfield{author}{\bibinfo{person}{Hiroyuki Ootomo}, \bibinfo{person}{Akira Naruse}, \bibinfo{person}{Corey~J. Nolet}, \bibinfo{person}{Ray Wang}, \bibinfo{person}{Tamas~B. Feh{\'e}r}, {and} \bibinfo{person}{Y. Wang}.} \bibinfo{year}{2023}\natexlab{}.
\newblock \showarticletitle{CAGRA: Highly Parallel Graph Construction and Approximate Nearest Neighbor Search for GPUs}.
\newblock \bibinfo{journal}{\emph{2024 IEEE 40th International Conference on Data Engineering (ICDE)}} (\bibinfo{year}{2023}), \bibinfo{pages}{4236--4247}.
\newblock


\bibitem[Pan et~al\mbox{.}(2024)]%
        {vector-search-survey}
\bibfield{author}{\bibinfo{person}{James~Jie Pan}, \bibinfo{person}{Jianguo Wang}, {and} \bibinfo{person}{Guoliang Li}.} \bibinfo{year}{2024}\natexlab{}.
\newblock \showarticletitle{Survey of vector database management systems}.
\newblock \bibinfo{journal}{\emph{{VLDB} J.}} \bibinfo{volume}{33}, \bibinfo{number}{5} (\bibinfo{year}{2024}), \bibinfo{pages}{1591--1615}.
\newblock


\bibitem[Patel et~al\mbox{.}(2024)]%
        {acorn}
\bibfield{author}{\bibinfo{person}{Liana Patel}, \bibinfo{person}{Peter Kraft}, \bibinfo{person}{Carlos Guestrin}, {and} \bibinfo{person}{Matei Zaharia}.} \bibinfo{year}{2024}\natexlab{}.
\newblock \showarticletitle{{ACORN:} Performant and Predicate-Agnostic Search Over Vector Embeddings and Structured Data}.
\newblock \bibinfo{journal}{\emph{Proc. {ACM} Manag. Data}} \bibinfo{volume}{2}, \bibinfo{number}{3} (\bibinfo{year}{2024}), \bibinfo{pages}{120}.
\newblock


\bibitem[Pinecone~Systems(2024)]%
        {pinecone}
\bibfield{author}{\bibinfo{person}{Inc. Pinecone~Systems}.} \bibinfo{year}{2024}\natexlab{}.
\newblock \bibinfo{title}{Overview}.
\newblock \bibinfo{howpublished}{\url{https://docs.pinecone.io/docs/overview}}.
\newblock
\newblock
\shownote{Accessed: 2025}.


\bibitem[Radford et~al\mbox{.}(2021)]%
        {clip}
\bibfield{author}{\bibinfo{person}{Alec Radford}, \bibinfo{person}{Jong~Wook Kim}, \bibinfo{person}{Chris Hallacy}, \bibinfo{person}{Aditya Ramesh}, \bibinfo{person}{Gabriel Goh}, \bibinfo{person}{Sandhini Agarwal}, \bibinfo{person}{Girish Sastry}, \bibinfo{person}{Amanda Askell}, \bibinfo{person}{Pamela Mishkin}, \bibinfo{person}{Jack Clark}, \bibinfo{person}{Gretchen Krueger}, {and} \bibinfo{person}{Ilya Sutskever}.} \bibinfo{year}{2021}\natexlab{}.
\newblock \showarticletitle{Learning Transferable Visual Models From Natural Language Supervision}. In \bibinfo{booktitle}{\emph{Proceedings of the 38th International Conference on Machine Learning, {ICML} 2021, 18-24 July 2021, Virtual Event}} \emph{(\bibinfo{series}{Proceedings of Machine Learning Research}, Vol.~\bibinfo{volume}{139})}. \bibinfo{publisher}{{PMLR}}, \bibinfo{pages}{8748--8763}.
\newblock


\bibitem[Ren et~al\mbox{.}(2020)]%
        {hm-ann}
\bibfield{author}{\bibinfo{person}{Jie Ren}, \bibinfo{person}{Minjia Zhang}, {and} \bibinfo{person}{Dong Li}.} \bibinfo{year}{2020}\natexlab{}.
\newblock \showarticletitle{{HM-ANN:} Efficient Billion-Point Nearest Neighbor Search on Heterogeneous Memory}. In \bibinfo{booktitle}{\emph{Advances in Neural Information Processing Systems 33: Annual Conference on Neural Information Processing Systems 2020, NeurIPS 2020, December 6-12, 2020, virtual}}.
\newblock


\bibitem[Sharma and Sharma(2024)]%
        {sharma2024comprehensive}
\bibfield{author}{\bibinfo{person}{Harshit Sharma} {and} \bibinfo{person}{Anmol Sharma}.} \bibinfo{year}{2024}\natexlab{}.
\newblock \showarticletitle{A Comprehensive Overview of GPU Accelerated Databases}.
\newblock \bibinfo{journal}{\emph{arXiv preprint arXiv:2406.13831}} (\bibinfo{year}{2024}).
\newblock


\bibitem[Simhadri(2025)]%
        {big-ann-repo}
\bibfield{author}{\bibinfo{person}{Harsha Simhadri}.} \bibinfo{year}{2025}\natexlab{}.
\newblock \bibinfo{title}{Big ANN Benchmarks}.
\newblock \bibinfo{howpublished}{\url{https://github.com/harsha-simhadri/big-ann-benchmarks}}.
\newblock
\newblock
\shownote{Accessed: 2025-01-18}.


\bibitem[Simhadri et~al\mbox{.}(2024)]%
        {big-ann-result}
\bibfield{author}{\bibinfo{person}{Harsha~Vardhan Simhadri}, \bibinfo{person}{Martin Aum{\"u}ller}, \bibinfo{person}{Amir Ingber}, \bibinfo{person}{Matthijs Douze}, \bibinfo{person}{George Williams}, \bibinfo{person}{Magdalen~Dobson Manohar}, \bibinfo{person}{Dmitry Baranchuk}, \bibinfo{person}{Edo Liberty}, \bibinfo{person}{Frank Liu}, \bibinfo{person}{Ben Landrum}, {et~al\mbox{.}}} \bibinfo{year}{2024}\natexlab{}.
\newblock \showarticletitle{Results of the Big ANN: NeurIPS'23 competition}.
\newblock \bibinfo{journal}{\emph{arXiv preprint arXiv:2409.17424}} (\bibinfo{year}{2024}).
\newblock


\bibitem[Singh et~al\mbox{.}(2021)]%
        {fresh-diskann}
\bibfield{author}{\bibinfo{person}{Aditi Singh}, \bibinfo{person}{Suhas~Jayaram Subramanya}, \bibinfo{person}{Ravishankar Krishnaswamy}, {and} \bibinfo{person}{Harsha~Vardhan Simhadri}.} \bibinfo{year}{2021}\natexlab{}.
\newblock \showarticletitle{FreshDiskANN: {A} Fast and Accurate Graph-Based {ANN} Index for Streaming Similarity Search}.
\newblock \bibinfo{journal}{\emph{CoRR}}  \bibinfo{volume}{abs/2105.09613} (\bibinfo{year}{2021}).
\newblock


\bibitem[Thomee et~al\mbox{.}(2016)]%
        {yfcc}
\bibfield{author}{\bibinfo{person}{Bart Thomee}, \bibinfo{person}{David~A. Shamma}, \bibinfo{person}{Gerald Friedland}, \bibinfo{person}{Benjamin Elizalde}, \bibinfo{person}{Karl Ni}, \bibinfo{person}{Douglas Poland}, \bibinfo{person}{Damian Borth}, {and} \bibinfo{person}{Li{-}Jia Li}.} \bibinfo{year}{2016}\natexlab{}.
\newblock \showarticletitle{{YFCC100M:} the new data in multimedia research}.
\newblock \bibinfo{journal}{\emph{Commun. {ACM}}} \bibinfo{volume}{59}, \bibinfo{number}{2} (\bibinfo{year}{2016}), \bibinfo{pages}{64--73}.
\newblock


\bibitem[V. et~al\mbox{.}(2024)]%
        {bang}
\bibfield{author}{\bibinfo{person}{Karthik V.}, \bibinfo{person}{Saim Khan}, \bibinfo{person}{Somesh Singh}, \bibinfo{person}{Harsha~Vardhan Simhadri}, {and} \bibinfo{person}{Jyothi Vedurada}.} \bibinfo{year}{2024}\natexlab{}.
\newblock \bibinfo{title}{BANG: Billion-Scale Approximate Nearest Neighbor Search using a Single GPU}.
\newblock
\newblock
\showeprint[arxiv]{2401.11324}~[cs.DC]
\urldef\tempurl%
\url{https://arxiv.org/abs/2401.11324}
\showURL{%
\tempurl}


\bibitem[Van~Gysel et~al\mbox{.}(2016)]%
        {product-search}
\bibfield{author}{\bibinfo{person}{Christophe Van~Gysel}, \bibinfo{person}{Maarten de Rijke}, {and} \bibinfo{person}{Evangelos Kanoulas}.} \bibinfo{year}{2016}\natexlab{}.
\newblock \showarticletitle{{Learning Latent Vector Spaces for Product Search}}. In \bibinfo{booktitle}{\emph{CIKM '16}}. \bibinfo{pages}{165--174}.
\newblock


\bibitem[Wang et~al\mbox{.}(2023)]%
        {nhq}
\bibfield{author}{\bibinfo{person}{Mengzhao Wang}, \bibinfo{person}{Lingwei Lv}, \bibinfo{person}{Xiaoliang Xu}, \bibinfo{person}{Yuxiang Wang}, \bibinfo{person}{Qiang Yue}, {and} \bibinfo{person}{Jiongkang Ni}.} \bibinfo{year}{2023}\natexlab{}.
\newblock \showarticletitle{An Efficient and Robust Framework for Approximate Nearest Neighbor Search with Attribute Constraint}. In \bibinfo{booktitle}{\emph{Advances in Neural Information Processing Systems 36: Annual Conference on Neural Information Processing Systems 2023, NeurIPS 2023, New Orleans, LA, USA, December 10 - 16, 2023}}.
\newblock


\bibitem[Wang et~al\mbox{.}(2021)]%
        {wang2021comprehensive}
\bibfield{author}{\bibinfo{person}{Mengzhao Wang}, \bibinfo{person}{Xiaoliang Xu}, \bibinfo{person}{Qiang Yue}, {and} \bibinfo{person}{Yuxiang Wang}.} \bibinfo{year}{2021}\natexlab{}.
\newblock \showarticletitle{A comprehensive survey and experimental comparison of graph-based approximate nearest neighbor search}.
\newblock \bibinfo{journal}{\emph{arXiv preprint arXiv:2101.12631}} (\bibinfo{year}{2021}).
\newblock


\bibitem[Weaviate(2022)]%
        {weaviate}
\bibfield{author}{\bibinfo{person}{Weaviate}.} \bibinfo{year}{2022}\natexlab{}.
\newblock \bibinfo{title}{Weaviate Documentation: Filters}.
\newblock \bibinfo{howpublished}{\url{https://weaviate.io/developers/weaviate/current/graphql-references/filters.html}}.
\newblock
\newblock
\shownote{Accessed: 2025}.


\bibitem[Wei et~al\mbox{.}(2020)]%
        {analytic-db-v}
\bibfield{author}{\bibinfo{person}{Chuangxian Wei}, \bibinfo{person}{Bin Wu}, \bibinfo{person}{Sheng Wang}, \bibinfo{person}{Renjie Lou}, \bibinfo{person}{Chaoqun Zhan}, \bibinfo{person}{Feifei Li}, {and} \bibinfo{person}{Yuanzhe Cai}.} \bibinfo{year}{2020}\natexlab{}.
\newblock \showarticletitle{AnalyticDB-V: a hybrid analytical engine towards query fusion for structured and unstructured data}.
\newblock \bibinfo{journal}{\emph{Proc. VLDB Endow.}} \bibinfo{volume}{13}, \bibinfo{number}{12} (\bibinfo{date}{Aug.} \bibinfo{year}{2020}), \bibinfo{pages}{3152–3165}.
\newblock
\showISSN{2150-8097}
\urldef\tempurl%
\url{https://doi.org/10.14778/3415478.3415541}
\showDOI{\tempurl}


\bibitem[Yang et~al\mbox{.}(2020)]%
        {pase}
\bibfield{author}{\bibinfo{person}{Wen Yang}, \bibinfo{person}{Tao Li}, \bibinfo{person}{Gai Fang}, {and} \bibinfo{person}{Hong Wei}.} \bibinfo{year}{2020}\natexlab{}.
\newblock \showarticletitle{PASE: PostgreSQL Ultra-High-Dimensional Approximate Nearest Neighbor Search Extension}. In \bibinfo{booktitle}{\emph{Proceedings of the 2020 ACM SIGMOD International Conference on Management of Data}} (Portland, OR, USA) \emph{(\bibinfo{series}{SIGMOD '20})}. \bibinfo{publisher}{Association for Computing Machinery}, \bibinfo{address}{New York, NY, USA}, \bibinfo{pages}{2241–2253}.
\newblock
\showISBNx{9781450367356}
\urldef\tempurl%
\url{https://doi.org/10.1145/3318464.3386131}
\showDOI{\tempurl}


\bibitem[Yu et~al\mbox{.}(2014)]%
        {dl-for-qa}
\bibfield{author}{\bibinfo{person}{Lei Yu}, \bibinfo{person}{Karl~Moritz Hermann}, \bibinfo{person}{Phil Blunsom}, {and} \bibinfo{person}{Stephen Pulman}.} \bibinfo{year}{2014}\natexlab{}.
\newblock \showarticletitle{{Deep Learning for Answer Sentence Selection}}.
\newblock \bibinfo{journal}{\emph{CoRR}}  \bibinfo{volume}{abs/1412.1632} (\bibinfo{year}{2014}).
\newblock


\bibitem[Yu et~al\mbox{.}(2022)]%
        {ganns}
\bibfield{author}{\bibinfo{person}{Yuanhang Yu}, \bibinfo{person}{Dong Wen}, \bibinfo{person}{Ying Zhang}, \bibinfo{person}{Lu Qin}, \bibinfo{person}{Wenjie Zhang}, {and} \bibinfo{person}{Xuemin Lin}.} \bibinfo{year}{2022}\natexlab{}.
\newblock \showarticletitle{GPU-accelerated Proximity Graph Approximate Nearest Neighbor Search and Construction}. In \bibinfo{booktitle}{\emph{38th {IEEE} International Conference on Data Engineering, {ICDE} 2022, Kuala Lumpur, Malaysia, May 9-12, 2022}}. \bibinfo{publisher}{{IEEE}}, \bibinfo{pages}{552--564}.
\newblock


\bibitem[Zamani et~al\mbox{.}(2018)]%
        {multi-field-neural-ranking}
\bibfield{author}{\bibinfo{person}{Hamed Zamani}, \bibinfo{person}{Bhaskar Mitra}, \bibinfo{person}{Xia Song}, \bibinfo{person}{Nick Craswell}, {and} \bibinfo{person}{Saurabh Tiwary}.} \bibinfo{year}{2018}\natexlab{}.
\newblock \showarticletitle{{Neural Ranking Models with Multiple Document Fields}}. In \bibinfo{booktitle}{\emph{WSDM '18}}.
\newblock


\bibitem[Zhang et~al\mbox{.}(2023)]%
        {vbase}
\bibfield{author}{\bibinfo{person}{Qianxi Zhang}, \bibinfo{person}{Shuotao Xu}, \bibinfo{person}{Qi Chen}, \bibinfo{person}{Guoxin Sui}, \bibinfo{person}{Jiadong Xie}, \bibinfo{person}{Zhizhen Cai}, \bibinfo{person}{Yaoqi Chen}, \bibinfo{person}{Yinxuan He}, \bibinfo{person}{Yuqing Yang}, \bibinfo{person}{Fan Yang}, \bibinfo{person}{Mao Yang}, {and} \bibinfo{person}{Lidong Zhou}.} \bibinfo{year}{2023}\natexlab{}.
\newblock \showarticletitle{{VBASE}: Unifying Online Vector Similarity Search and Relational Queries via Relaxed Monotonicity}. In \bibinfo{booktitle}{\emph{17th USENIX Symposium on Operating Systems Design and Implementation (OSDI 23)}}. \bibinfo{publisher}{USENIX Association}, \bibinfo{address}{Boston, MA}, \bibinfo{pages}{377--395}.
\newblock
\showISBNx{978-1-939133-34-2}
\urldef\tempurl%
\url{https://www.usenix.org/conference/osdi23/presentation/zhang-qianxi}
\showURL{%
\tempurl}


\bibitem[Zhang et~al\mbox{.}(2024)]%
        {rummy}
\bibfield{author}{\bibinfo{person}{Zili Zhang}, \bibinfo{person}{Fangyue Liu}, \bibinfo{person}{Gang Huang}, \bibinfo{person}{Xuanzhe Liu}, {and} \bibinfo{person}{Xin Jin}.} \bibinfo{year}{2024}\natexlab{}.
\newblock \showarticletitle{Fast Vector Query Processing for Large Datasets Beyond $\{$GPU$\}$ Memory with Reordered Pipelining}. In \bibinfo{booktitle}{\emph{21st USENIX Symposium on Networked Systems Design and Implementation (NSDI 24)}}. \bibinfo{pages}{23--40}.
\newblock


\bibitem[Zhao et~al\mbox{.}(2020)]%
        {song}
\bibfield{author}{\bibinfo{person}{Weijie Zhao}, \bibinfo{person}{Shulong Tan}, {and} \bibinfo{person}{Ping Li}.} \bibinfo{year}{2020}\natexlab{}.
\newblock \showarticletitle{{SONG:} Approximate Nearest Neighbor Search on {GPU}}. In \bibinfo{booktitle}{\emph{36th {IEEE} International Conference on Data Engineering, {ICDE} 2020, Dallas, TX, USA, April 20-24, 2020}}. \bibinfo{publisher}{{IEEE}}, \bibinfo{pages}{1033--1044}.
\newblock


\end{thebibliography}


\end{document}